\documentclass[12pt]{article}
\usepackage{setspace} 
\usepackage[utf8]{inputenc}
\usepackage[english]{babel}
\usepackage[dvipsnames]{xcolor}
\usepackage[]{appendix}
\usepackage{booktabs}
\usepackage{tabularx}
\usepackage{caption}
\usepackage{array}
\usepackage{float}
\usepackage{booktabs}

\usepackage{tikz}
\usetikzlibrary{arrows.meta, positioning, shapes.geometric, fit, backgrounds}

\tikzstyle{startstop} = [rectangle, rounded corners, minimum width=3.5cm, minimum height=1cm, text centered, draw=black, fill=blue!10]
\tikzstyle{decision} = [diamond, aspect=2, text centered, draw=black, fill=green!20, inner sep=1pt]
\tikzstyle{process} = [rectangle, minimum width=3.5cm, minimum height=1cm, text centered, draw=black, fill=blue!20]
\tikzstyle{arrow} = [thick,->,>=stealth]

\usepackage{subcaption}
\usepackage{graphicx}
\usetikzlibrary{arrows.meta}

\makeatletter
  \newcommand*\l@authors{\@dottedtocline{1}{0pt}{0pt}}
\makeatother

\usepackage[a4paper,top=2cm,bottom=2cm,left=2cm,right=2cm,marginparwidth=2cm]{geometry}

\usepackage[parfill]{parskip}

\usepackage{graphicx}
\usepackage[colorlinks=true, allcolors=blue]{hyperref}
\usepackage{pdflscape}
\usepackage{rotating}
\usepackage{multirow}
\usepackage{caption}
\usepackage{subcaption}
\usepackage[flushleft]{threeparttable}
\usepackage{multicol}
\usepackage{wrapfig}
\renewcommand{\arraystretch}{1.5}
\usepackage{amsmath}
\usepackage{amssymb}

\usepackage{amsthm}

\newtheorem{theorem}{Theorem}

\theoremstyle{definition}

\newtheorem{assumption}{Assumption}
\theoremstyle{definition}

\newtheorem{proposition}{Proposition}
\theoremstyle{definition}

\theoremstyle{definition}

\usepackage{enumerate}

\usepackage{comment}

\usepackage{booktabs}
\usepackage{makecell}
\usepackage{float}

\usepackage{pifont}
\usepackage{placeins}

\usepackage[]{natbib}
\setcitestyle{authoryear,open={(},close={)}}
\title{Testing Common Causality of Covariates: Evidence from Private-Sector Ban-the-Box Policies}
\author{Sunny Karim\thanks{Carleton University, Sunny.Karim@cmail.carleton.ca } }
\date{\today}

\begin{document}
\maketitle

\begin{abstract}
When including covariates in Difference-in-Differences, researchers must consider not only \textit{which} covariates to include, but also \textit{how} to model them. The Common Causality of Covariates (CCC) assumption requires that covariate coefficients in the untreated potential-outcome model remain constant across regions and periods. Violations can bias DID estimators even when the relevant covariates are correctly included. I develop a CCC test based on Wald tests from an additive model that uses untreated observations in pre-treatment periods to assess whether covariate coefficients vary across regions, periods, or both. Motivated by this specification problem, I show theoretically that well-specified Integrated DID estimators are consistent, whereas overly restrictive models can introduce bias and, under conditional homoskedasticity, unnecessarily flexible models can reduce efficiency. Monte Carlo simulations show that the CCC test has good size and power. An application to private-sector Ban-the-Box policies finds variation across both regions and periods, with positive estimated effects on wages and occupational quality among employed private-sector workers.
\end{abstract}

\section{Introduction}

In many empirical settings, Difference-in-differences (DID) is widely used to estimate the Average Treatment Effect on the Treated (ATT) when the treatment is not randomly assigned. However, this approach for identifying the ATT depends on the \textit{parallel trends assumption}, which states that, in the absence of treatment, the average outcome of both the treated and control regions would have evolved in parallel. To improve the plausibility of parallel trends, researchers often include covariates in DID to account for observable characteristics that may differ across regions and be related to the outcome of interest. When covariates are included, identification instead relies on the conditional parallel trends assumption, which states that the untreated potential outcomes of the treated and untreated regions would have evolved similarly over time, conditional on the observed characteristics. This paper first establishes how alternative restrictions on covariate effects across regions and over time affect the bias and efficiency of difference-in-differences estimators, and then develops a formal statistical test for assessing which restrictions are supported by the data. To motivate these contributions, it is useful to first consider the challenges that arise when covariates are incorporated into DID models.

Care must be taken when incorporating covariates into DID models, as both the selection and the way they are included in the model can affect identification. In particular, researchers should not include \textit{bad controls}, or covariates that are affected by treatment. This is why researchers often choose time-invariant covariates, which are predetermined and less susceptible to being affected by treatment. Including time-varying covariates in DID analyses can make the analysis more complicated. For instance, \cite{caetano2024difference} showed that the standard two-way fixed effects (TWFE) regression can suffer from \textit{hidden linearity bias}, leading to biased ATT estimates.

Choosing the appropriate covariates for DID is only one part of the challenge. Researchers must also decide how to incorporate these covariates into the estimating equation. \cite{karim2024good} formalize this problem by using the \textit{common causality of covariates (CCC)} assumption, which is implicit in the standard TWFE and in a few modern difference-in-differences estimators, such as the Callaway and Sant’Anna DID (CS-DID) estimator. In its strongest form, CCC requires that the causal effect of a covariate on untreated potential outcomes be stable across regions and over time. A CCC violation occurs when the coefficient of a covariate varies by region, by period, or by the interaction between region and period.

In this paper, I examine cases where the CCC assumption is likely to be violated. When CCC is violated, regression-based approaches such as TWFE can be modified by allowing covariate coefficients to vary across regions or periods through interactions with region and/or period dummies. However, if the imposed coefficient structure is more restrictive than the underlying pattern of the CCC violation, the resulting TWFE specification is misspecified and may result in biased estimates. In contrast, DID estimators based on propensity scores, such as Inverse Probability Weighting \citep{abadie2005semiparametric} and Doubly Robust DID \citep{sant2020doubly}, are generally unable to accommodate CCC violations \citep{karim2024good}. This limitation can also extend to staggered-adoption estimators such as CS-DID, which relies on these propensity-score-based methods to estimate the ATT. 

\cite{karim2024good} show that TWFE identifies the ATT only when it does not vary with the covariates. By contrast, CS-DID can identify the ATT even when it varies with the covariates, though it still requires the covariate distribution to remain stable over time. These concerns motivate the Integrated DID (DID-INT) estimator, which can identify the ATT under CCC violations, while allowing treatment effects to vary with the covariates and permitting the covariate distribution to change over time. The DID-INT estimator relies on a weaker assumption than conditional parallel trends, which they refer to as the integrated parallel trends (IPT) assumption. Unlike conditional parallel trends, the IPT requires parallel trends to hold only after the treated and control groups are evaluated at the same covariate distribution of the treated region and period. \cite{karim2024good} further show that violations of CCC can generally undermine conditional parallel trends. Although conditional parallel trends may continue to hold under CCC violations, this requires the strong necessary and sufficient condition that the pre-post differences in covariate effects are the same in the treated and control regions. 

The DID-INT estimator can be implemented using four alternative covariate specifications, depending on the type of CCC assumption that is violated. When covariate effects remain constant across regions and periods, \cite{karim2024good} recommend using the homogeneous specification. When the effects remain constant across regions but vary over time, they recommend the time-varying specification, in which the covariates are interacted with time indicators. Conversely, when the effects remain constant over time but vary across regions, they recommend the region-varying specification, in which the covariates are interacted with region indicators. Finally, when covariate effects vary across both regions and periods, they recommend the two-way specification, in which the covariates are interacted with both region and time indicators. Throughout the paper, the correct DID-INT specification refers to the specification that imposes the appropriate structure on the covariate effects for the corresponding IPT assumption.

Through theoretical proofs, I show that when the effect of the covariates is homogeneous, all four DID-INT specifications are consistent. When covariate effects vary only across regions, the region-varying and two-way DID-INT estimators remain consistent because the region-varying specification is nested within the two-way specification, whereas the homogeneous and time-varying specifications are generally inconsistent since they impose incorrect restrictions on the covariate effects. Analogously, when covariate effects vary only over time, the time-varying and two-way DID-INT estimators are consistent, while the homogeneous and region-varying specifications are generally inconsistent. Finally, when covariate effects vary across both regions and periods, only the two-way DID-INT estimator is consistent.

Among the consistent estimators, I further show that under conditional homoskedasticity, correctly imposing the more restrictive covariate structure improves efficiency relative to the two-way DID-INT estimator. However, under heteroskedasticity, the variance ranking between the correctly specified restrictive estimator and the two-way DID-INT estimator is generally ambiguous. More broadly, allowing covariate effects to vary across regions and periods can substantially increase the number of parameters that must be estimated as the number of regions, periods and covariates grows, which can substantially increase computational burden. These considerations motivate the need for a formal statistical test which can serve as a preliminary diagnostic to determine the correct covariate coefficient structure supported by the data before implementing DID estimators more generally. 

To address this gap, I introduce the CCC test, a formal Wald-test-based method to assess whether covariate effects vary across regions, over time, or both. The CCC test proceeds in two stages. In the first stage, I test for two-way CCC violations, where the covariate effects vary by both region and period. If the test fails to reject the null hypothesis, the evidence supports a homogeneous covariate specification with a common slope across regions and periods. Therefore, the covariates can be incorporated without interactions. However, if the null is rejected, the procedure proceeds to the second stage, where region- and time-varying violations are tested separately. If the region-varying null is rejected but the time-varying null is not, the evidence favors a structure in which the covariate coefficients vary across regions. In this case, the covariates should be interacted with region indicators. Conversely, if the time-varying null is rejected but the region-varying null is not, the evidence supports a structure in which the covariate coefficients vary across periods. In this case, the covariates should be interacted with period indicators. If both null hypotheses are rejected, the evidence supports a structure in which the covariate coefficients vary across both regions and periods. In this case, the covariates should be interacted with both region and period indicators. In this paper, I use the CCC test to guide the choice among alternative DID-INT specifications in both the Monte Carlo simulations and the empirical application. Although the analyses focus on the DID-INT estimator, I recommend using the CCC test more broadly as a preliminary diagnostic before applying other DID methods with covariates, particularly as a robustness check. Next, I evaluate the statistical properties of the CCC test using a Monte Carlo simulation study and show that it exhibits good size and power. I further show that using the CCC test to inform the choice of DID-INT specification can improve efficiency under conditional homoskedasticity and, in the heteroskedastic design considered in my simulations, can also yield efficiency gains. 

To demonstrate the utility of the CCC test, I explore a new empirical question: whether private-sector Ban-the-Box policies improve job quality for those already employed in the U.S. I combine American Community Survey (ACS) data from 2010 to 2023, after adjusting for the income reference period, with the Bureau of Labor Statistics (BLS) O*NET Job Zones ranking data, allowing me to examine whether these policies affect the quality of jobs held by employed individuals. I estimate the policy effects using TWFE, CS-DID, and DID-INT methods, where the appropriate DID-INT specification is based on the results of the CCC test. This approach helps me compare how the estimated effects vary across different methods, especially between conventional DID approaches and those that better handle staggered adoption and CCC violations. 

I first examine whether Ban-the-Box policies in the private sector led to higher wages for employed individuals, and then investigate whether this wage increase is associated with a shift into higher-ranked occupations. Both the TWFE and CS-DID methods show a positive and statistically significant effect of these policies on wages. The CS-DID approach is more suitable here because the policies are implemented in a staggered fashion. Nonetheless, the CCC test indicates a two-way CCC violation and selects the two-way DID-INT specification. Using the CCC-selected specification, the estimated effect on wages is positive and statistically significant at the 5\% level. For the High Job Zones outcome, the TWFE and CS-DID estimates are statistically insignificant, whereas the CCC-selected DID-INT estimate is positive and statistically significant at the 10\% level. This suggests that Ban-the-Box policies may improve access to higher-quality jobs. Additional analyses using both individuals who are unemployed and those employed in the private sector show a small and statistically significant decline in private-sector employment and no significant effect on the unconditional probability of employment in a high Job Zone, while the positive effect on occupational quality remains when using the continuous Job Zone measure. Subgroup analyses provide little evidence that these effects are concentrated within a particular racial or ethnic group.

The remainder of the paper proceeds as follows. Section \ref{section: assumptions} introduces the CCC assumptions and the additional assumptions required for identification. Section \ref{section: ccctestsection} derives the main theoretical results on bias and efficiency across alternative DID-INT specifications. It also presents the CCC test introduced in this paper and describes its two stages in more detail. Section \ref{section: mc} presents the design and results of the Monte Carlo simulation study used to evaluate the statistical properties of the CCC test. Lastly, Section \ref{section: Empiricalexercise} presents the results of the empirical application and the estimated effects of private-sector Ban-the-Box policies on wages and job quality.

\section{Common Causality of Covariates}
\label{section: assumptions}

In this section, I introduce the notation, the setup and the assumptions used throughout the paper. Assume that there are $s= \{1,2,\cdots, S\}$ regions observed over $t = \{1,2,\cdots, T\}$ periods in a staggered adoption setting, where different regions first receive treatment in different periods. For a treated region $s$, let $s'$ be a relevant control group in period $t$, which can either be never-treated or not-yet-treated. Let $S^\prime$ denote the total number of relevant control regions for region $s$ in period $t$. Finally, $t_s$ denotes the first period in which region $s$ is treated, and $t_s - 1$ denotes the period right before the region is treated.

\begin{assumption}[Random Sampling]
\label{as: randomsampling}
For each region-period cell $(s,t)$, the observations
\begin{equation}
    \left\{Y_{i,s,t},X_{i,s,t}\right\}_{i=1}^{N_{s,t}}
\end{equation}
are independently and identically distributed draws from the corresponding population. The analysis below maintains this sampling framework; extensions of the CCC test to clustered dependence are discussed in Section \ref{section: ccctestsection} but are not developed in the current paper.
\end{assumption}

\begin{assumption}[Covariates Unaffected by Treatment]
\label{as: covariateunaffected}
Treatment does not affect the covariates. That is, for all individuals, regions, periods, and covariates,
\begin{equation}
    X^k_{i,s,t}(1)=X^k_{i,s,t}(0)=X^k_{i,s,t},
    \qquad k=1,\ldots,K.
\end{equation}
\end{assumption}
Assumption \ref{as: covariateunaffected} implies that $\mathbb{E}\biggr[X_{t}(0) \mid D_s = 1 \biggr] = \mathbb{E}\biggr[X_{t}(1) \mid D_s = 1 \biggr]$, and rules out \textit{bad controls}, or controls which are affected by treatment. For simplicity, I omit the $(0)$ since the covariate is unaffected by treatment due to Assumption \ref{as: covariateunaffected}.

\begin{assumption}[No Anticipation]
\label{as: noanticipation}
Treatment does not affect the potential outcome prior to treatment. That is,
\begin{equation}
    Y_{i,s,t}(1)=Y_{i,s,t}(0)
    \qquad \text{a.s. for all } t<t_s.
\end{equation}
\end{assumption}

\begin{assumption}[Linearity]
\label{as: linearity}
Within each region-period cell, the conditional mean of the untreated potential outcome is a linear combination of the covariates. Therefore, the untreated potential outcome of individual $i$ in region $s$ at period $t$ is given by:
\begin{equation}
    Y_{i,s,t}(0)
    =
    \theta^{\circ}_{s,t}
    +
    \sum_{k=1}^{K}
    \gamma^{\circ}_{k,s,t}X^k_{i,s,t}
    +
    \epsilon_{i,s,t},
\end{equation}
where
\begin{equation}
    \mathbb{E}
    \left[
    \epsilon_{i,s,t}
    \mid
    X_{i,s,t},S=s,T=t
    \right]
    =0.
\end{equation}
\end{assumption}

The parameters $\gamma^{\circ}_{k,s,t}$ characterize the relationship between covariate $k$ and the untreated potential outcome in region $s$ and period $t$. The CCC assumptions impose alternative restrictions on how these coefficients may vary across regions and periods.

\begin{assumption}[Two-way Common Causality of Covariates]
\label{assumption: twowayCCC}
For each covariate $k$, the covariate effect is constant across regions and periods:
\begin{equation}
    \gamma^{\circ}_{k,s,t}
    =
    \gamma^{\circ}_{k,s',t'}
    \qquad
    \forall s,s',t,t'.
\end{equation}
\end{assumption}

Under Assumption \ref{assumption: twowayCCC}, each covariate has a common coefficient across all region-period cells.

\begin{assumption}[Region-invariant Common Causality of Covariates]
\label{as2: regioninvariantCCC}
For each covariate $k$, the covariate effect is common across regions within a given period, but may vary across periods:
\begin{equation}
    \gamma^{\circ}_{k,s,t}
    =
    \gamma^{\circ}_{k,s',t}
    \qquad
    \forall s,s',t.
\end{equation}
\end{assumption}

Under Assumption \ref{as2: regioninvariantCCC}, covariate effects may change over time, but within each period they are common across regions.

\begin{assumption}[Time-invariant Common Causality of Covariates]
\label{as2: timeinvariantCCC}
For each covariate $k$, the covariate effect is constant over time within a given region, but may vary across regions:
\begin{equation}
    \gamma^{\circ}_{k,s,t}
    =
    \gamma^{\circ}_{k,s,t'}
    \qquad
    \forall s,t,t'.
\end{equation}
\end{assumption}

Under Assumption \ref{as2: timeinvariantCCC}, covariate effects may differ across regions, but remain constant over time within each region. The identifying assumptions for DID can then be stated in terms of how untreated potential outcomes evolve across treated and control regions. Conditional Parallel Trends requires this restriction to hold pointwise at each value of the covariates.

\begin{assumption}[Conditional Parallel Trends]
\label{as: cpt}
For every treated region $s$, post-treatment period $t\geq t_s$, valid control region
$s'$, and covariate value $x$,
\begin{equation}
\begin{aligned}
&\mathbb{E}
\left[
Y_{i,s,t}(0)
\mid
S=s,X_{i,s,t}=x
\right]
-
\mathbb{E}
\left[
Y_{i,s,t_s-1}(0)
\mid
S=s,X_{i,s,t_s-1}=x
\right]
\\
={}&
\mathbb{E}
\left[
Y_{i,s',t}(0)
\mid
S=s',X_{i,s',t}=x
\right]
-
\mathbb{E}
\left[
Y_{i,s',t_s-1}(0)
\mid
S=s',X_{i,s',t_s-1}=x
\right].
\end{aligned}
\end{equation}
\end{assumption}

Conditional Parallel Trends therefore requires the untreated outcome trends of treated and control regions to be equal at every value of the covariates. The DID-INT estimator relies instead on the weaker Integrated Parallel Trends assumption, which requires these trends to be equal only after both regions are evaluated over the same covariate distribution.

\begin{assumption}[Integrated Parallel Trends]
\label{as: ipt}
For every treated region $s$, post-treatment period $t\geq t_s$, and valid control region
$s'$,
\begin{equation}
\begin{aligned}
&
\int
\Bigg[
\mathbb{E}
\left[
Y_{i,s,t}(0)
\mid
S=s,X=x
\right]
-
\mathbb{E}
\left[
Y_{i,s,t_s-1}(0)
\mid
S=s,X=x
\right]
\Bigg]
dF_{X\mid S=s,T=t}(x)
\\
={}&
\int
\Bigg[
\mathbb{E}
\left[
Y_{i,s',t}(0)
\mid
S=s',X=x
\right]
-
\mathbb{E}
\left[
Y_{i,s',t_s-1}(0)
\mid
S=s',X=x
\right]
\Bigg]
dF_{X\mid S=s,T=t}(x).
\end{aligned}
\end{equation}
\end{assumption}

Unlike Conditional Parallel Trends, Integrated Parallel Trends does not require equality of untreated outcome trends at every value of $X_{i,s,t}$. Instead, it requires the treated and control trends to be equal after both are evaluated over the covariate distribution of the treated region in period $t$. Consequently, Conditional Parallel Trends implies Integrated Parallel Trends, whereas Integrated Parallel Trends may hold even when Conditional Parallel Trends is violated \citep{karim2024good}.

\section{A Statistical Test for CCC violations}
\label{section: ccctestsection}

This section motivates the need for the CCC test as a formal statistical test for selecting the appropriate DID-INT specification. It first establishes the statistical properties of the alternative DID-INT specifications under different covariate coefficient structures. When the true specification is region-varying, the region-varying and the two-way DID-INT estimates remain consistent, while the time-varying DID-INT estimates are generally inconsistent because they impose an incorrect restriction on the covariate effects. Among the correctly specified estimators, the region-varying DID-INT estimator is more efficient than the two-way DID-INT estimator under conditional homoskedasticity. Conversely, if the true model is the time-varying DID-INT, both the time-varying and two-way DID-INT estimators are consistent, while the region-varying DID-INT estimator is generally inconsistent. However, under conditional homoskedasticity, the time-varying estimator is more efficient than the two-way DID-INT estimator. When the two-way model is true, neither homogeneous, region-varying, nor time-varying estimators can identify the ATT. However, under heteroskedasticity, the variance ranking between the correctly specified restrictive DID-INT estimator and the two-way DID-INT estimator is generally ambiguous.

In empirical applications, the true model that describes how covariate effects change across regions and over time is usually unknown. Therefore, researchers require a systematic method to identify which DID-INT specification offers enough flexibility for the available data. A further practical consideration is that the two-way DID-INT model becomes highly parameterized as the number of regions, periods, and covariates increases. For example, with 6 regions, 6 periods, and 6 covariates, the model interacts each covariate with region-by-time indicators, resulting in 216 covariate terms. In contrast, the homogeneous model only produces six covariate terms. Consequently, the two-way DID-INT estimator may require somewhat greater computational power, particularly in larger datasets. More broadly, other estimators such as TWFE may also require more flexible covariate specifications to identify the ATT under CCC violations. The resulting increase in the number of covariates can impose additional memory requirements, making it very resource-intensive. As a result, the CCC test is valuable because it gives researchers a guideline for selecting a more efficient, less computationally demanding specification.

\subsection{Statistical Properties of the Alternative DID-INT Specifications}
\label{sec: theory}

This subsection examines the statistical properties of the alternative DID-INT specifications under various true CCC violations. First, consider the case where the homogeneous model is true. In this context, the covariate effects are constant across regions and periods, meaning the two-way CCC assumption is satisfied. Therefore, the true untreated potential outcome model is given by:
\begin{equation}
    Y^{\circ}_{i,s,t}(0) = \theta^{\circ}_{s,t} + \sum_{k}\gamma^{\circ}_{k} X^k_{i,s,t} + \epsilon_{i,s,t}. 
\end{equation}
Here, $\gamma^{\circ}_{k}$ does not vary by regions or periods. Under this data-generating process (DGP), all DID-INT specifications contain the true untreated potential outcome model. Let $\Gamma_H$ be the coefficient structure permitted by the homogeneous DID-INT model, $\Gamma_R$ for the region-varying DID-INT model, $\Gamma_T$ for the time-varying model, and $\Gamma_{TW}$ for the two-way model.
\begin{align}
\Gamma_H
&=
\left\{
\gamma_{k,s,t} :
\gamma_{k,s,t} = \gamma_k
\quad \forall k,s,t
\right\}, \\
\Gamma_R
&=
\left\{
\gamma_{k,s,t} :
\gamma_{k,s,t} = \gamma_{k,s}
\quad \forall k,s,t
\right\}, \\
\Gamma_T
&=
\left\{
\gamma_{k,s,t} :
\gamma_{k,s,t} = \gamma_{k,t}
\quad \forall k,s,t
\right\}, \\
\Gamma_{TW}
&=
\left\{
\gamma_{k,s,t} :
\gamma_{k,s,t}
\text{ is unrestricted across regions and periods}
\right\}.
\end{align}
Also, $\gamma^{*}_{k,s,t}$ denotes the probability limit of the estimated coefficient of the $kth$ covariate for region $s$ in period $t$ from the two-way DID-INT specification.

\begin{theorem}[Nesting of alternative DID-INT specifications] The homogeneous, region-varying and time-varying DID-INT specifications are nested within the two-way DID-INT specification.
\label{theorem: nesting}
 \begin{equation}
     \Gamma_H \subset \Gamma_R \subset \Gamma_{TW}, \qquad \Gamma_H \subset \Gamma_T \subset \Gamma_{TW}.
 \end{equation}   
\end{theorem}
A proof of Theorem \ref{theorem: nesting} is shown in Appendix \ref{app3: nesting}.

\begin{theorem}[Identification under alternative DID-INT specifications under conditional parallel trends or integrated parallel trends.] Suppose Assumption \ref{assumption: twowayCCC} holds, such that the homogeneous DID-INT is the true specification. Under Assumptions \ref{as: randomsampling}, \ref{as: covariateunaffected}, \ref{as: noanticipation}, \ref{as: linearity} and either Assumption \ref{as: cpt} or Assumption \ref{as: ipt}:
\label{theorem: equivalencehom}
\begin{equation}
    \tau^{{H}^*}_{s,t} = \tau^{{R}^*}_{s,t} = \tau^{{T}^*}_{s,t} = \tau^{{TW}^*}_{s,t} = ATT(s,t). 
\end{equation}
\end{theorem}
Here, $\tau^{{H}^*}_{s,t}$, $\tau^{{R}^*}_{s,t}$, $\tau^{{T}^*}_{s,t}$ and $\tau^{{TW}^*}_{s,t}$ denote the probability limit of the estimates of the $ATT(s,t)$ from the homogeneous, region-varying, time-varying and two-way  DID-INT specifications, respectively.  A proof of Theorem \ref{theorem: equivalencehom} is shown in \cite{karim2024good}. Next, when Assumption \ref{assumption: twowayCCC} holds, the homogeneous DID-INT model is shown to be more efficient than the region-varying DID-INT model and, consequently, more efficient than the two-way DID-INT model. Let $\widehat{\boldsymbol{\gamma}}^{TW}$ denote the $(KST) \times 1$ vector of covariate coefficient estimates from the two-way DID-INT specification. \begin{equation}
    \widehat{\boldsymbol{\gamma}}^{TW}
    =
    \left(
    \widehat{\gamma}^{TW}_{1,1,1},
    \ldots,
    \widehat{\gamma}^{TW}_{1,S,T},
    \ldots,
    \widehat{\gamma}^{TW}_{K,1,1},
    \ldots,
    \widehat{\gamma}^{TW}_{K,S,T}
    \right)'
\end{equation}
Similarly, let $\widehat{\boldsymbol{\gamma}}^{R}$, $\widehat{\boldsymbol{\gamma}}^{T}$ and $\widehat{\boldsymbol{\gamma}}^{H}$ denote the covariate coefficient estimates from the region-varying, time-varying and homogeneous DID-INT, respectively. These vectors have dimensions $(KS) \times 1$, $(KT) \times 1$ and $(K) \times 1$, respectively. 

\begin{theorem}[Efficiency of the homogeneous DID-INT estimator under two-way CCC]
\label{theorem: hom_efficiency}

Suppose Assumption \ref{assumption: twowayCCC} holds. Under the OLS regularity conditions and homoskedastic error terms, the conditional variance of the homogeneous DID-INT estimator of $ATT(s,t)$ is weakly smaller than that of the region-varying, time-varying, and the two-way DID-INT estimators. The conditional variance of the region-varying and the time-varying DID-INT estimators is weakly smaller than that of the two-way DID-INT estimator.
\label{theorem: efficiency}
\begin{equation}
\begin{gathered}
    \operatorname{Var}\left(\widehat{\tau}^{H}_{s,t} \mid \boldsymbol{X}^{TW}\right) \leq \operatorname{Var}\left(\widehat{\tau}^{R}_{s,t} \mid \boldsymbol{X}^{TW}\right) \leq \operatorname{Var}\left(\widehat{\tau}^{TW}_{s,t} \mid \boldsymbol{X}^{TW} \right), \\ \quad            \operatorname{Var}\left(\widehat{\tau}^{H}_{s,t} \mid \boldsymbol{X}^{TW}\right) \leq \operatorname{Var}\left(\widehat{\tau}^{T}_{s,t} \mid \boldsymbol{X}^{TW}\right) \leq \operatorname{Var}\left(\widehat{\tau}^{TW}_{s,t} \mid \boldsymbol{X}^{TW} \right).
\end{gathered}
\end{equation}
\end{theorem}

\begin{proof}[Proof of Theorem \ref{theorem: hom_efficiency}]

The proof begins by showing that the conditional variance of the homogeneous DID-INT estimator is weakly smaller than that of the region-varying DID-INT estimator under homoskedasticity. Let $\boldsymbol{X}^{TW}$ denote the $N\times(ST+KST)$ matrix of regressors for the two-way DID-INT first-stage regression:
\begin{equation}
    \boldsymbol{X}^{TW}
    =
    \begin{pmatrix}
        (\boldsymbol{X}^{TW}_{1})' \\
        (\boldsymbol{X}^{TW}_{2})' \\
        \vdots \\
        (\boldsymbol{X}^{TW}_{N})'
    \end{pmatrix}.
\end{equation}
where $\boldsymbol{X}_i^{TW}$ is the observation-specific vector of regressors for the two-way DID-INT specification of dimension $(ST + KST) \times 1$. Here, 
\begin{equation}
    Z_{i,s,t}
    =
    1\{S_i=s,T_i=t\}.
\end{equation}
Therefore, $\boldsymbol{X}_i^{TW}$ can be expressed as:
\begin{equation}
\begin{aligned}
    \boldsymbol{X}_i^{TW}
    =
    \big(&
    Z_{i,1,1},\ldots,Z_{i,S,1},
    Z_{i,1,2},\ldots,Z_{i,S,T},
    \\
    &
    Z_{i,1,1}X_{1,i},\ldots,
    Z_{i,1,1}X_{K,i},
    \\
    &
    \ldots,
    \\
    &
    Z_{i,S,1}X_{1,i},\ldots,
    Z_{i,S,1}X_{K,i},
    \\
    &
    Z_{i,1,2}X_{1,i},\ldots,
    Z_{i,1,2}X_{K,i},
    \\
    &
    \ldots,
    \\
    &
    Z_{i,S,T}X_{1,i},\ldots,
    Z_{i,S,T}X_{K,i}
    \big)'.
\end{aligned}
\end{equation}
Let $\boldsymbol{X}^{R}$ denote the $N\times(ST+KS)$ matrix of regressors for the region-varying DID-INT first-stage regression.
\begin{equation}
    \boldsymbol{X}^{R}
    =
    \begin{pmatrix}
        (\boldsymbol{X}^{R}_{1})' \\
        (\boldsymbol{X}^{R}_{2})' \\
        \vdots \\
        (\boldsymbol{X}^{R}_{N})'
    \end{pmatrix},
\end{equation}
Therefore, $\boldsymbol{X_i^R}$ vector of dimension $(ST + KS) \times 1$ can be denoted as:
\begin{equation}
\begin{aligned}
    \boldsymbol{X}_i^R
    =
    \big(&
    Z_{i,1,1},\ldots,Z_{i,S,1},
    Z_{i,1,2},\ldots,Z_{i,S,T},
    \\
    &
    1\{S_i=1\}X_{1,i},\ldots,
    1\{S_i=1\}X_{K,i},
    \\
    &
    \ldots,
    \\
    &
    1\{S_i=S\}X_{1,i},\ldots,
    1\{S_i=S\}X_{K,i}
    \big)'.
\end{aligned}
\end{equation}

Similarly, the observation-specific vector of regressors for the homogeneous model $\boldsymbol{X_i^H}$ of dimension $(ST + K) \times 1$ can be expressed as:
\begin{equation}
\begin{aligned}
    \boldsymbol{X}_i^H
    =
    \big(&
    Z_{i,1,1},\ldots,Z_{i,S,1},
    Z_{i,1,2},\ldots,Z_{i,S,T},
    X_{1,i},\ldots,
    X_{K,i}\big)'.
\end{aligned}
\end{equation}
Therefore, the regressor matrix for the homogeneous model $\boldsymbol{X}^{H}$ of dimension $N\times(ST+K)$ can be expressed as: 
\begin{equation}
    \boldsymbol{X}^{H}
    =
    \begin{pmatrix}
        (\boldsymbol{X}^{H}_{1})' \\
        (\boldsymbol{X}^{H}_{2})' \\
        \vdots \\
        (\boldsymbol{X}^{H}_{N})'
    \end{pmatrix},
\end{equation}

For every pair of nested DID-INT specifications, let $\boldsymbol{X^{C}}$ denote the regressor matrix of the constrained specification and $\boldsymbol{X^{U}}$ denote the regressor matrix of the corresponding unrestricted specification. For instance, when comparing the homogeneous and region-varying DID-INT specifications, the homogeneous model can be represented as a restricted version of the region-varying model under the two-way CCC restriction. Accordingly, the homogeneous specification is the constrained model, while the region-varying specification is the corresponding unrestricted model. 

Since the constrained and unrestricted regressor matrices have different dimensions, a mapping matrix $\boldsymbol{M}$ can be used to express the constrained specification in the parameter space of the unrestricted specification. When comparing the homogeneous and region-varying DID-INT specifications, the corresponding mapping matrix is: 
\begin{equation}
    \boldsymbol{M}
    =
    \begin{pmatrix}
        \boldsymbol{I}_{ST} & \boldsymbol{0} \\
        \boldsymbol{0} & \boldsymbol{C}
    \end{pmatrix},
\end{equation}
and,
\begin{equation}
    \boldsymbol{C}
    =
    \boldsymbol{1}_{S} \otimes \boldsymbol{I}_{K}.
\end{equation}
More generally, the relationship between the constrained and unrestricted regressor matrices can be written as:
\begin{gather}
\label{eq: generic_transformation}
\boldsymbol{X}^{C}
=
\boldsymbol{X}^{U}\boldsymbol{M},
\end{gather}
For the homogeneous and the region-varying comparison, this becomes:
\begin{gather}
\label{eq: transformation}
\boldsymbol{X}^H = \boldsymbol{X}^R\boldsymbol{M}.
\end{gather}
A short proof of Equation \eqref{eq: transformation} is shown in Appendix \ref{appendix: A}. Analogous nesting relationships hold for other nested DID-INT specifications.  This relationship allows a common conditioning set to be used throughout the theorem. Homoskedasticity implies:
\begin{gather}
    \operatorname{Var}(\boldsymbol{u} \mid \boldsymbol{X}^{TW}) = \sigma^2 \boldsymbol{I}_N.
\end{gather}
Here, $\boldsymbol{u} = (u_1,u_2,\cdots,u_N)'$ is the vector of population residuals of the DID-INT specification. Let the full coefficient vectors of the homogeneous and region-varying DID-INT specifications be
\begin{equation}
    \boldsymbol{\widehat{T}}^{H}
    =
    \begin{pmatrix}
        \boldsymbol{\widehat{\theta}}^{H} \\
        \boldsymbol{\widehat{\gamma}}^{H}
    \end{pmatrix},
    \qquad
    \boldsymbol{\widehat{T}}^{R}
    =
    \begin{pmatrix}
        \boldsymbol{\widehat{\theta}}^{R} \\
        \boldsymbol{\widehat{\gamma}}^{R}
    \end{pmatrix}.
\end{equation}
Here, $\boldsymbol{\widehat{\theta}}^H = (\widehat{\theta}_{1,1}, \cdots, \widehat{\theta}_{S,1},\widehat{\theta}_{1,2},\cdots,\widehat{\theta}_{S,2}, \cdots,\widehat{\theta}_{1,T}, \cdots \widehat{\theta}_{S,T})'$ is a $(ST \times 1)$ vector containing the region-by-period intercepts of the homogeneous model, and $\boldsymbol{\widehat{\gamma}}^H = (\widehat{\gamma}_{1}, \cdots, \widehat{\gamma}_{K})'$ is a $(K \times 1)$ vector of the coefficients of the $K$ covariates. Similarly, $\boldsymbol{\widehat{\theta}}^R$ is a $(ST \times 1)$ vector containing the region-by-period intercepts of the region-varying model, and $\boldsymbol{\widehat{\gamma}}^R = (\widehat{\gamma}_{1,1}, \cdots, \widehat{\gamma}_{K,1}, \widehat{\gamma}_{1,2},$ $ \cdots, \widehat{\gamma}_{K,2}, \cdots, \widehat{\gamma}_{1,S}, \cdots, \widehat{\gamma}_{K,S})'$ is a $(KS \times 1)$ vector of coefficients of the $K$ covariates across $S$ regions. Since these are finite-sample estimates rather than true population parameters, $\boldsymbol{\widehat{\theta}}^H$ and $\boldsymbol{\widehat{\theta}}^R$ are not identical. Since the first-stage DID-INT specifications are estimated using OLS, the regressors are expressed in matrix form and additional notation is introduced to facilitate the analysis.

The estimate of the $ATT(s,t)$ from the region-varying model is derived as a linear combination of the region-and-period intercepts and involves an integral over the covariate coefficients. Here, $ATT(s,t)$ denotes the region-period average treatment effect on the treated for region $s$ in period $t$. This differs from the cohort-period $ATT(g,t)$ defined in \cite{callaway2021difference}, where $g$ indexes treatment cohorts defined by the period in which units are first treated. Each cohort $g$ can include multiple regions $s$ if all regions are first treated in the same period. The benefits of doing so are shown in \citet*{austin2026policy}. Let $s'$ be a relevant control group for group $s$ that is never treated or not yet treated. Region $s$ in period $t$ may have multiple relevant control regions, and their contributions are aggregated using a weighted average in the estimation of the $ATT(s,t)$. Let the total number of control groups for region $s$ be $S'$, and the corresponding weights for region $s'$ are given by $\omega_{s'}$, which sum to one. So the $ATT(s,t)$ for the region-varying DID-INT model can be written as:
\begin{gather}
    \widehat{\tau}^R_{s,t} = \int \biggr[ \widehat{\theta}^R_{s,t} + \sum_{k=1}^K \widehat{\gamma}^R_{k,s} x^k \biggr]d\widehat{F}_{X|S=s, T=t}(x) - \int \biggr[ \widehat{\theta}^R_{s,t^s-1} + \sum_{k=1}^K \widehat{\gamma}^R_{k,s} x^k \biggr]d\widehat{F}_{X|S=s, T=t}(x) \nonumber \\  - \sum_{s'=1}^{S'} \omega_{s'} \biggr\{ \int \biggr[ \widehat{\theta}^R_{s',t} + \sum_{k=1}^K \widehat{\gamma}^R_{k,s'} x^k \biggr]d\widehat{F}_{X|S=s, T=t}(x) - \int \biggr[ \widehat{\theta}^R_{s',t^s-1} + \sum_{k=1}^K \widehat{\gamma}^R_{k,s'} x^k \biggr]d\widehat{F}_{X|S=s, T=t}(x)\biggr\}.
\end{gather}
Since the covariate coefficients are the same in period $t$ and $t^s-1$, the covariate terms cancel out. So the $ATT(s,t)$ can be simplified to:
\begin{gather}
    \widehat{\tau}^R_{s,t} =  \biggr( \widehat{\theta}^R_{s,t} -  \widehat{\theta}^R_{s,t^s-1} \biggr) - \sum_{s'=1}^{S'} \omega_{s'}\biggr( \widehat{\theta}^R_{s',t} -  \widehat{\theta}^R_{s',t^s-1} \biggr).
\end{gather}

For simplicity of notation, the $ATT(s,t)$ is expressed as a linear combination of the corresponding coefficient vector using a weighting vector. Therefore, the scalar $\widehat{\tau}^R_{s,t}$ can be derived from a vector multiplication of the weighting vector $\boldsymbol{W}$ and $\boldsymbol{\widehat{T}^R}$, where $\boldsymbol{W} = (1,-1,-\omega_{s_1'},\allowbreak
\omega_{s_2'},\allowbreak
\cdots,\allowbreak
\omega_{s_S'},\allowbreak
\omega_{s_S'},\allowbreak
0,\cdots,0)$ is a vector of dimension $(ST + KS)$.    
\begin{equation}
\label{equation: regvartransform}
    \widehat{\tau}^R_{s,t} = \boldsymbol{W}  \boldsymbol{\widehat{T}^R}.
\end{equation}
The variance of $\widehat{\tau}^R_{s,t}$ conditional on $\boldsymbol{X^{TW}}$ can be expressed as:
\begin{equation}
    \operatorname{Var}(\widehat{\tau}^R_{s,t} \mid \boldsymbol{X}^{TW}) = \operatorname{Var}(\boldsymbol{W}  \boldsymbol{\widehat{T}^R} \mid \boldsymbol{X}^{TW}) = \boldsymbol{W} \operatorname{Var}( \boldsymbol{\widehat{T}^R}\mid \boldsymbol{X}^{TW})\boldsymbol{W}'.
\end{equation}
Similarly, the $ATT(s,t)$ for the homogeneous DID-INT model can be written as:
\begin{gather}
    \widehat{\tau}^H_{s,t} = \int \biggr[ \widehat{\theta}^H_{s,t} + \sum_{k=1}^K \widehat{\gamma}^H_{k} x^k \biggr]d\widehat{F}_{X|S=s, T=t}(x) - \int \biggr[ \widehat{\theta}^H_{s,t^s-1} + \sum_{k=1}^K \widehat{\gamma}^H_{k} x^k \biggr]d\widehat{F}_{X|S=s, T=t}(x) \nonumber \\  - \sum_{s'=1}^{S'} \omega_{s'} \biggr\{ \int \biggr[ \widehat{\theta}^H_{s',t} + \sum_{k=1}^K \widehat{\gamma}^H_{k} x^k \biggr]d\widehat{F}_{X|S=s, T=t}(x) - \int \biggr[ \widehat{\theta}^H_{s',t^s-1} + \sum_{k=1}^K \widehat{\gamma}^H_{k} x^k \biggr]d\widehat{F}_{X|S=s, T=t}(x)\biggr\}.
\end{gather}
Since the covariate coefficients are the same for all regions and periods, the covariate terms cancel out. So the $ATT(s,t)$ can be simplified to:
\begin{gather}
    \widehat{\tau}^H_{s,t} =  \biggr( \widehat{\theta}^H_{s,t} -  \widehat{\theta}^H_{s,t^s-1} \biggr) - \sum_{s'=1}^{S'} \omega_{s'}\biggr( \widehat{\theta}^H_{s',t} -  \widehat{\theta}^H_{s',t^s-1} \biggr).
\end{gather}
Since the vector $\boldsymbol{\mathbf{\widehat{T}}}^{H}$ is of dimension $(ST + K) \times 1$, it is non-conformable with vector $\boldsymbol{W}$ and $\boldsymbol{\widehat{T}^R}$. A mapping matrix $\boldsymbol{M}$ is therefore used to map the vector $\boldsymbol{\mathbf{\widehat{T}}}^{H}$ into the coefficient space of $\boldsymbol{\mathbf{\widehat{T}}}^{R}$, such that the vector $\boldsymbol{M}\boldsymbol{\mathbf{\widehat{T}}}^{H}$ is of dimension $(ST + KS) \times 1$. Therefore, the scalar $\widehat{\tau}^H_{s,t}$ can be derived as:
\begin{equation}
    \widehat{\tau}^H_{s,t} = \boldsymbol{W}\boldsymbol{M}  \boldsymbol{\widehat{T}^H},
\end{equation}
and its conditional variance can be expressed as:
\begin{equation}
    \operatorname{Var}(\widehat{\tau}^H_{s,t}\mid \boldsymbol{X}^{TW}) =  \boldsymbol{W}\boldsymbol{M} \operatorname{Var}( \boldsymbol{\widehat{T}^H}\mid \boldsymbol{X}^{TW})\boldsymbol{M}'\boldsymbol{W}'.
\end{equation}
The mapping matrix $\boldsymbol{M}$ and weighting vector $\boldsymbol{W}$ are defined for the particular pair of nested DID-INT specifications being compared and therefore change accordingly across the comparisons below. The analysis continues to condition on $\boldsymbol{X^{TW}}$ due to the mapping established in Equation \eqref{eq: generic_transformation}. Since the homogeneous DID-INT specification is contained within the region-varying DID-INT specification when Assumption \ref{assumption: twowayCCC} holds, Hausman equality \cite[pg. 202, Equation 8.18]{hansen2022econometrics} can be used to show that the homogeneous DID-INT first-stage regression is more efficient than the region-varying DID-INT specification under homoskedasticity. Hausman equality implies:
\begin{gather}
    \operatorname{Var}(\boldsymbol{\widehat{T}}^R \mid \boldsymbol{X}^{TW}) - \operatorname{Var}(\boldsymbol{\boldsymbol{M}\widehat{T}}^H \mid \boldsymbol{X}^{TW}) \geq 0. \\
    \operatorname{Var}(\boldsymbol{\widehat{T}}^R \mid \boldsymbol{X}^{TW}) - \boldsymbol{M}\operatorname{Var}(\boldsymbol{\widehat{T}}^H \mid \boldsymbol{X}^{TW})\boldsymbol{M}' \geq 0.\\
    \operatorname{Var}(\boldsymbol{\widehat{T}}^R \mid \boldsymbol{X}^{TW}) \geq \boldsymbol{M}\operatorname{Var}(\boldsymbol{\widehat{T}}^H \mid \boldsymbol{X}^{TW})\boldsymbol{M}' \label{eq: Hausmanequality} .
\end{gather}
Pre-multiplying both sides of the inequality in Equation \eqref{eq: Hausmanequality} by $\boldsymbol{W}$ and post-multiplying by $\boldsymbol{W}'$ yields:
\begin{gather}
    \boldsymbol{W}\operatorname{Var}(\boldsymbol{\widehat{T}}^R \mid \boldsymbol{X}^{TW})\boldsymbol{W}' \geq \boldsymbol{W}\boldsymbol{M}\operatorname{Var}(\boldsymbol{\widehat{T}}^H \mid \boldsymbol{X}^{TW})\boldsymbol{M}'\boldsymbol{W}' \\
    = \boxed{ \operatorname{Var}(\widehat{\tau}^R_{s,t} \mid \boldsymbol{X}^{TW}) \geq \operatorname{Var}(\widehat{\tau}^H_{s,t} \mid \boldsymbol{X}^{TW}).} \label{eq: variance_region_homogeneous}
\end{gather}
The proof next establishes that the conditional variance of the homogeneous DID-INT estimator is weakly smaller than the time-varying DID-INT estimator under homoskedasticity. So the $ATT(s,t)$ for the time-varying DID-INT model can be written as:
\begin{gather}
    \widehat{\tau}^T_{s,t} = \int \biggr[ \widehat{\theta}^T_{s,t} + \sum_{k=1}^K \widehat{\gamma}^T_{k,t} x^k \biggr]d\widehat{F}_{X|S=s, T=t}(x) - \int \biggr[ \widehat{\theta}^T_{s,t^s-1} + \sum_{k=1}^K \widehat{\gamma}^T_{k,t^s-1} x^k \biggr]d\widehat{F}_{X|S=s, T=t}(x) \nonumber \\  - \sum_{s'=1}^{S'} \omega_{s'} \biggr\{ \int \biggr[ \widehat{\theta}^T_{s',t} + \sum_{k=1}^K \widehat{\gamma}^T_{k,t} x^k \biggr]d\widehat{F}_{X|S=s, T=t}(x) - \int \biggr[ \widehat{\theta}^T_{s',t^s-1} + \sum_{k=1}^K \widehat{\gamma}^T_{k,t^s-1} x^k \biggr]d\widehat{F}_{X|S=s, T=t}(x)\biggr\}.
\end{gather}
Since the time-varying coefficients are common across regions for each period, and the fitted outcomes of the control regions are evaluated at the covariate distribution of the treated group, and the weights $\omega_s'$ sum to one, these covariate terms cancel out. Therefore the $ATT(s,t)$ simplifies to:
\begin{gather}
    \widehat{\tau}^T_{s,t} =  \biggr( \widehat{\theta}^T_{s,t} -  \widehat{\theta}^T_{s,t^s-1} \biggr) - \sum_{s'=1}^{S'} \omega_{s'}\biggr( \widehat{\theta}^T_{s',t} -  \widehat{\theta}^T_{s',t^s-1} \biggr).
\end{gather}
Since the homogeneous DID-INT is contained within the time-varying DID-INT under the two-way CCC restriction, the same argument used above for the region-varying DID-INT applies. Hence,
\begin{gather}
\boxed{ \operatorname{Var}(\widehat{\tau}^T_{s,t} \mid \boldsymbol{X}^{TW}) \geq \operatorname{Var}(\widehat{\tau}^H_{s,t} \mid \boldsymbol{X}^{TW}).}
\end{gather}
Lastly, it follows that the conditional variances of the region-varying and time-varying DID-INT estimators are weakly smaller than those of the two-way DID-INT estimator under homoskedasticity. The $ATT(s,t)$ for the two-way DID-INT estimator can be written as:
\begin{gather}
    \widehat{\tau}^{TW}_{s,t} = \int \biggr[ \widehat{\theta}^{TW}_{s,t} + \sum_{k=1}^K \widehat{\gamma}^{TW}_{k,s,t} x^k \biggr]d\widehat{F}_{X|S=s, T=t}(x) - \int \biggr[ \widehat{\theta}^{TW}_{s,t^s-1} + \sum_{k=1}^K \widehat{\gamma}^{TW}_{k,s,t^s-1} x^k \biggr]d\widehat{F}_{X|S=s, T=t}(x) \nonumber \\  - \sum_{s'=1}^{S'} \omega_{s'} \biggr\{ \int \biggr[ \widehat{\theta}^{TW}_{s',t} + \sum_{k=1}^K \widehat{\gamma}^{TW}_{k,s',t} x^k \biggr]d\widehat{F}_{X|S=s, T=t}(x) - \int \biggr[ \widehat{\theta}^{TW}_{s',t^s-1} + \sum_{k=1}^K \widehat{\gamma}^{TW}_{k,s',t^s-1} x^k \biggr]d\widehat{F}_{X|S=s, T=t}(x)\biggr\}.
\end{gather}
Unlike the homogeneous, region-varying, and time-varying DID-INT models, the covariates no longer cancel out. Therefore, the $ATT(s,t)$ remains the same. Under the two-way DID-INT specification, the weighting vector $\boldsymbol{W}^{TW}$ contains non-zero weights on the covariate coefficients since they do not cancel out. The weights on the corresponding covariate coefficients are given by these same DID weights multiplied by the relevant moments $\int x^k d\widehat{F}_{X|S=s,T=t}(x)$. 
\begin{gather}
\boldsymbol{W}^{TW}
=
\biggr(
1,-1,-\omega_{s'_1},\omega_{s'_1},\ldots,
-\omega_{s'_{S'}},\omega_{s'_{S'}},
\nonumber\\
\int x^1 d\widehat{F},\ldots,\int x^K d\widehat{F},
-\int x^1 d\widehat{F},\ldots,-\int x^K d\widehat{F},
\nonumber\\
-\omega_{s'_1}\int x^1 d\widehat{F},\ldots,
-\omega_{s'_1}\int x^K d\widehat{F},
\omega_{s'_1}\int x^1 d\widehat{F},\ldots,
\omega_{s'_1}\int x^K d\widehat{F},
\nonumber\\
\ldots,
-\omega_{s'_{S'}}\int x^1 d\widehat{F},\ldots,
-\omega_{s'_{S'}}\int x^K d\widehat{F},
\omega_{s'_{S'}}\int x^1 d\widehat{F},\ldots,
\omega_{s'_{S'}}\int x^K d\widehat{F},
0,\ldots,0
\biggr),
\end{gather}
where $\boldsymbol{W}^{TW}$ is of dimension $1\times(ST+KST)$.
Therefore the $ATT(s,t)$ can be expressed as:
\begin{gather}
\label{equation: twtransformation}
\widehat{\tau}^{TW}_{s,t}
=
\boldsymbol{W}^{TW}\boldsymbol{\widehat{T}}^{TW}.
\end{gather}
Here, the full coefficient vector of the two-way DID-INT specification is given by: 
\begin{equation}
    \boldsymbol{\widehat{T}}^{TW}
    =
    \begin{pmatrix}
        \boldsymbol{\widehat{\theta}}^{TW} \\
        \boldsymbol{\widehat{\gamma}}^{TW}
    \end{pmatrix}.
\end{equation}

Since the region-varying DID-INT specification is nested within the two-way DID-INT specification, the region-varying coefficient vector can be mapped into the two-way parameter space. Applying the same constrained-versus-unrestricted variance argument yields:
\begin{gather}
\label{eq: transitivityargument1}
\boxed{ \operatorname{Var}(\widehat{\tau}^{TW}_{s,t} \mid \boldsymbol{X}^{TW}) \geq \operatorname{Var}(\widehat{\tau}^{R}_{s,t} \mid \boldsymbol{X}^{TW}).}
\end{gather}
Finally, since the time-varying DID-INT specification is also nested within the two-way DID-INT specification, it follows that:
\begin{gather}
\label{eq: transitivityargument2}
\boxed{ \operatorname{Var}(\widehat{\tau}^{TW}_{s,t} \mid \boldsymbol{X}^{TW}) \geq \operatorname{Var}(\widehat{\tau}^{T}_{s,t} \mid \boldsymbol{X}^{TW}).}
\end{gather}
Combining the inequalities in \eqref{eq: transitivityargument1} and \eqref{eq: transitivityargument2} yields:
\begin{gather}
\boxed{ \operatorname{Var}(\widehat{\tau}^{TW}_{s,t} \mid \boldsymbol{X}^{TW}) \geq \operatorname{Var}(\widehat{\tau}^{C}_{s,t} \mid \boldsymbol{X}^{TW}).}
\end{gather}
\end{proof}

The variance orderings established in Theorem \ref{theorem: efficiency} rely on conditional homoskedasticity. The DID-INT method can also accommodate heteroskedasticity and support cluster-robust inference. Under conditional homoskedasticity, \cite[pg. 202]{hansen2022econometrics} establishes the following relationship:
\begin{equation}
\operatorname{Cov}\left(\widehat{\boldsymbol{T}}^{U}-\boldsymbol{M}\widehat{\boldsymbol{T}}^{C},\boldsymbol{M}\widehat{\boldsymbol{T}}^{C}\mid\boldsymbol{X}^{TW}\right)=\boldsymbol{0}.
\end{equation}
Here, $\widehat{\boldsymbol{T}}^{U}$ and $\widehat{\boldsymbol{T}}^{C}$ are the full coefficient vectors of the unrestricted model and the constrained model, respectively.
\begin{gather}
\operatorname{Cov}\left(\widehat{\boldsymbol{T}}^{U}-\boldsymbol{M}\widehat{\boldsymbol{T}}^{C},\boldsymbol{M}\widehat{\boldsymbol{T}}^{C}\mid\boldsymbol{X}^{TW}\right)=\boldsymbol{0},\\
\operatorname{Cov}\left(\widehat{\boldsymbol{T}}^{U},\boldsymbol{M}\widehat{\boldsymbol{T}}^{C}\mid\boldsymbol{X}^{TW}\right)-\operatorname{Cov}\left(\boldsymbol{M}\widehat{\boldsymbol{T}}^{C},\boldsymbol{M}\widehat{\boldsymbol{T}}^{C}\mid\boldsymbol{X}^{TW}\right)=\boldsymbol{0},\\
\operatorname{Cov}\left(\widehat{\boldsymbol{T}}^{U},\boldsymbol{M}\widehat{\boldsymbol{T}}^{C}\mid\boldsymbol{X}^{TW}\right)-\operatorname{Var}\left(\boldsymbol{M}\widehat{\boldsymbol{T}}^{C}\mid\boldsymbol{X}^{TW}\right)=\boldsymbol{0},\\
\operatorname{Cov}\left(\widehat{\boldsymbol{T}}^{U},\boldsymbol{M}\widehat{\boldsymbol{T}}^{C}\mid\boldsymbol{X}^{TW}\right)=\operatorname{Var}\left(\boldsymbol{M}\widehat{\boldsymbol{T}}^{C}\mid\boldsymbol{X}^{TW}\right).
\end{gather}
To compare the efficiency of the unrestricted estimator with that of the constrained estimator, consider the variance of the difference between the unrestricted coefficient vector and the constrained coefficient vector mapped into the parameter space of the unrestricted specification. Since the conditional variance is positive semidefinite,
\begin{gather}
\operatorname{Var}\left(\widehat{\boldsymbol{T}}^R-\boldsymbol{M}\widehat{\boldsymbol{T}}^H\mid\boldsymbol{X}^{TW}\right)
=
\operatorname{Var}\left(\widehat{\boldsymbol{T}}^R\mid\boldsymbol{X}^{TW}\right)
+
\operatorname{Var}\left(\boldsymbol{M}\widehat{\boldsymbol{T}}^H\mid\boldsymbol{X}^{TW}\right) - 2\operatorname{Cov}\left(\widehat{\boldsymbol{T}}^R,\boldsymbol{M}\widehat{\boldsymbol{T}}^H\mid\boldsymbol{X}^{TW}\right)
 \\
=
\operatorname{Var}\left(\widehat{\boldsymbol{T}}^R\mid\boldsymbol{X}^{TW}\right)
+ 
\operatorname{Var}\left(\boldsymbol{M}\widehat{\boldsymbol{T}}^H\mid\boldsymbol{X}^{TW}\right) - 2 \operatorname{Var}\left(\boldsymbol{M}\widehat{\boldsymbol{T}}^H\mid\boldsymbol{X}^{TW}\right)\\
=
\operatorname{Var}\left(\widehat{\boldsymbol{T}}^R\mid\boldsymbol{X}^{TW}\right)
-
\operatorname{Var}\left(\boldsymbol{M}\widehat{\boldsymbol{T}}^H\mid\boldsymbol{X}^{TW}\right) \geq \boldsymbol{0} \label{eq: usedequation}.
\end{gather}
The inequality in Equation \eqref{eq: usedequation} is the expression used to establish the variance ordering of the estimators. However, under heteroskedasticity,
\begin{gather}
    \operatorname{Cov}\left(\widehat{\boldsymbol{T}}^{U}-\boldsymbol{M}\widehat{\boldsymbol{T}}^{C},\boldsymbol{M}\widehat{\boldsymbol{T}}^{C}\mid\boldsymbol{X}^{TW}\right) \neq \boldsymbol{0},
\end{gather}
Therefore, an unambiguous variance ordering between the estimators can no longer be established under heteroskedasticity.   

Before analyzing the cases in which region-invariant, time-invariant, and two-way CCC structures arise, a general expression is first established for the bias between the probability limit of each alternative DID-INT estimator and the true ATT. This is necessary because not all DID-INT specifications produce a consistent ATT estimate, as some may be misspecified. The expression is derived for the most flexible two-way DID-INT specification. The results for the intermediate region-varying and time-varying DID-INT models are derived from the same reasoning, with the necessary restrictions applied to the covariate coefficients. 

For a potentially misspecified DID-INT specification $j\in\{H,R,T,TW\}$, let $\boldsymbol{T}^{j*}$ denote the probability limits of the coefficients under specification $j$. The elements $\theta^{j*}_{s,t}$ and $\gamma^{j*}_{k,s,t}$ therefore denote the corresponding probability limits of the intercept and covariate coefficients under specification $j$.

\begin{proposition}[Misspecification bias of the alternative DID-INT estimators]
\label{proposition: general_bias}

Let $j$ denote a potentially misspecified DID-INT specification. The difference between the probability limit of the DID-INT estimator under specification $j$ and the true $ATT(s,t)$ is given by
\begin{equation}
    \tau^{j*}_{s,t}
    =
    ATT(s,t)
    +
    Bias^{j}_{s,t},
\end{equation}
where,
\begin{equation}
\footnotesize
\begin{aligned}
Bias^{j}_{s,t} = & \int\left[\bigr(\theta^{j*}_{s,t}-\theta^{\circ}_{s,t} \bigr)+\sum_{k=1}^{K}\bigr(\gamma^{j*}_{k,s,t}-\gamma^{\circ}_{k,s,t}\bigr)x^k\right]dF_{X\mid S=s,T=t}(x) \\ - & \int \left[\bigr(\theta^{j*}_{s,t_s-1}-\theta^{\circ}_{s,t_s-1}\bigr)+\sum_{k=1}^{K}\bigr(\gamma^{j*}_{k,s,t_s-1}-\gamma^{\circ}_{k,s,t_s-1}\bigr)x^k\right]dF_{X\mid S=s,T=t}(x)
\\ &- \sum_{s'\in\mathcal{C}_{s,t}}
\omega_{s',s,t} \Bigg\{\int\left[\bigr(\theta^{j*}_{s',t}-\theta^{\circ}_{s',t}\bigr)+\sum_{k=1}^{K}\bigr(\gamma^{j*}_{k,s',t}-\gamma^{\circ}_{k,s',t}\bigr)x^k
\right]dF_{X\mid S=s,T=t}(x) \\ - & \int
\left[\bigr(\theta^{j*}_{s',t_s-1}-\theta^{\circ}_{s',t_s-1}\bigr) + 
\sum_{k=1}^{K} \bigr(\gamma^{j*}_{k,s',t_s-1}-\gamma^{\circ}_{k,s',t_s-1}\bigr)x^k \right]dF_{X\mid S=s,T=t}(x)
\Bigg\}.
\end{aligned}
\end{equation}
\end{proposition}
\begin{proof}[Proof of Proposition \ref{proposition: general_bias}]
Under the $j$ DID-INT estimator, the $ATT(s,t)$ is given by:
\begin{equation}
\footnotesize
\begin{aligned}
\widehat{\tau}^{j}_{s,t} = & \int\left[\widehat{\theta}^{j}_{s,t}+\sum_{k=1}^{K}\widehat{\gamma}^{j}_{k,s,t}x^k\right]d\widehat{F}_{X\mid S=s,T=t}(x) - \int \left[\widehat{\theta}^{j}_{s,t_s-1}+\sum_{k=1}^{K}\widehat{\gamma}^{j}_{k,s,t_s-1}x^k\right]d\widehat{F}_{X\mid S=s,T=t}(x)
\\ &- \sum_{s'\in\mathcal{C}_{s,t}}
\omega_{s',s,t} \Bigg\{\int\left[\widehat{\theta}^{j}_{s',t}+\sum_{k=1}^{K}\widehat{\gamma}^{j}_{k,s',t}x^k
\right]d\widehat{F}_{X\mid S=s,T=t}(x) \\ - & \int
\left[\widehat{\theta}^{j}_{s',t_s-1} + 
\sum_{k=1}^{K} \widehat{\gamma}^{j}_{k,s',t_s-1}x^k \right]d\widehat{F}_{X\mid S=s,T=t}(x)
\Bigg\}.
\end{aligned}
\end{equation}
The true $ATT(s,t)$ evaluated over the same treated region-period covariate distribution is given by:
\begin{equation}
\footnotesize
\begin{aligned}
ATT(s,t) = & \int\left[\theta^{\circ}_{s,t}+\sum_{k=1}^{K}\gamma^{\circ}_{k,s,t}x^k\right]dF_{X\mid S=s,T=t}(x) - \int \left[\theta^{\circ}_{s,t_s-1}+\sum_{k=1}^{K}\gamma^{\circ}_{k,s,t_s-1}x^k\right]dF_{X\mid S=s,T=t}(x)
\\ &- \sum_{s'\in\mathcal{C}_{s,t}}
\omega_{s',s,t} \Bigg\{\int\left[\theta^{\circ}_{s',t}+\sum_{k=1}^{K}\gamma^{\circ}_{k,s',t}x^k
\right]dF_{X\mid S=s,T=t}(x) \\ - & \int
\left[\theta^{\circ}_{s',t_s-1} + 
\sum_{k=1}^{K} \gamma^{\circ}_{k,s',t_s-1}x^k \right]dF_{X\mid S=s,T=t}(x)
\Bigg\}.
\end{aligned}
\end{equation}
The expression of the bias is given by:
\begin{equation}
\footnotesize
\begin{aligned}
\tau^{j*}_{s,t} - ATT(s,t) = & \int\left[\bigr(\theta^{j*}_{s,t}-\theta^{\circ}_{s,t} \bigr)+\sum_{k=1}^{K}\bigr(\gamma^{j*}_{k,s,t} -\gamma^{\circ}_{k,s,t}\bigr)x^k\right]dF_{X\mid S=s,T=t}(x) \\ - & \int \left[\bigr(\theta^{j*}_{s,t_s-1}-\theta^{\circ}_{s,t_s-1}\bigr)+\sum_{k=1}^{K}\bigr(\gamma^{j*}_{k,s,t_s-1}-\gamma^{\circ}_{k,s,t_s-1}\bigr)x^k\right]dF_{X\mid S=s,T=t}(x)
\\ &- \sum_{s'\in\mathcal{C}_{s,t}}
\omega_{s',s,t} \Bigg\{\int\left[\bigr(\theta^{j*}_{s',t}-\theta^{\circ}_{s',t}\bigr)+\sum_{k=1}^{K}\bigr(\gamma^{j*}_{k,s',t}-\gamma^{\circ}_{k,s',t}\bigr)x^k
\right]dF_{X\mid S=s,T=t}(x) \\ - & \int
\left[\bigr(\theta^{j*}_{s',t_s-1}-\theta^{\circ}_{s',t_s-1}\bigr) + 
\sum_{k=1}^{K} \bigr(\gamma^{j*}_{k,s',t_s-1}-\gamma^{\circ}_{k,s',t_s-1}\bigr)x^k \right]dF_{X\mid S=s,T=t}(x)
\Bigg\}.
\end{aligned}
\end{equation}
Rearranging yields:
\begin{equation}
    \tau^{j*}_{s,t}
    =
    ATT(s,t)
    +
    Bias^{j}_{s,t},
\end{equation}
where,
\begin{equation}
\footnotesize
\begin{aligned}
Bias^{j}_{s,t} = & \int\left[\bigr(\theta^{j*}_{s,t}-\theta^{\circ}_{s,t} \bigr)+\sum_{k=1}^{K}\bigr(\gamma^{j*}_{k,s,t}-\gamma^{\circ}_{k,s,t}\bigr)x^k\right]dF_{X\mid S=s,T=t}(x) \\ - & \int \left[\bigr(\theta^{j*}_{s,t_s-1}-\theta^{\circ}_{s,t_s-1}\bigr)+\sum_{k=1}^{K}\bigr(\gamma^{j*}_{k,s,t_s-1}-\gamma^{\circ}_{k,s,t_s-1}\bigr)x^k\right]dF_{X\mid S=s,T=t}(x)
\\ &- \sum_{s'\in\mathcal{C}_{s,t}}
\omega_{s',s,t} \Bigg\{\int\left[\bigr(\theta^{j*}_{s',t}-\theta^{\circ}_{s',t}\bigr)+\sum_{k=1}^{K}\bigr(\gamma^{j*}_{k,s',t}-\gamma^{\circ}_{k,s',t}\bigr)x^k
\right]dF_{X\mid S=s,T=t}(x) \\ - & \int
\left[\bigr(\theta^{j*}_{s',t_s-1}-\theta^{\circ}_{s',t_s-1}\bigr) + 
\sum_{k=1}^{K} \bigr(\gamma^{j*}_{k,s',t_s-1}-\gamma^{\circ}_{k,s',t_s-1}\bigr)x^k \right]dF_{X\mid S=s,T=t}(x)
\Bigg\}.
\end{aligned}
\end{equation}
\end{proof}
In this setting, the effect of each covariate varies across regions but remains constant over time within each region. Therefore, Assumption \ref{as2: timeinvariantCCC} holds, while Assumption \ref{as2: regioninvariantCCC} is violated.
\begin{equation}
\label{eq:region_varying_structure}
    \gamma^{\circ}_{k,s,t}
    =
    \gamma^{\circ}_{k,s}
    \qquad
    \forall k,s,t.
\end{equation}

\begin{theorem}[Misspecification under region-varying CCC]
\label{theorem: region_bias}

 Suppose Assumption \ref{as2: timeinvariantCCC} holds and Assumption 
\ref{as2: regioninvariantCCC} is violated, such that the true covariate 
coefficient structure is region-varying as defined in Equation 
\eqref{eq:region_varying_structure}. Under Assumptions \ref{as: randomsampling},  \ref{as: covariateunaffected}, \ref{as: noanticipation}, \ref{as: linearity} and either Assumption \ref{as: cpt} or \ref{as: ipt}, the homogeneous and time-varying DID-INT estimators of $ATT(s,t)$ are generally inconsistent. Therefore,
\begin{equation}
    \tau^{H*}_{s,t}
    =
    ATT(s,t)
    +
    Bias^{H}_{s,t},
    \qquad
    \tau^{T*}_{s,t}
    =
    ATT(s,t)
    +
    Bias^{T}_{s,t},
\end{equation}
where $Bias^{H}_{s,t}$ and $Bias^{T}_{s,t}$ are generally non-zero.
\end{theorem}

\begin{proof}[Proof of Theorem \ref{theorem: region_bias}]

Under the true region-varying specification,
\begin{equation}
    \gamma^{\circ}_{k,s,t}
    =
    \gamma^{\circ}_{k,s}.
\end{equation}
The homogeneous DID-INT specification restricts the covariate coefficient to be common across regions and periods:
\begin{equation}
    \gamma^{H*}_{k,s,t}
    =
    \gamma^{H*}_{k}
    \qquad
    \forall s,t,
\end{equation}
while the time-varying DID-INT specification allows the covariate coefficient to vary over time but restricts it to be common across regions:
\begin{equation}
    \gamma^{T*}_{k,s,t}
    =
    \gamma^{T*}_{k,t}
    \qquad
    \forall s.
\end{equation}
Since the true coefficients vary across regions, neither restriction generally reproduces $\gamma^{\circ}_{k,s}$. Substituting these restrictions into the general bias expression in Proposition \ref{proposition: general_bias} shows that $Bias^{H}_{s,t}$ and $Bias^{T}_{s,t}$ are generally non-zero, except in special cases where the misspecification terms cancel. Hence,
\begin{equation}
    \tau^{H*}_{s,t}
    \neq
    ATT(s,t),
    \qquad
    \tau^{T*}_{s,t}
    \neq
    ATT(s,t)
\end{equation}
in general.
\end{proof}

\begin{theorem}[Identification under region-varying CCC]
\label{theorem: region_identification}

Suppose Assumption \ref{as2: timeinvariantCCC} holds and Assumption
\ref{as2: regioninvariantCCC} is violated, such that the true covariate
coefficient structure is region-varying as defined in Equation
\eqref{eq:region_varying_structure}. Under Assumptions \ref{as: randomsampling},  \ref{as: covariateunaffected}, \ref{as: noanticipation}, \ref{as: linearity} and either Assumption \ref{as: cpt} or \ref{as: ipt}, the region-varying and two-way DID-INT estimators identify $ATT(s,t)$:
\begin{equation}
    \tau^{R*}_{s,t}
    =
    \tau^{TW*}_{s,t}
    =
    ATT(s,t).
\end{equation}
\end{theorem}
Since the region-varying model is correctly specified, the proof that the region-varying DID-INT estimator is consistent for the ATT follows directly from the proof established in \cite{karim2024good}. Likewise, since the region-varying DID-INT model is nested within the two-way DID-INT model, it is also consistent for the ATT. 
Consequently, the region-varying DID-INT estimator is weakly more efficient than the two-way DID-INT estimator under the conditional homoskedasticity assumption. 

\begin{theorem}[Efficiency of the region-varying DID-INT estimator under region-varying CCC]
\label{theorem: region_efficiency}

Suppose Assumption \ref{as2: timeinvariantCCC} holds and Assumption
\ref{as2: regioninvariantCCC} is violated, such that the true covariate
coefficient structure is region-varying as defined in Equation
\eqref{eq:region_varying_structure}. Under the OLS regularity conditions and homoskedastic error terms, the conditional variance of the region-varying DID-INT estimator of $ATT(s,t)$ is weakly smaller than that of the two-way DID-INT estimator:
\begin{equation}
\operatorname{Var}\left(\widehat{\tau}^{R}_{s,t} \mid \boldsymbol{X}^{TW}\right)
    \leq
    \operatorname{Var}\left(\widehat{\tau}^{TW}_{s,t}\mid \boldsymbol{X}^{TW}\right).
\end{equation}
\end{theorem}
The proof of Theorem \ref{theorem: region_efficiency} follows directly
from the proof established in Theorem \ref{theorem: hom_efficiency}.

Next, consider the case where the time-varying DID-INT is the true
specification. In this setting, the effect of each covariate varies
across periods but remains constant across regions within each period.
Therefore, Assumption \ref{as2: regioninvariantCCC} holds, while
Assumption \ref{as2: timeinvariantCCC} is violated.
\begin{equation}
\label{eq:time_varying_structure}
    \gamma^{\circ}_{k,s,t}
    =
    \gamma^{\circ}_{k,t}
    \qquad
    \forall k,s,t.
\end{equation}

\begin{theorem}[Misspecification under time-varying CCC]
\label{theorem: time_bias}

Suppose Assumption \ref{as2: regioninvariantCCC} holds and Assumption
\ref{as2: timeinvariantCCC} is violated, such that the true covariate
coefficient structure is time-varying as defined in Equation
\eqref{eq:time_varying_structure}. Under Assumptions \ref{as: randomsampling},  \ref{as: covariateunaffected}, \ref{as: noanticipation}, \ref{as: linearity} and either Assumption \ref{as: cpt} or \ref{as: ipt}, the homogeneous and region-varying DID-INT estimators of $ATT(s,t)$ are generally inconsistent. Therefore,
\begin{equation}
    \tau^{H*}_{s,t}
    =
    ATT(s,t)
    +
    Bias^{H}_{s,t},
    \qquad
    \tau^{R*}_{s,t}
    =
    ATT(s,t)
    +
    Bias^{R}_{s,t},
\end{equation}
where $Bias^{H}_{s,t}$ and $Bias^{R}_{s,t}$ are generally non-zero.
\end{theorem}

\begin{proof}[Proof of Theorem \ref{theorem: time_bias}]

Under the true time-varying specification,
\begin{equation}
    \gamma^{\circ}_{k,s,t}
    =
    \gamma^{\circ}_{k,t}.
\end{equation}
The homogeneous DID-INT specification restricts the covariate coefficient to be common across regions and periods:
\begin{equation}
    \gamma^{H*}_{k,s,t}
    =
    \gamma^{H*}_{k}
    \qquad
    \forall s,t,
\end{equation}
while the region-varying DID-INT specification allows the covariate coefficient to vary across regions but restricts it to remain constant over time:
\begin{equation}
    \gamma^{R*}_{k,s,t}
    =
    \gamma^{R*}_{k,s}
    \qquad
    \forall t.
\end{equation}
Since the true coefficients vary across periods, neither restriction generally reproduces $\gamma^{\circ}_{k,t}$. Substituting these restrictions into the general bias expression in Proposition \ref{proposition: general_bias} shows that $Bias^{H}_{s,t}$ and $Bias^{R}_{s,t}$ are generally non-zero, except in special cases where the misspecification terms cancel. Hence,
\begin{equation}
    \tau^{H*}_{s,t}
    \neq
    ATT(s,t),
    \qquad
    \tau^{R*}_{s,t}
    \neq
    ATT(s,t)
\end{equation}
in general.
\end{proof}

\begin{theorem}[Identification under time-varying CCC]
\label{theorem: time_identification}

Suppose Assumption \ref{as2: regioninvariantCCC} holds and Assumption
\ref{as2: timeinvariantCCC} is violated, such that the true covariate
coefficient structure is time-varying as defined in Equation
\eqref{eq:time_varying_structure}. Under Assumptions \ref{as: randomsampling},  \ref{as: covariateunaffected}, \ref{as: noanticipation}, \ref{as: linearity} and either Assumption \ref{as: cpt} or \ref{as: ipt}, the time-varying and two-way DID-INT estimators identify $ATT(s,t)$:
\begin{equation}
    \tau^{T*}_{s,t}
    =
    \tau^{TW*}_{s,t}
    =
    ATT(s,t).
\end{equation}
\end{theorem}

Since the time-varying model is correctly specified, the proof that the time-varying DID-INT estimator is consistent for the ATT follows directly from the proof established in \cite{karim2024good}. Likewise, since the time-varying DID-INT estimator is nested within the two-way DID-INT model, it is also consistent for the ATT. 
Consequently, the time-varying DID-INT estimator is weakly more efficient than the two-way DID-INT estimator under the conditional homoskedasticity assumption. 

\begin{theorem}[Efficiency of the time-varying DID-INT under time-varying CCC]
\label{theorem: time_efficiency}

Suppose Assumption \ref{as2: regioninvariantCCC} holds and Assumption
\ref{as2: timeinvariantCCC} is violated, such that the true covariate
coefficient structure is time-varying as defined in Equation
\eqref{eq:time_varying_structure}. Under the OLS regularity conditions and homoskedastic error terms, the conditional variance of the time-varying DID-INT estimator of $ATT(s,t)$ is weakly smaller than that of the two-way DID-INT estimator:
\begin{equation}
\operatorname{Var}\left(\widehat{\tau}^{T}_{s,t} \mid \boldsymbol{X}^{TW}\right)
    \leq
    \operatorname{Var}\left(\widehat{\tau}^{TW}_{s,t}\mid \boldsymbol{X}^{TW}\right).
\end{equation}
\end{theorem}
The proof of Theorem \ref{theorem: time_efficiency} follows directly
from the proof established in Theorem \ref{theorem: hom_efficiency}. Finally, consider the case where the true covariate coefficient
structure is two-way, such that the covariate effects vary jointly
across regions and periods and cannot be represented by a homogeneous,
region-varying, or time-varying specification. Therefore, Assumptions
\ref{assumption: twowayCCC}, \ref{as2: regioninvariantCCC}, and
\ref{as2: timeinvariantCCC} do not provide a valid restricted
representation of the true covariate coefficient structure. Therefore, Assumption \ref{assumption: twowayCCC} is violated, and $\gamma^{\circ}_{k,s,t}$ varies across both regions and periods.

\begin{theorem}[Misspecification under two-way CCC violations]
\label{theorem: twoway_bias}

Suppose Assumptions \ref{as2: regioninvariantCCC} and
\ref{as2: timeinvariantCCC} are both violated, such that the true
covariate coefficient structure varies jointly across regions and
periods and cannot be represented by the homogeneous, region-varying,
or time-varying DID-INT specifications. Under Assumptions \ref{as: randomsampling},  \ref{as: covariateunaffected}, \ref{as: noanticipation}, \ref{as: linearity} and either Assumption \ref{as: cpt} or \ref{as: ipt}, the homogeneous, region-varying, and time-varying DID-INT estimators of $ATT(s,t)$ are generally inconsistent. Therefore,
\begin{equation}
    \tau^{H*}_{s,t}
    =
    ATT(s,t)
    +
    Bias^{H}_{s,t},
\end{equation}
\begin{equation}
    \tau^{R*}_{s,t}
    =
    ATT(s,t)
    +
    Bias^{R}_{s,t},
\end{equation}
and
\begin{equation}
    \tau^{T*}_{s,t}
    =
    ATT(s,t)
    +
    Bias^{T}_{s,t},
\end{equation}
where $Bias^{H}_{s,t}$, $Bias^{R}_{s,t}$, and $Bias^{T}_{s,t}$ are generally non-zero.
\end{theorem}

\begin{proof}[Proof of Theorem \ref{theorem: twoway_bias}]

Under the true two-way specification, the covariate coefficients are given by
\begin{equation}
    \gamma^{\circ}_{k,s,t},
\end{equation}
and vary jointly across regions and periods. The homogeneous DID-INT specification restricts the covariate coefficient to be common across regions and periods:
\begin{equation}
    \gamma^{H*}_{k,s,t}
    =
    \gamma^{H*}_{k}.
\end{equation}
Under the region-varying DID-INT specification, the covariate coefficient is allowed to vary across regions but is restricted to remain constant over time:
\begin{equation}
    \gamma^{R*}_{k,s,t}
    =
    \gamma^{R*}_{k,s}.
\end{equation}
Under the time-varying DID-INT specification, the covariate coefficient is allowed to vary over time but is restricted to be common across regions:
\begin{equation}
    \gamma^{T*}_{k,s,t}
    =
    \gamma^{T*}_{k,t}.
\end{equation}
None of these restricted coefficient structures generally reproduces $\gamma^{\circ}_{k,s,t}$ under two-way CCC violations. Substituting each restriction into Proposition \ref{proposition: general_bias} therefore yields generally non-zero misspecification bias, except in special cases where the bias terms cancel. Hence,
\begin{equation}
    \tau^{H*}_{s,t}
    \neq
    ATT(s,t),
    \qquad
    \tau^{R*}_{s,t}
    \neq
    ATT(s,t),
    \qquad
    \tau^{T*}_{s,t}
    \neq
    ATT(s,t)
\end{equation}
in general.
\end{proof}

\begin{theorem}[Identification of the two-way DID-INT estimator under two-way CCC violations]
Suppose the covariate effects vary jointly across regions and periods, such that the two-way DID-INT specification is correctly specified. Under Assumptions 1--4 and either Assumption 8 or Assumption 9,
\begin{equation}
    \tau^{TW*}_{s,t}=ATT(s,t).
\end{equation}
\end{theorem}

\begin{proof}
Since the two-way DID-INT estimator allows the covariate coefficients to vary freely across region-period cells, the true coefficient structure is contained in the two-way parameter space. Therefore, the probability limits of the estimated coefficients equal their corresponding population coefficients. Substituting these coefficients into the DID-INT estimand eliminates the misspecification terms in Proposition 1, yielding
\[
\tau^{TW*}_{s,t}=ATT(s,t).
\]
\end{proof}

In summary, the results in this subsection prove that the two-way DID-INT estimator is consistent across all DGPs, including homogeneous, region-varying, time-varying, and two-way scenarios. In contrast, other specifications are generally inconsistent when they impose incorrect restrictions. The homogeneous and time-varying DID-INT specifications are generally inconsistent when covariate coefficients differ across regions but are constant over time. Similarly, the homogeneous and region-varying DID-INT specifications are generally inconsistent when covariate coefficients differ across periods but stay the same across regions. All three specifications are generally inconsistent when covariate coefficients vary across both regions and periods. Although the two-way DID-INT specification avoids such restrictions by allowing covariate coefficients to vary freely across regions and periods, it requires estimating more nuisance parameters. Under homoskedasticity, I show that a more flexible model can unambiguously increase variance and, consequently, reduce precision. However, under heteroskedasticity, an unambiguous variance ordering can no longer be derived.

Let $\boldsymbol{\widehat{T}}^{j}$ be the coefficient matrix of specification $j$, where $j = \{H,R,T,TW\}$. Using OLS, the estimator $\boldsymbol{\widehat{T}}^{j}$ can be decomposed into the true parameter $\boldsymbol{{T}}^{j}$ and the stochastic component $\boldsymbol{({X}}^{j^\prime}{X}^{j})^{-1}{X}^{j^\prime}u$ \citep[p.g 100, Equation 4.6]{hansen2022econometrics}. $\boldsymbol{{X}}^{j}$ is the regressor matrix of specification $j$.
\begin{equation}
    \boldsymbol{\widehat{T}}^{j} = \boldsymbol{T}^{j} + \boldsymbol{({X}}^{j^\prime}\boldsymbol{X}^{j})^{-1}\boldsymbol{X}^{j^\prime}u.
\end{equation}
Under heteroskedasticity, the variance of $\boldsymbol{\widehat{T}}^{j}$ can be expressed as:
\begin{gather}
\operatorname{Var}(\boldsymbol{\widehat{T}}^{j} \mid \boldsymbol{X}^{j}) = \boldsymbol{({X}}^{j^\prime}{X}^{j})^{-1}\boldsymbol{{X}}^{j^\prime}\Omega_{s,t}\boldsymbol{X}^{j}\boldsymbol{({X}}^{j^\prime}\boldsymbol{X}^{j})^{-1}.
\end{gather}
Provided the restricted model is correctly specified, it pools information by estimating fewer coefficients than the unconstrained model. Under heteroskedasticity, both the $(\boldsymbol{X}^{j\prime}\boldsymbol{X}^j)^{-1}$ terms and the middle $\boldsymbol{X}^{j\prime}\Omega_{s,t} \boldsymbol{X}^j$ term change when restrictions are imposed. Consequently, nesting alone does not imply an unambiguous variance ordering. Intuitively, the restricted model pools covariate information across cells, enabling each shared coefficient to be estimated with more observations and covariate variation. In contrast, the unrestricted model distributes the same sample data among many nuisance parameters. Therefore, using a restricted model may increase precision, although such gains may not arise under heteroskedasticity. 

\subsection{CCC test}
\label{section: ccctest}

This subsection outlines the CCC test, which can be used more broadly to select an appropriate covariate specification. In this paper, I highlight its utility for selecting the appropriate DID-INT specification. More generally, the test can also be used to guide covariate specification in other estimators, such as TWFE. The analytical results in the previous subsection indicate that the two-way DID-INT estimator can yield a consistent estimate of the ATT whether the CCC assumption is violated or holds. However, using a more flexible specification decreases efficiency and may increase computational costs. Additionally, adopting a more restrictive model may introduce misspecification bias. 

The CCC test described in this section will be used to assess whether covariate coefficients are equal across regions, across periods, or both. Since I do not observe the untreated potential outcomes of the treated region in the post-intervention period, I only do this for a subset of $s$ and $t$, which includes never-treated or not-yet-treated observations. 
This test can only detect violations of the CCC assumption and guide researchers in incorporating covariates into the regression. The CCC test is \textit{not a statistical test for conditional parallel trends or integrated parallel trends}, as introduced in Assumptions \ref{as: cpt} and \ref{as: ipt} respectively. The CCC test is a joint test of linear restrictions on covariate coefficients by comparing an unrestricted auxiliary regression with a restricted version that imposes the relevant CCC restriction. The first step of the CCC test is to check whether the coefficients are the same or vary jointly across both regions and periods. This gatekeeping test checks whether the two-way CCC holds or is violated. To do so, the auxiliary regression shown in Equation \eqref{eq:auxiliary1}, which includes covariates interacted with region and period dummies, is estimated. The CCC test is estimated using observations from periods in which the corresponding regions remain untreated, including never-treated and not-yet-treated observations. Let $U$ denote the total number of untreated region-period cells in the CCC test. For each untreated cell $u=1,\ldots,U$, let $s^u$ and $t^u$ denote the corresponding region and period, respectively. For these observations, the observed outcome satisfies $Y_{i,s,t}=Y_{i,s,t}(0)$.

\begin{equation}
\label{eq:auxiliary1}
Y_{i,s,t}
=
\sum_{u=1}^{U}
\theta^{\circ}_{s^u,t^u}
Z_{i,s^u,t^u}
+
\sum_{k=1}^{K}
\sum_{u=1}^{U}
\gamma^{\circ,k}_{s^u,t^u}
I(s^u)I(t^u)X^k_{i,s,t}
+
\epsilon_{i,s,t}.
\end{equation}

Here, $Z_{i,s,t}$ denotes a full set of region-by-time indicators for the $U$ untreated region-period cells used in the CCC test, and $\gamma_{s,t}^{\circ,k}$ is the coefficient on covariate $k$ in region $s$ and year $t$. The null hypothesis for the first stage is:
\begin{equation}
\label{eq:null1}
H^0_{tw}:
\gamma_{s^1,t^1}^{\circ,k}
=
\cdots
=
\gamma_{s^2,t^2}^{\circ,k}
=
\cdots
=
\gamma_{s^U,t^U}^{\circ,k}
=
\gamma^{\circ,k},
\qquad
\forall k=1,\ldots,K.
\end{equation}
In the above equation, $S^u$ denotes the total number of regions in the dataset, and $T^u$ denotes the total number of years in the dataset. The number of restrictions is given by:
\begin{equation}
    Q_{tw} = K(U - 1).
\end{equation}
The restricted model in this case will be:
\begin{equation}
Y_{i,s,t} = \sum_{u=1}^{U} \theta^{\circ}_{s^u,t^u}
Z_{i,s^u,t^u} + \sum_{k=1}^{K} \gamma^{\circ,k}\left[\sum_{u=1}^{U}I(s^u)I(t^u)X^k_{i,s,t}\right]+\epsilon_{i,s,t}.
\end{equation}

If the null hypothesis shown in Equation \eqref{eq:null1} is not rejected at level $\alpha$, then there is no evidence of two-way CCC violations and the homogeneous DID-INT specification is used. If the null hypothesis is rejected, the data provide evidence that the covariate coefficients are not constant across regions and periods. In this case, the homogeneous DID-INT specification may impose invalid restrictions on the covariate coefficients and may therefore produce an inconsistent estimate of the ATT, as shown in the previous subsection.

Following a rejection in the first stage, researchers could use the two-way DID-INT specification, since this specification allows the covariate coefficients to vary freely across regions and periods. However, the analytical results in the previous subsection show that the two-way DID-INT estimator may be less efficient when the true coefficient structure is more restrictive. Therefore, the procedure proceeds to a second stage to determine whether the coefficient heterogeneity detected in the first stage can be characterized by region-specific or time-specific variation. The region and time hypotheses are tested within the same second stage rather than sequentially. It is important to note that these are two alternative possibilities within the same stage, rather than sequential ones. The test for time-invariant CCC violations is not conditional on first detecting region-invariant CCC violations. Instead, the region-invariant and time-invariant hypotheses are tested separately within the second stage to determine whether the two-way violation reflects variation across regions, over time, or both. The auxiliary regression in the second stage is given by Equation \eqref{eq:auxiliary2}, which imposes an additive region and year structure on the covariate coefficients.
\begin{equation}
\label{eq:auxiliary2}
Y_{i,s,t} = \sum_{u=1}^{U} \theta^{\circ}_{s^u,t^u}
Z_{i,s^u,t^u} + \sum_{k=1}^{K} \sum_{s^u=1}^{S^u} \gamma^{\circ,k}_{s^u}
I(s^u)X^k_{i,s,t} + \sum_{k=1}^{K} \sum_{t^u=2}^{T^u} \gamma^{\circ,k}_{t^u}
I(t^u)X^k_{i,s,t} + \epsilon_{i,s,t}.
\end{equation}
In this specification, the effect of each covariate is allowed to vary by region and by year, but only in an additive manner. For identification, Period 1 is used as the reference
categories for the time-varying component to avoid multicollinearity.
The null hypothesis that there is no specific region-invariant CCC violation is shown in Equation \eqref{eq:null2}, and the null that there is no specific time-invariant CCC violation is shown in Equation \eqref{eq:null3}. 

\begin{equation}
\label{eq:null2}
H^0_R: \gamma^{\circ,k}_{1} = \gamma^{\circ,k}_{2} = \cdots = \gamma^{\circ,k}_{S^u} = \gamma^{\circ,k}_R, \qquad \forall k=1,\ldots,K.
\end{equation}

The number of joint restrictions for the region null is:
\begin{equation}
    Q_{\text{R}} = K(S^u-1)
\end{equation}
The null hypothesis to check whether the time CCC holds is:
\begin{equation}
\label{eq:null3}
H^{0}_{\text{T}}:
\gamma^{\circ,k}_{2}
= \gamma^{\circ,k}_{3}
= \cdots = \gamma^{\circ,k}_{T^u} = 0, \qquad
\forall k=1,\ldots,K.
\end{equation}
Since one time-specific coefficient is dropped to avoid multicollinearity, the time null tests whether the time-specific deviations from the period 1 coefficient are jointly equal to zero. This differs from the region null, where none of the region-specific coefficients are omitted. The number of joint restrictions for the time null is:
\begin{equation}
   Q_{\text{T}} = K(T^u-1) 
\end{equation}
The restricted model for the no region-invariant CCC violation is:

\begin{equation}
\label{eq:restricted_group}
Y_{i,s,t}
=
\sum_{u=1}^{U}
\theta^{\circ}_{s^u,t^u}
Z_{i,s^u,t^u}
+
\sum_{k=1}^{K}
\gamma^{\circ,k}_{R}X^k_{i,s,t}
+
\sum_{k=1}^{K}
\sum_{t^u=2}^{T^u}
\gamma^{\circ,k}_{t^u}
I(t^u)X^k_{i,s,t}
+
\epsilon_{i,s,t}.
\end{equation}
Similarly, the restricted model for the no time-invariant CCC violation is:
\begin{equation}
\label{eq:restricted_year}
Y_{i,s,t}
=
\sum_{u=1}^{U}
\theta^{\circ}_{s^u,t^u}
Z_{i,s^u,t^u}
+
\sum_{k=1}^{K}
\gamma^{\circ,k}_TX^k_{i,s,t}
+
\sum_{k=1}^{K}
\sum_{s^u=2}^{S^u}
\gamma^{\circ,k}_{s^u}
I(s^u)X^k_{i,s,t}
+
\epsilon_{i,s,t}.
\end{equation}

The null hypotheses in both stages of the CCC test can be expressed as joint linear restrictions that impose equality among the relevant covariate coefficients. Let $\widehat{\boldsymbol{T}}_j$ denote the OLS coefficient vector from the relevant auxiliary regression, where $j \in \{R,T,TW\}$, and let $\boldsymbol{R}_j$ denote a $Q_j \times P_j$ restriction matrix with full row rank. The rows of $\boldsymbol{R}_j$ are constructed to impose equality among the relevant covariate coefficients. Depending on the normalization of the auxiliary regression, these equality restrictions can be represented in different ways. For the region null, they are expressed as differences between the region-specific coefficients. For the time null, they are expressed as deviations from the Period 1 reference coefficient being equal to zero. Here, $P_j$ is the number of parameters, including the region-by-time indicators.  The null hypothesis can then be written as:
\begin{equation}
  H^0_j:\boldsymbol{R}_j\boldsymbol{T}_j=\boldsymbol{0}.
\end{equation}
Following \citet[p.g. 202]{hansen2022econometrics}, these restrictions are tested using the Wald statistic
\begin{equation}
\label{eq:wald_ccc}
    W_j
    =
    \left(
    \boldsymbol{R}_j\widehat{\boldsymbol{T}}_j
    \right)'
    \left[
    \boldsymbol{R}_j
    \widehat{\boldsymbol{V}}_j
    \boldsymbol{R}_j'
    \right]^{-1}
    \left(
    \boldsymbol{R}_j\widehat{\boldsymbol{T}}_j
    \right),
\end{equation}
where $\widehat{\boldsymbol{V}}_j$ is the covariance matrix estimator of $\widehat{\boldsymbol{T}}_j$. Under the null hypothesis and standard regularity conditions,
\begin{equation}
    W_j
    \xrightarrow{d}
    \chi^2_{Q_j}.
\end{equation}
The corresponding p-values are obtained from the $\chi^2_{Q_j}$ distribution \citep[p. 184, Theorem 7.13]{hansen2022econometrics}, and the null hypothesis is rejected when the p-value is less than the chosen significance level. If only the region null is rejected, the evidence favors the region-varying DID-INT specification, whereas if only the time null is rejected, the evidence favors the time-varying DID-INT specification. If both nulls are rejected, then the two-way DID-INT specification is recommended. Since the region and time null hypotheses are tested simultaneously in the second stage, conducting both tests at the same significance level can increase the probability of Type I error. To prevent false positives in the second stage and avoid incorrect model selection, the Holm's step-down procedure, described in Section \ref{holm}, is implemented. This is especially important because the final DID-INT specification depends on the results of these tests.

The asymptotic analysis in this paper does not develop a cluster-robust version of the CCC testing procedure. Extending the joint Wald tests to clustered dependence is non-trivial in many regional DID applications because the number of available clusters can be relatively small. For example, a U.S. state-level application can contain at most 50 state clusters, while an analysis at the provincial level in Canada may contain only 10 provincial clusters, or 13 provincial and territorial jurisdictions if the territories are included. At the same time, the number of restrictions imposed by the CCC test increases with the number of regions, periods, and covariates. Consequently, the number of restrictions can become large relative to the number of clusters, making conventional cluster-robust Wald approximations unreliable or, in sufficiently high-dimensional settings, difficult to implement. For this reason, the current paper focuses on the sampling framework stated in Assumption \ref{as: randomsampling}. Developing cluster-robust versions of the CCC test, including procedures designed for few-cluster settings, is left for future research.

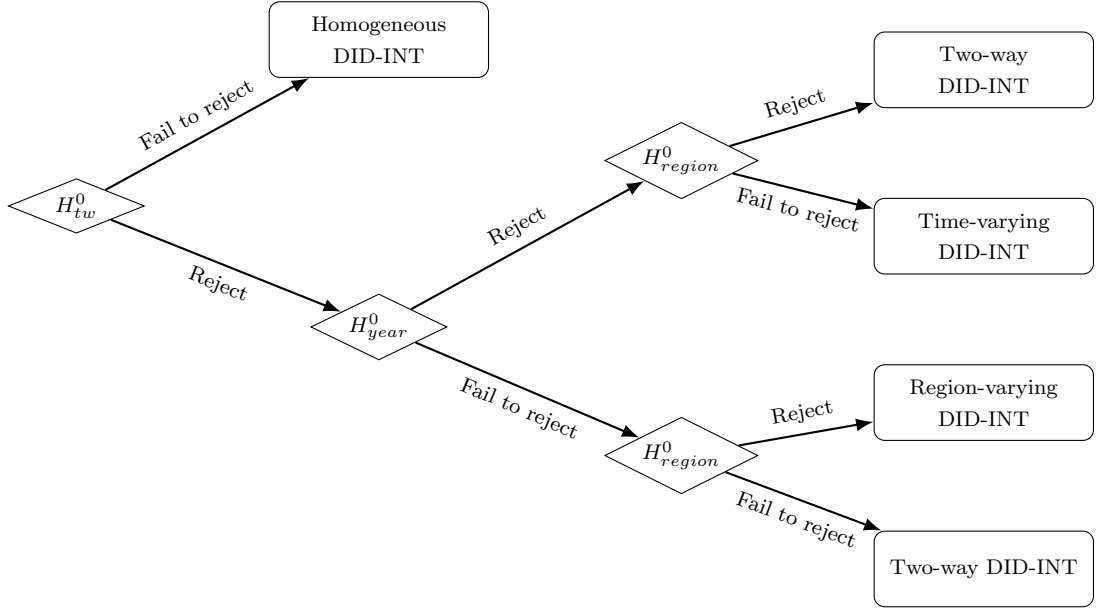
\begin{figure}[H]
\centering
\begin{tikzpicture}[
    x=1cm, y=1cm,
    decision/.style={
        diamond,
        draw,
        aspect=2,
        align=center,
        inner sep=1.2pt,
        font=\scriptsize,
        minimum width=1.8cm
    },
    model/.style={
        rectangle,
        draw,
        rounded corners,
        align=center,
        minimum width=2.9cm,
        minimum height=1cm,
        font=\scriptsize
    },
    lab/.style={font=\scriptsize},
    >=Latex
]

\node[decision] (tw) at (0,0) {\(H^0_{tw}\)};
\node[model] (hom) at (4,2.2) {Homogeneous\\ DID-INT};
\node[decision] (yr) at (4,-1.6) {\(H^0_{year}\)};

\node[decision] (reg1) at (8,0.6) {\(H^0_{region}\)};
\node[decision] (reg2) at (8,-3.3) {\(H^0_{region}\)};

\node[model] (two) at (12,1.8) {Two-way\\ DID-INT};
\node[model] (time) at (12,-0.4) {Time-varying\\ DID-INT};
\node[model] (region) at (12,-2.6) {Region-varying\\ DID-INT};
\node[model] (und) at (12,-4.8) {Two-way DID-INT};

\draw[->, thick] (tw) -- node[lab, above, sloped] {Fail to reject} (hom);
\draw[->, thick] (tw) -- node[lab, below, sloped] {Reject} (yr);

\draw[->, thick] (yr) -- node[lab, above, sloped] {Reject} (reg1);
\draw[->, thick] (yr) -- node[lab, below, sloped] {Fail to reject} (reg2);

\draw[->, thick] (reg1) -- node[lab, above, sloped] {Reject} (two);
\draw[->, thick] (reg1) -- node[lab, below, sloped] {Fail to reject} (time);

\draw[->, thick] (reg2) -- node[lab, above, sloped] {Reject} (region);
\draw[->, thick] (reg2) -- node[lab, below, sloped] {Fail to reject} (und);

\end{tikzpicture}
\caption{Sequential Decision Rule for the CCC Test}
\label{fig:ccc_sequential_flowchart}
\end{figure}

The decision rules for the CCC test are shown in Figure \ref{fig:ccc_sequential_flowchart}, with a more formal presentation in Table \ref{tab:sequential_f_decision_rule}. If $H^0_{tw}$ is failed to be rejected in the first stage, the second stage is not conducted and the homogeneous DID-INT specification is selected. However, if $H^0_{tw}$ is rejected in the first stage, the procedure moves on to the second stage, where both $H^0_{region}$ and $H^0_{year}$ are tested within the same stage. If $H^0_{region}$ is rejected, but $H^0_{year}$ is not rejected, the region-varying DID-INT specification is used. Similarly, if $H^0_{year}$ is rejected, but $H^0_{region}$ is not rejected, the time-varying DID-INT specification is used. If both $H^0_{region}$ and $H^0_{year}$ are rejected in the second stage, then the two-way DID-INT specification is used. Finally, if $H^0_{tw}$ is rejected in the first stage, but both $H^0_{region}$ and $H^0_{year}$ are not rejected in the second stage, the two-way DID-INT specification is used. This outcome can occur when covariate coefficients show non-additive variation across regions and time, which the first-stage test detects but cannot be assigned solely to either the additive region or time components analyzed in the second stage. Although theoretically possible, it did not arise in the simulation studies conducted in Section \ref{section: mc}.

\begin{table}[H]
\centering
\scriptsize
\setlength{\tabcolsep}{4pt}
\renewcommand{\arraystretch}{1.25}
\begin{tabular}{p{2.8cm}p{3.0cm}p{3.0cm}p{3.2cm}}
\toprule
\makecell[l]{\textbf{First Stage}\\ \textbf{Decision}}
&
\makecell[l]{\textbf{Second Stage}\\ \textbf{Region Null}\\ \(H^0_{region}\)}
&
\makecell[l]{\textbf{Second Stage}\\ \textbf{Year Null}\\ \(H^0_{year}\)}
&
\makecell[l]{\textbf{Selected}\\ \textbf{Specification}} \\
\midrule

\makecell[l]{Fail to reject\\ \(H^0_{tw}\)}
&
\makecell[l]{Not tested}
&
\makecell[l]{Not tested}
&
\makecell[l]{Homogeneous\\ DID-INT} \\

\addlinespace

\makecell[l]{Reject\\ \(H^0_{tw}\)}
&
\makecell[l]{Reject\\ \(H^0_{region}\)}
&
\makecell[l]{Fail to reject\\ \(H^0_{year}\)}
&
\makecell[l]{Region-varying\\ DID-INT} \\

\addlinespace

\makecell[l]{Reject\\ \(H^0_{tw}\)}
&
\makecell[l]{Fail to reject\\ \(H^0_{region}\)}
&
\makecell[l]{Reject\\ \(H^0_{year}\)}
&
\makecell[l]{Time-varying\\ DID-INT} \\

\addlinespace

\makecell[l]{Reject\\ \(H^0_{tw}\)}
&
\makecell[l]{Reject\\ \(H^0_{region}\)}
&
\makecell[l]{Reject\\ \(H^0_{year}\)}
&
\makecell[l]{Two-way\\ DID-INT} \\

\addlinespace

\makecell[l]{Reject\\ \(H^0_{tw}\)}
&
\makecell[l]{Fail to reject\\ \(H^0_{region}\)}
&
\makecell[l]{Fail to reject\\ \(H^0_{year}\)}
&
\makecell[l]{Two-way DID-INT} \\

\bottomrule
\end{tabular}
\caption{Decision Rule for the CCC Test}
\label{tab:sequential_f_decision_rule}
\end{table}

\subsection{Holm's Step Down Procedure}
\label{holm}

The first stage of the CCC test evaluates violations of the two-way CCC assumption using a single hypothesis, so no multiple-testing correction is needed. In the second phase, both region and time null hypotheses are tested simultaneously, conditional on rejecting the first-stage null. Performing both second-stage tests at level $\alpha$ can raise the probability of at least one Type I error, so the \textit{family-wise error rate} may exceed $\alpha$ without adjustment \citep{holm1979simple}. To address this issue, a multiple-testing correction is implemented for the two hypotheses in the second stage. The Bonferroni method splits the overall significance level $\alpha$ by the number of tests, so each is conducted at level $\alpha/2$. Nonetheless, the Bonferroni adjustment may be overly conservative, reducing statistical power \citep{holm1979simple,shaffer1995multiple}. 

To ensure that the family-wise error rate does not exceed $\alpha$, Holm's step-down multiple testing procedure is used, which is less conservative than the Bonferroni procedure. Suppose that, at a given decision stage, there are $m$ hypotheses to be tested. Holm's procedure first re-orders the p-values from smallest to largest. The smallest p-value is compared to the strictest cutoff $\alpha/m$. If the null is rejected, the procedure moves to the next smallest p-value and compares it to the less strict cutoff $\alpha/(m-1)$. This process continues sequentially, with the critical value becoming larger at each step, until a null hypothesis fails to be rejected, at which point the procedure stops. 

Let $p_{tw}$ denote the p-value of the CCC test from the first stage, and let $p_{region}$ and $p_{time}$ denote the p-values associated with the region and time null hypotheses in the second stage. The CCC test is conducted at a significance level of 0.05. In the first stage, if $p_{tw} > 0.05$, the null hypothesis is not rejected and the homogeneous DID-INT specification is used. The first stage involves only a single hypothesis test and therefore does not require adjustment. In contrast, if $p_{tw} < 0.05$, the null hypothesis is rejected and the procedure moves to the second stage. Let $p_1$ denote the smallest p-value between $p_{region}$ and $p_{time}$. Here, $m=2$ since two tests are conducted within the same stage. 
\begin{equation}
    p_1 = \min\{p_{region},p_{time}\}
\end{equation}
Similarly, $p_2$ denotes the largest p-value between $p_{region}$ and $p_{time}$. 
\begin{equation}
    p_2 = \max\{p_{region},p_{time}\}
\end{equation}
If $p_1 \leq \alpha/2$ and $p_2 > \alpha$, the null hypothesis associated with the smaller p-value is rejected but the other null is not rejected. In this case, the corresponding one-way DID-INT specification associated with the smallest p-value is selected, which can either be the region-varying DID-INT specification if $p_{region} = p_1$, or the time-varying DID-INT specification if $p_{time} = p_1$. If $p_1 \leq \alpha/2$ and $p_2 \leq \alpha$, both null hypotheses are rejected, and the two-way DID-INT is selected. Finally, if $p_1 > \alpha/2$, neither null hypothesis is rejected in the second stage. Since the first-stage null has already been rejected, the two-way DID-INT specification is also selected in this case. 

\section{Monte Carlo Simulations}
\label{section: mc}

\subsection{Data Generating Process}
\label{sec: mcdgp}

This subsection outlines the design of the Monte Carlo Simulation Study, which assesses the performance of the proposed CCC test introduced in Section \ref{section: ccctest}. The DGPs feature a balanced region-period structure with staggered adoption. There are $S=6$ regions and $T=6$ periods, and the number of observations per region-period cell varies across 25, 50, 100, 200, and 500. This corresponds to total sample sizes of 900, 1800, 3600, 7200 and 18000, respectively. Regions 1 and 2 are treated in period 3; regions 3 and 4 are treated in period 5; and regions 5 and 6 are never treated. The true ATT is set to zero. The untreated potential outcome is generated using the following formula:
\begin{equation}
    \label{equation: baselineuntreatpo}
        Y^0_{i,s,t} = 3s + t + \epsilon_{i,s,t}, \qquad \epsilon_{i,s,t} \sim N(0,5^2)
\end{equation}
This ensures that the regions start with varying initial outcome levels while maintaining a common upward trend over time. $\epsilon_{i,s,t}$ is the independently and identically distributed (i.i.d.) error term, which is independent of the covariate $X_{i,s,t}$. The standard deviation of the idiosyncratic error term is set to 5 to ensure sufficient residual variation relative to the systematic component. This way, the simulated outcome regressions target a moderate Adjusted $R^2$. Consistent with the maintained sampling framework, the Monte Carlo simulations use independently distributed idiosyncratic errors and do not introduce region-level common shocks. With only six regions in the Monte Carlo design, conventional cluster-robust inference may be unreliable due to the small number of clusters \citep{mackinnon2018wild}.

For simplicity, the simulation generates a single time-varying covariate $X_{i,s,t}$ to better reflect how the distributio of these covariates can change over time. The formula used for generating these covariates is detailed in Equation \eqref{equation: covariatedgp}. 
\begin{equation}
    \label{equation: covariatedgp}
    X_{i,s,t} = \mu^X_{s,t} + \nu_{i,s,t},
    \qquad
    \nu_{i,s,t} \sim N(0,2^2),
\end{equation}
where the region-period-specific covariate mean is given by
\begin{equation}
    \mu^X_{s,t}
    =
    \begin{cases}
        0, & s=1,2,3,4,\\
        -1.25c_t, & s=5,\\
        \phantom{-}1.25c_t, & s=6,
    \end{cases}
    \qquad
    (c_1,c_2,c_3)=(0.5,1,1.5).
\end{equation}

In applied settings, the covariate distributions are likely to vary across regions and over time. To account for this, the covariate means are allowed to differ between the two never-treated regions and across different periods. For simplicity, covariates in the treated regions are drawn from the same distribution with a mean of zero, as the primary focus is comparing distributions between treated and control regions rather than differences among treated regions. Although the control regions have different covariate means, their aggregate mean remains zero. This formula produces distinct covariate distributions across regions and across periods while keeping the overall mean constant between treated and control groups. The term $\nu_{i,s,t}$ represents an i.i.d. error with a standard deviation of 2. 

Next, four outcome DGPs are generated to examine the CCC test across various types of CCC violations. The first DGP features a homogeneous covariate effect, satisfying the two-way CCC assumption. The second involves covariate effects that vary by region, while the third includes effects that change over time. The final DGP incorporates covariate effects that vary both by region and time.   Equations \eqref{eq:homfull} to \eqref{eq:twowayfull} list the four DGPs.
\begin{equation}
\label{eq:homfull}
\mbox{\textbf{DGP 1 (Homogeneous):}} \;\; Y^1_{i,s,t} = Y^{0}_{i,s,t} + 1*X_{i,s,t}.
\end{equation}
\begin{equation}
\label{eq:statefull}
\mbox{\textbf{DGP 2 (Region-Varying):}} \;\; Y^2_{i,s,t} = Y^{0}_{i,s,t} + a_s*X_{i,s,t}.
\end{equation}
\begin{equation}
\label{eq:timefull}
\mbox{\textbf{DGP 3 (Time-Varying):}} \;\; Y^3_{i,s,t} = Y^{0}_{i,s,t} + b_t*X_{i,s,t}.
\end{equation}
\begin{equation}
\label{eq:twowayfull}
\mbox{\textbf{DGP 4 (Two-Way):}} \;\; Y^4_{i,s,t} = Y^{0}_{i,s,t} + a_s*b_t*X_{i,s,t}.
\end{equation}
Here, $a_s = (0.5,0.9,1.5,1.1,0.7,1.3)$ denotes the region-specific coefficient sequence, and $b_t = (0.5,1.3,0.9,1.5,0.7,1.1)$ denotes the time-specific coefficient sequence. These coefficients are calibrated to have an average value of one, making them comparable to the homogeneous case where the coefficient is one. They also follow a non-monotonic pattern, meaning there is no systematic order across regions and periods. These coefficients are kept constant across all Monte Carlo replications.

\subsection{Efficiency Gains from Specification Selection}
\label{sec: truevstwoway}

This section compares the two-way DID-INT estimator with the correctly specified estimator, which uses the correctly specified covariate coefficient structure for each DGP. I refer to the correctly specified estimator as the ``true" estimator. In applied settings, the true model is unknown and must be identified using the CCC test. Since it is known in this simulation, it can be used to validate the statistical properties discussed in Section \ref{sec: theory}. After generating the data as outlined in Section \ref{sec: mcdgp}, the ATT is estimated using both the true model and the two-way DID-INT estimator, repeating this process 1000 times. The Bias, Monte Carlo standard deviation, the root mean squared errors (RMSE), and relative efficiency (RE) are then computed for each specification $j$ across various sample sizes  ($n_{\text{cell}}$).
A relative efficiency greater than one therefore indicates that the two-way DID-INT estimator has a larger Monte Carlo variance than the true specification. The simulations use 1,000 replications for each DGP and sample size. The RE measures how large the two-way DID-INT estimator is relative to the true model. If the RE is greater than 1, then the sampling variance of the two-way estimator is greater than the sampling variance of the true estimator. The formulas for Bias, Monte Carlo standard deviation, RMSE, and RE are provided below.
\begin{gather}
    \operatorname{Bias^{(j)}}(\widehat{\tau}^{(j)}) = \overline{\widehat{\tau}^{(j)}}-\tau^{\circ}, \label{eq: biasmc}\\
    \operatorname{MCSD}(\widehat{\tau}^{(j)}) = \sqrt{\frac{1}{N_r-1}\sum_{r=1}^{N_r} \left(\widehat{\tau}^{(j)}_r-\overline{\widehat{\tau}}^{(j)}\right)^2}, \\
    \operatorname{RMSE}(\widehat{\tau}^{(j)}) = \sqrt{\frac{1}{N_r}\sum_{r=1}^{N_r} \left(\widehat{\tau}^{(j)}_r-\tau^{\circ}\right)^2}, \\
    \operatorname{RE} = \frac{\operatorname{MCSD}(\widehat{\tau}^{TW})^2}{\operatorname{MCSD}(\widehat{\tau}^{(j)})^2}.
\end{gather}
Here, $N_r$ denote the number of Monte Carlo replications, $\widehat{\tau}_r$ denote the estimated ATT in replication $r$, $\tau^{\circ}$ denote the true ATT which is set to zero, and $\overline{\widehat{\tau}}=\frac{1}{N_r}\sum_{r=1}^{N_r}\widehat{\tau}_r$. 

Table \ref{tab:oracle_twoway} reports the results of the Monte Carlo simulation under homoskedasticity. Both the true and two-way specifications exhibit very low bias close to zero, with bias magnitude declining as sample size increases. This provides evidence that both estimators remain approximately unbiased across all DGPs, even when the true coefficient structure is more restrictive. The more substantive results lie in the comparison between the Monte Carlo standard deviation, RMSE, and the RE. Both the Monte Carlo standard deviation and the RMSE of the true model are lower than those of the two-way DID-INT estimator across all sample sizes, although both measures decline as sample size increases. The RE is also greater than one, indicating that the sampling variance of the two-way estimator exceeds that of the true estimator. These findings align with the theory, indicating that both the true estimator and the two-way estimators are unbiased across various DGPs. However, the two-way estimator exhibits less precision under homoskedasticity. Table \ref{tab:oracle_twoway_hetero} presents the simulation results for heteroskedasticity. The simulation design remains the same, with the only difference being that the errors are now produced as:
\begin{gather}
\epsilon_{i,s,t}
=
\sigma_s u_{i,s,t},
\qquad
u_{i,s,t} \sim N(0,1),
\qquad
\sigma_s {\sim} U(3,7).
\end{gather}

\begin{table}[H]
\centering
\caption{True DID-INT Specification versus Two-Way DID-INT Specification under Homoskedasticity}
\label{tab:oracle_twoway}
\resizebox{\textwidth}{!}{
\begin{tabular}{llccccccc}
\toprule
& & \multicolumn{2}{c}{Bias}
& \multicolumn{2}{c}{Monte Carlo SD}
& \multicolumn{2}{c}{RMSE}
&  \\
\cmidrule(lr){3-4}
\cmidrule(lr){5-6}
\cmidrule(lr){7-8}
DGP
& $n_{\text{cell}}$
& True
& Two-way
& True
& Two-way
& True
& Two-way
& Relative Efficiency \\
\midrule

\multirow{5}{*}{Homogeneous}
& 25  &  0.048 &  0.040 & 0.820 & 0.919 & 0.821 & 0.919 & 1.255 \\
& 50  & -0.004 & -0.001 & 0.579 & 0.642 & 0.579 & 0.641 & 1.228 \\
& 100 &  0.006 &  0.004 & 0.405 & 0.440 & 0.404 & 0.440 & 1.181 \\
& 200 &  0.005 &  0.007 & 0.306 & 0.343 & 0.306 & 0.343 & 1.260 \\
& 500 & -0.003 & -0.001 & 0.189 & 0.204 & 0.189 & 0.204 & 1.167 \\

\midrule

\multirow{5}{*}{Region-varying}
& 25  &  0.046 &  0.040 & 0.823 & 0.919 & 0.824 & 0.919 & 1.246 \\
& 50  & -0.007 & -0.001 & 0.581 & 0.642 & 0.581 & 0.641 & 1.220 \\
& 100 &  0.004 &  0.004 & 0.406 & 0.440 & 0.406 & 0.440 & 1.172 \\
& 200 &  0.003 &  0.007 & 0.306 & 0.343 & 0.306 & 0.343 & 1.255 \\
& 500 & -0.003 & -0.001 & 0.190 & 0.204 & 0.190 & 0.204 & 1.158 \\

\midrule

\multirow{5}{*}{Time-varying}
& 25  &  0.048 &  0.040 & 0.823 & 0.919 & 0.824 & 0.919 & 1.246 \\
& 50  & -0.004 & -0.001 & 0.583 & 0.642 & 0.583 & 0.641 & 1.212 \\
& 100 &  0.007 &  0.004 & 0.404 & 0.440 & 0.404 & 0.440 & 1.183 \\
& 200 &  0.004 &  0.007 & 0.306 & 0.343 & 0.306 & 0.343 & 1.260 \\
& 500 & -0.003 & -0.001 & 0.189 & 0.204 & 0.189 & 0.204 & 1.167 \\

\bottomrule
\end{tabular}
}
\end{table}

The group-specific standard deviations, \(\sigma_s\), are randomly drawn once from a uniform distribution on the interval $[3,7]$, and then left fixed for the rest of the simulation. The interval $[3,7]$ is selected to ensure the distribution is centered at 5, corresponding to the standard deviation in the homoskedastic case. Under heteroskedasticity, the bias remains near zero, and the RE exceeds one. Therefore, the two-way estimator remains unbiased but is still less efficient than the true estimator in the current simulation design, although this ordering may not hold generally. 

\subsection{Performance of CCC-Based Selected DID-INT Specification}
\label{sec: selectvstwoway}

This section assesses the statistical performance of the selected DID-INT specification after conducting the CCC test. First, this specification is compared with the two-way DID-INT specification, focusing on Bias, Monte Carlo standard deviation, RMSE, and RE. Second, the analysis evaluates whether the selected DID-INT specification from the CCC test preserves valid inference. To do so, the estimated ATT, the jackknife standard error, and the corresponding jackknife p-value are recorded for the selected and the two-way specification for each replication. Given the small number of regions in the simulation design, I rely on jackknife inference rather than conventional cluster-robust inference, which can perform poorly when the number of clusters is limited \citep{mackinnon2018wild}. The empirical rejection frequencies of the selected model and the two-way DID-INT estimators are then compared to check whether the inference procedure remains correctly sized after model selection. Additional Monte Carlo evidence on the finite-sample performance of jackknife inference is provided in \citet*{austin2026policy}. Given that the true ATT is zero across all DGPs, a well-specified model should theoretically reject the null hypothesis of no treatment effect approximately 5 percent of the time. The average confidence interval widths between the selected model and the two-way DID-INT estimators are also compared. These widths are calculated by deriving a 95\% confidence interval for each ATT estimate. The difference between the upper and lower confidence bounds is then computed and averaged across replications. The average confidence interval width helps assess whether the selected model can produce narrower intervals without compromising size. In this simulation, the number of Monte Carlo replications is increased from 1,000 to 5,000 to evaluate the selected model's inference capabilities. The CCC testing procedure continues to use the same homoskedastic inference setup as in the preceding simulations, while inference for the estimated ATT is based on the jackknife standard errors and p-values described above. 

Table \ref{tab:selected_twoway} presents the Bias, Monte Carlo SD, RMSE, and RE results for the selected specification from the CCC test. These findings align with the comparison between the true estimator and the two-way estimator in the preceding subsection. Across all four DGPs, the selected estimator's bias remains very close to zero, similar to that of the two-way estimator. However, the selected estimator has a lower Monte Carlo standard deviation and RMSE than the two-way estimator, and both statistics decrease as sample size increases. The relative efficiency also remains greater than one, which implies that the selected specification largely preserves the efficiency gains observed in the preceding section. In the two-way DGP, the bias, Monte Carlo standard deviation, RMSE, and RE do not match the two-way estimator because the CCC procedure can sometimes choose an incorrect model in certain replications. However, these differences disappear when the sample size increases.

\begin{table}[H]
\centering
\caption{Selected DID-INT versus Two-Way DID-INT}
\label{tab:selected_twoway}
\resizebox{\textwidth}{!}{
\begin{tabular}{llccccccc}
\toprule
& & \multicolumn{2}{c}{Bias}
& \multicolumn{2}{c}{Monte Carlo SD}
& \multicolumn{2}{c}{RMSE}
& \\
\cmidrule(lr){3-4}
\cmidrule(lr){5-6}
\cmidrule(lr){7-8}
DGP
& $n_{\text{cell}}$
& Selected
& Two-way
& Selected
& Two-way
& Selected
& Two-way
& Relative Efficiency \\
\midrule

\multirow{5}{*}{Homogeneous}
& 25  & -0.003 & -0.004 & 0.855 & 0.925 & 0.855 & 0.925 & 1.169 \\
& 50  & -0.007 & -0.007 & 0.591 & 0.647 & 0.591 & 0.647 & 1.197 \\
& 100 &  0.001 &  0.001 & 0.418 & 0.454 & 0.418 & 0.454 & 1.182 \\
& 200 &  0.003 &  0.004 & 0.299 & 0.329 & 0.299 & 0.329 & 1.215 \\
& 500 &  0.001 &  0.000 & 0.187 & 0.203 & 0.187 & 0.203 & 1.184 \\

\midrule

\multirow{5}{*}{Region-varying}
& 25  &  0.027 & -0.004 & 0.855 & 0.925 & 0.855 & 0.925 & 1.171 \\
& 50  &  0.007 & -0.007 & 0.592 & 0.647 & 0.592 & 0.647 & 1.194 \\
& 100 &  0.002 &  0.001 & 0.414 & 0.454 & 0.414 & 0.454 & 1.203 \\
& 200 &  0.003 &  0.004 & 0.298 & 0.329 & 0.298 & 0.329 & 1.225 \\
& 500 &  0.002 &  0.000 & 0.186 & 0.203 & 0.186 & 0.203 & 1.190 \\

\midrule

\multirow{5}{*}{Time-varying}
& 25  & -0.003 & -0.004 & 0.847 & 0.925 & 0.847 & 0.925 & 1.192 \\
& 50  & -0.007 & -0.007 & 0.588 & 0.647 & 0.588 & 0.647 & 1.212 \\
& 100 & -0.000 &  0.001 & 0.414 & 0.454 & 0.414 & 0.454 & 1.203 \\
& 200 &  0.003 &  0.004 & 0.297 & 0.329 & 0.297 & 0.329 & 1.230 \\
& 500 &  0.002 &  0.000 & 0.185 & 0.203 & 0.185 & 0.203 & 1.201 \\

\midrule

\multirow{5}{*}{Two-way}
& 25  &  0.083 & -0.005 & 0.903 & 0.925 & 0.907 & 0.925 & 1.049 \\
& 50  &  0.010 & -0.007 & 0.648 & 0.647 & 0.648 & 0.647 & 0.998 \\
& 100 &  0.001 &  0.001 & 0.454 & 0.454 & 0.454 & 0.454 & 0.999 \\
& 200 &  0.004 &  0.004 & 0.329 & 0.329 & 0.329 & 0.329 & 1.000 \\
& 500 &  0.000 &  0.000 & 0.203 & 0.203 & 0.203 & 0.203 & 1.000 \\

\bottomrule
\end{tabular}
}
\end{table}

Table \ref{tab:selected_twoway_inference} presents the results of inference under the selected specification. The empirical rejection frequencies for both the selected and two-way specifications are below the nominal 5\% level, reflecting the conservativeness of the jackknife procedure in finite samples. The more notable finding is that average confidence interval widths decrease as the region-by-period sample size increases for both estimators. Crucially, the confidence intervals for the selected DID-INT estimator are always narrower than those for the two-way estimator, indicating that the efficiency gains from the model-selection procedure lead to more precise inference. 

\begin{table}[H]
\centering
\caption{Cluster-jackknife Inference for Selected and Two-Way DID-INT Estimators}
\label{tab:selected_twoway_inference}
\resizebox{0.80\textwidth}{!}{
\begin{tabular}{llcccc}
\toprule
& &
\multicolumn{2}{c}{Empirical Rejection Frequency}
& \multicolumn{2}{c}{Average CI Width} \\
\cmidrule(lr){3-4}
\cmidrule(lr){5-6}
DGP
& $n_{\text{cell}}$
& Selected
& Two-way
& Selected
& Two-way \\
\midrule

\multirow{5}{*}{Homogeneous}
& 25  & 0.034 & 0.035 & 5.009 & 5.638 \\
& 50  & 0.036 & 0.034 & 3.442 & 3.877 \\
& 100 & 0.040 & 0.038 & 2.421 & 2.720 \\
& 200 & 0.043 & 0.040 & 1.698 & 1.925 \\
& 500 & 0.039 & 0.037 & 1.082 & 1.219 \\

\midrule

\multirow{5}{*}{Region-varying}
& 25  & 0.032 & 0.035 & 5.052 & 5.638 \\
& 50  & 0.033 & 0.034 & 3.430 & 3.877 \\
& 100 & 0.037 & 0.038 & 2.427 & 2.720 \\
& 200 & 0.040 & 0.040 & 1.706 & 1.925 \\
& 500 & 0.037 & 0.037 & 1.090 & 1.219 \\

\midrule

\multirow{5}{*}{Time-varying}
& 25  & 0.028 & 0.035 & 5.268 & 5.638 \\
& 50  & 0.032 & 0.034 & 3.597 & 3.877 \\
& 100 & 0.035 & 0.038 & 2.487 & 2.720 \\
& 200 & 0.037 & 0.040 & 1.753 & 1.925 \\
& 500 & 0.035 & 0.037 & 1.115 & 1.219 \\

\midrule

\multirow{5}{*}{Two-way}
& 25  & 0.032 & 0.035 & 5.506 & 5.639 \\
& 50  & 0.035 & 0.034 & 3.859 & 3.879 \\
& 100 & 0.038 & 0.038 & 2.721 & 2.721 \\
& 200 & 0.041 & 0.041 & 1.926 & 1.926 \\
& 500 & 0.037 & 0.037 & 1.219 & 1.219 \\

\bottomrule
\end{tabular}
}
\end{table}

\subsection{Power of the CCC Testing Procedure}
\label{sec: power}

This section examines the size and power properties of the CCC test. The underlying DGPs remain mostly the same, and homoskedastic standard errors continue to be used. The primary change is that the extent of CCC violation is now adjusted. Larger violations are produced by increasing the difference in covariate coefficients across regions, periods, or region-by-period cells, depending on the DGP. To quantify this difference, a vector $\delta$ is defined, where higher values indicate a greater level of CCC violation. Since this simulation's goal is to assess the inference properties of the CCC test, 5000 replications are performed. 

\begin{equation}
    \delta = \begin{pmatrix}
     0 & 0.25 & 0.50 & 0.75 & 1.00 & 1.50
    \end{pmatrix}.
\end{equation}
For each value of $\delta$, the covariate coefficients are then scaled according to the following formula:
\begin{equation}
 1+\delta\left(\gamma^{*}_{s,t}-1\right).
\end{equation}
Here, the $\gamma^{*}_{s,t}$ is specified by $a_s$ for the region-varying DID-INT specification, $b_t$ for the time-varying DID-INT specification, and $a_s \times b_t$ for the two-way DID-INT specification, as defined in Section \ref{sec: mcdgp}. For $\delta = 0$, the covariate coefficients translate directly to the homogeneous case, with a coefficient of 1 across region and period cells. This allows the size of the CCC test to be assessed. As $\delta$ increases, the coefficients move progressively further away from the common value of 1, and generate larger differences across regions, periods, or region-by-period cells. The number of observations per region-period cell is also varied for each value of $\delta$, with $n_{cell} = 25, 50, 100, 200, 500$.

\begin{figure}[H]
    \centering
    \includegraphics[width=0.7\linewidth]{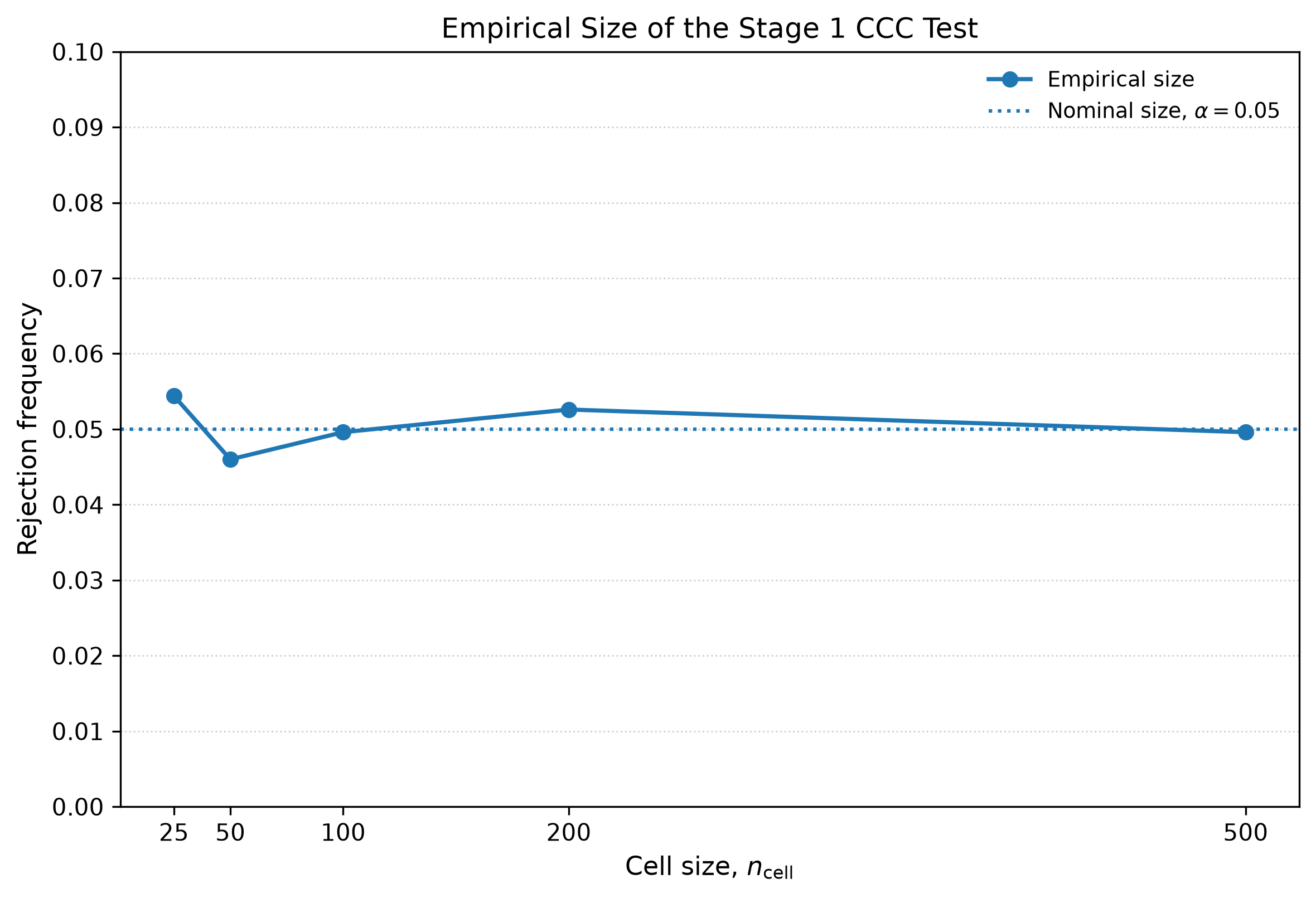}
    \caption{Empirical Size of the Stage 1 CCC Test.}
    \label{fig:empiricalsize}
\end{figure}

Figure \ref{fig:empiricalsize} reports the empirical rejection frequency of the first stage of the CCC test when $\delta = 0$. At this value, the covariate coefficients are one across all region-by-period cells, ensuring the two-way CCC holds by design. Consequently, the empirical rejection frequency measures the probability of falsely rejecting the null hypothesis that the covariate effects are the same across regions and periods, effectively measuring the test's size. Across all region-by-period sample sizes, the empirical rejection frequency remains close to the nominal 5\% level, thus indicating that the CCC test is well sized under the null. 

\begin{figure}[H]
    \centering
    \includegraphics[width=0.75\linewidth]{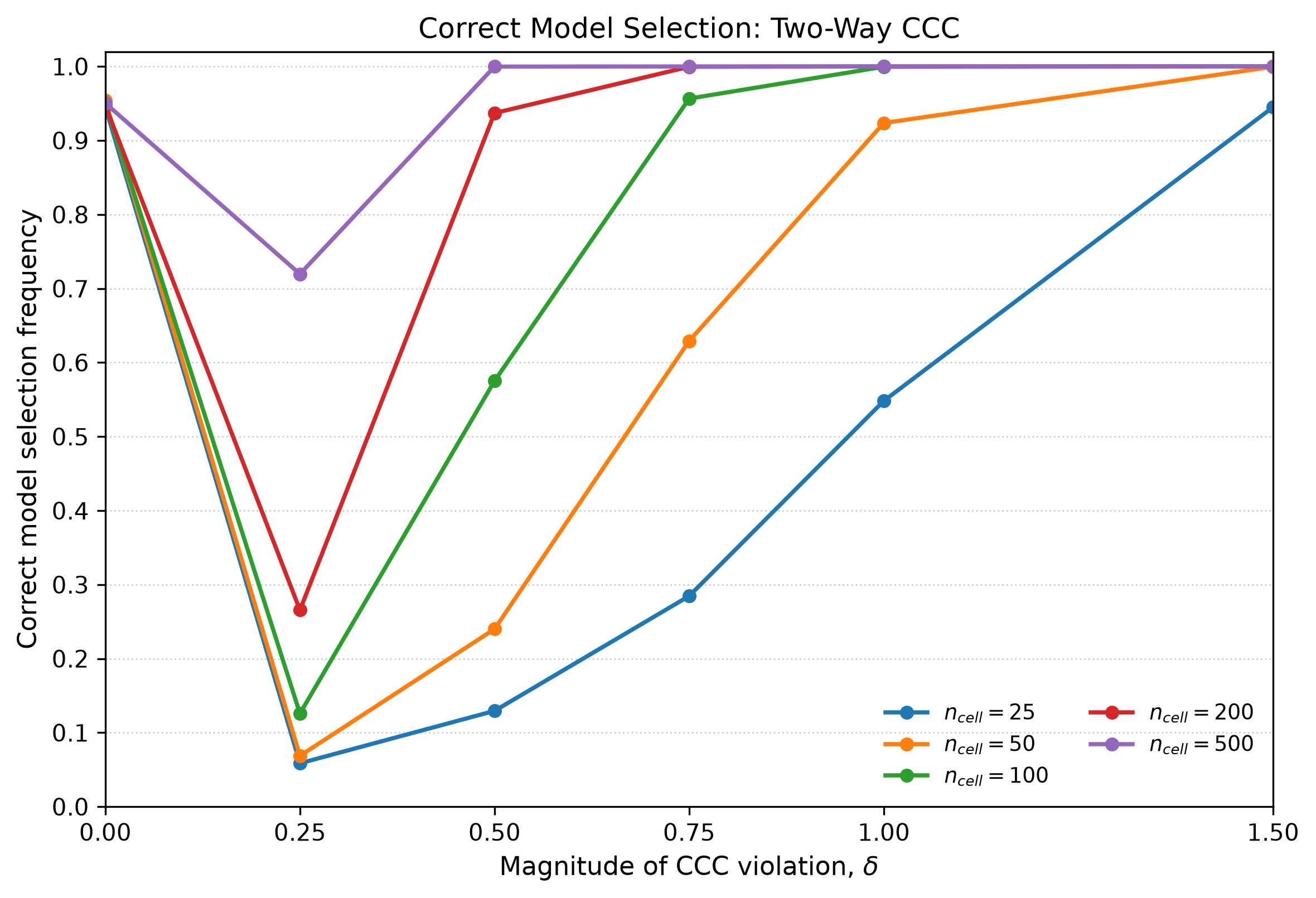}
    \caption{Frequency of Correct CCC Specification Selection: Two-Way DGP}
    \label{fig:correctselection}
\end{figure}

Figure \ref{fig:correctselection} illustrates how often the CCC test correctly identifies the specification across various sample sizes and degrees of CCC violations. When $\delta = 0$, the correct model selection frequency is around 95 \% across all three DGPs. This is because, under $\delta = 0$, the homogeneous specification holds true. Consequently, the CCC test falsely rejects the null around 5\% of the time, aligning with the 5\% significance level used in the first stage. Therefore, the test correctly chooses the correct model about 95 \% of the time. 

When $\delta \neq 0$, the correct model depends on the DGP. For the two-way DGP, three main patterns emerge. First, increasing $\delta$ increases the differences in covariate coefficients across regions, periods, or both, thereby improving the ability of the CCC test to select the correct model. Second, larger sample sizes allow the CCC test to detect weaker violations, making model selection more sensitive to small increases in $\delta$. As a result, the correct specification curves shift upward and to the left.  Third, the selection frequency for the region-varying and time-varying DGPs approaches 95\%, whereas the selection frequency for the two-way DGP approaches 100\% as sample size and degree of CCC violation increase. In the two-way case, both $H^0_{R}$ and $H^0_{T}$ are false, so the probability of selecting the two-way model approaches 100 \%. The corresponding results for the region-varying and time-varying DGPs are reported in Appendix \ref{app:additional_ccc_power}.

\begin{figure}[H]
    \centering

    \begin{subfigure}[t]{0.47\linewidth}
        \centering
        \includegraphics[width=\linewidth]{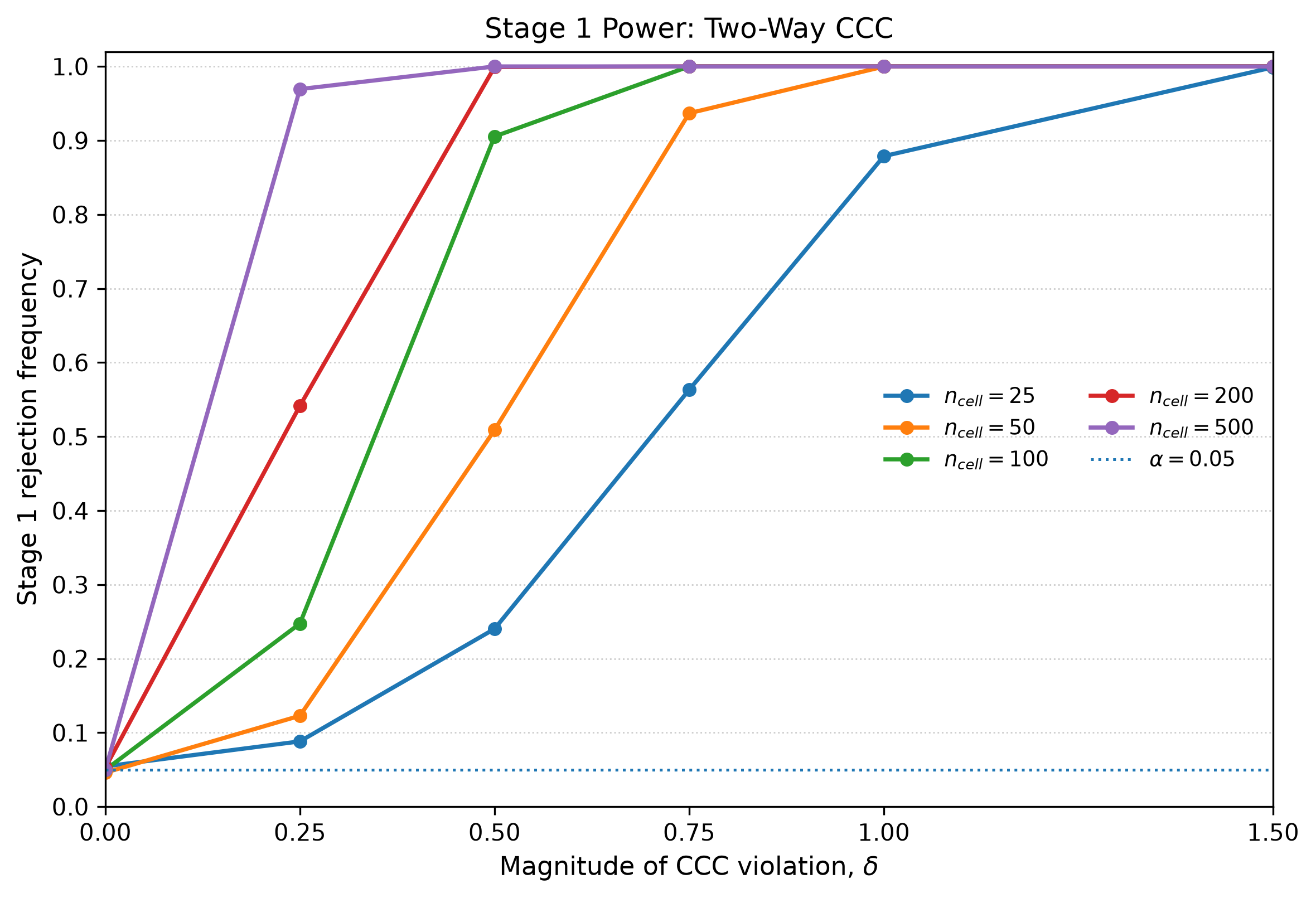}
        \caption{Stage 1}
        \label{fig:ccc_power_stage1_twoway}
    \end{subfigure}
    \hfill
    \begin{subfigure}[t]{0.47\linewidth}
        \centering
        \includegraphics[width=\linewidth]{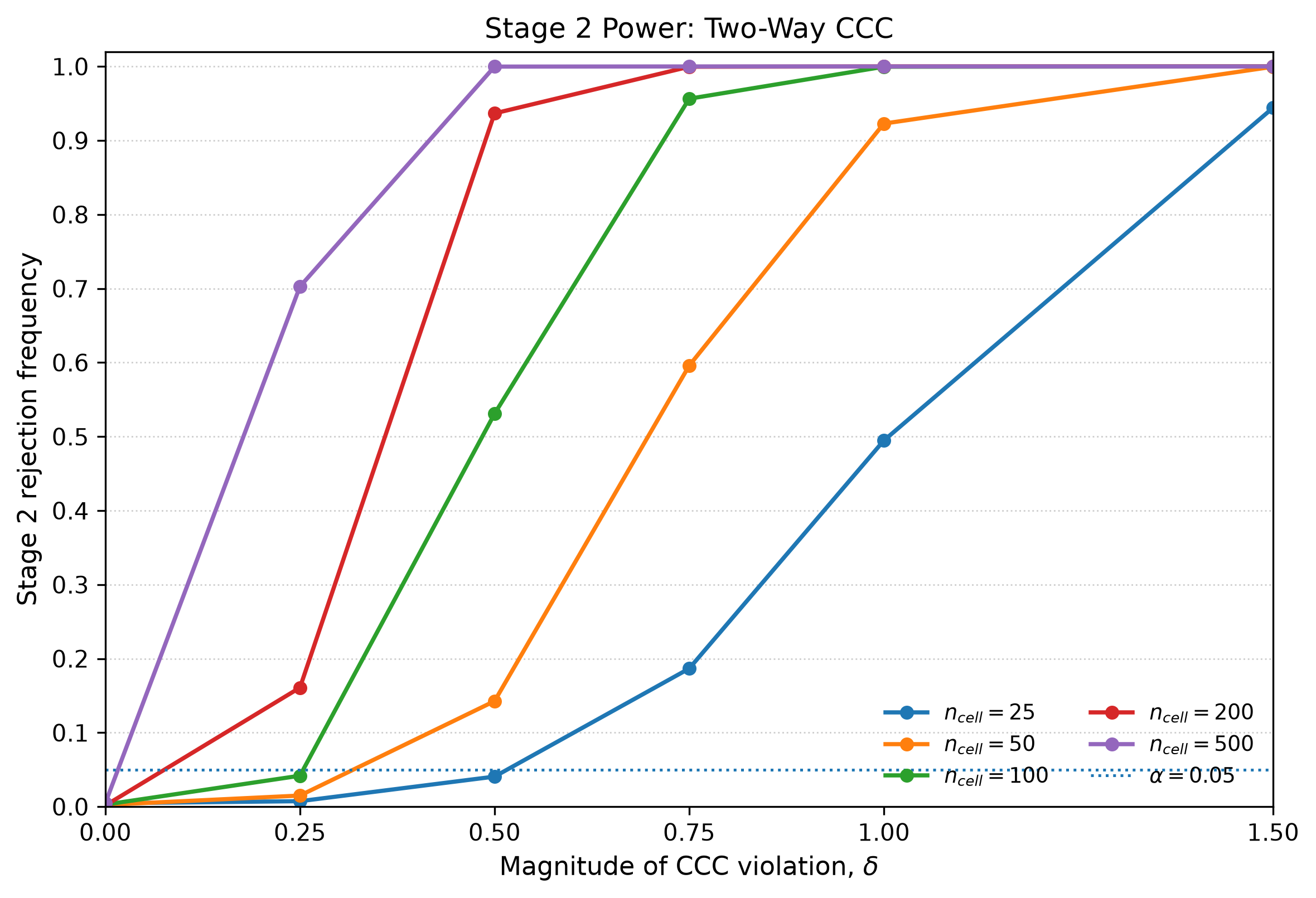}
        \caption{Stage 2}
        \label{fig:ccc_power_stage2_twoway}
    \end{subfigure}

    \caption{Power of the Sequential CCC Tests: Two-Way DGP}
    \label{fig:ccc_power}
\end{figure}

Figure \ref{fig:ccc_power} illustrates the power curves for Stage 1 and Stage 2 CCC tests for the two-way DGPs. The power curves for Stage 1 and Stage 2 display the expected pattern. In Stage 1, the test evaluates whether the two-way null hypothesis $H^0_{TW}$ holds across all three DGPs. Since this null is false under the two-way DGP, the probability of rejection approaches one as sample size and degree of CCC violation increase. In stage 2, both $H^0_{R}$ and $H^0_{T}$ are false, so the rejection probability for both tests also approaches one. In both stages, larger CCC violations increase the ability of the tests to detect violations of the null. As a result, the power curves slope upward and approach one as delta increases. In addition, larger sample sizes allow weaker violations to be detected more reliably, causing the power curves to shift upward and to the left as the number of observations per region-by-period cell increases. The corresponding results for the region-varying and time-varying DGPs are reported in Appendix \ref{app:additional_ccc_power}. 

\subsection{Split-Sample Selection and Estimation}
\label{sec: split}

In Section \ref{sec: selectvstwoway}, the DID-INT specification is
selected using the CCC test, and the ATT is then estimated using the
same sample. Because the selected specification is itself data-dependent,
using the same observations for both specification selection and ATT
estimation raises a potential post-model-selection inference concern.
However, Table \ref{tab:selected_twoway_inference} directly evaluates
the finite-sample inference behavior of this full-sample procedure and
finds no evidence of over-rejection in the DGPs considered. As an
additional exercise, I examine sample splitting as a way to separate
the observations used for specification selection from those used for
ATT estimation. I split the sample within each region-period cell
randomly at the individual level into two mutually exclusive subsamples. The \textit{selection} sample is used only to conduct the CCC test and select the DID-INT specification, while the \textit{estimation} sample is used to estimate the ATT under the specification selected using the selection sample. Separating these two steps prevents the individual observations used for ATT estimation from also determining the selected specification. I consider several selection-estimation sample splits, allocating 10\%, 20\%, 40\%, 50\%, 60\%, 80\%, and 90\% of the sample to the CCC-based model-selection step, with the remaining observations used for ATT estimation.

Although individual-level sample splitting separates the observations
used for specification selection and ATT estimation, it may not fully
separate the two stages when individuals within a region-period cell
are correlated due to within-cluster dependence. Ideally, the sample should be split at the region level, so that the regions used for model selection are different from those used for ATT estimation. However, in many DID applications, the number of treated and control regions is relatively small. Therefore, splitting the available regions across the two subsamples may reduce the power of the CCC test and the subsequent ATT estimates, while also making jackknife standard errors less reliable. Furthermore, the Monte Carlo design considered in this sample does not include within-cluster correlation. As a result, the individual-level split I use in this exercise is less problematic. In empirical settings with within-region dependence, however, a region-level split may be preferable when feasible. How best to implement sample splitting in such settings, particularly when the number of available regions is small, remains an important area for future research.

\begin{table}[H]
\centering
\caption{Split-Sample DID-INT Performance Relative to Full-Sample Two-Way DID-INT}
\label{tab:test_train_split}
\resizebox{\textwidth}{!}{
\begin{tabular}{llcccccccc}
\toprule
& &
\multicolumn{7}{c}{Selection/Estimation Split}
& Full-Sample \\
\cmidrule(lr){3-9}
DGP
& Statistic
& 10/90
& 20/80
& 40/60
& 50/50
& 60/40
& 80/20
& 90/10
& Two-way \\
\midrule

\multirow{4}{*}{Homogeneous}
& Bias
& -0.002
& 0.001
& 0.002
& 0.001
& 0.001
& 0.000
& 0.024
& -0.001 \\

& Relative Efficiency
& 1.127
& 0.980
& 0.737
& 0.616
& 0.479
& 0.255
& 0.125
& 1.000 \\

& Empirical Rejection Frequency
& 0.037
& 0.040
& 0.041
& 0.035
& 0.044
& 0.035
& 0.040
& 0.039 \\

& Correct Selection Frequency
& 0.946
& 0.945
& 0.944
& 0.947
& 0.950
& 0.947
& 0.949
& -- \\

\midrule

\multirow{4}{*}{Region-varying}
& Bias
& 0.032
& 0.003
& 0.003
& 0.000
& 0.001
& 0.003
& 0.025
& -0.001 \\

& Relative Efficiency
& 1.018
& 0.962
& 0.721
& 0.591
& 0.470
& 0.251
& 0.125
& 1.000 \\

& Empirical Rejection Frequency
& 0.048
& 0.046
& 0.039
& 0.041
& 0.047
& 0.036
& 0.044
& 0.039 \\

& Correct Selection Frequency
& 0.695
& 0.936
& 0.933
& 0.933
& 0.949
& 0.952
& 0.950
& -- \\

\midrule

\multirow{4}{*}{Time-varying}
& Bias
& -0.004
& 0.001
& 0.003
& 0.001
& 0.000
& 0.001
& 0.025
& -0.001 \\

& Relative Efficiency
& 1.121
& 0.990
& 0.746
& 0.614
& 0.483
& 0.254
& 0.126
& 1.000 \\

& Empirical Rejection Frequency
& 0.029
& 0.039
& 0.038
& 0.033
& 0.038
& 0.036
& 0.040
& 0.039 \\

& Correct Selection Frequency
& 0.823
& 0.949
& 0.949
& 0.948
& 0.947
& 0.943
& 0.948
& -- \\

\midrule

\multirow{4}{*}{Two-way}
& Bias
& 0.024
& 0.005
& 0.010
& 0.008
& 0.008
& 0.006
& 0.021
& -0.002 \\

& Relative Efficiency
& 0.832
& 0.840
& 0.627
& 0.507
& 0.404
& 0.209
& 0.100
& 1.000 \\

& Empirical Rejection Frequency
& 0.049
& 0.033
& 0.044
& 0.037
& 0.034
& 0.042
& 0.059
& 0.040 \\

& Correct Selection Frequency
& 0.911
& 0.999
& 1.000
& 1.000
& 1.000
& 1.000
& 1.000
& -- \\

\bottomrule
\end{tabular}
}
\end{table}

The remaining features of the simulation design, including the number of regions, periods, and observations per region-period cell, are the same as in Section \ref{sec: mcdgp}. In these simulations, I conduct 1,000 Monte Carlo replications and set the true ATT equal to zero. The results are reported in Table \ref{tab:test_train_split}. The results in Table \ref{tab:test_train_split} show that sample
splitting generally does not improve efficiency relative to estimating
the two-way DID-INT on the full sample, since the relative efficiency
measures are below one for most sample splits. This contrasts with the full-sample selection-and-estimation procedure, which can be more efficient than the full-sample two-way specification when the true model is more parsimonious. Accordingly, the empirical application uses the full sample for both
CCC-based specification selection and ATT estimation. The Monte Carlo
evidence in Section \ref{sec: selectvstwoway}, particularly the
empirical rejection frequencies reported in Table
\ref{tab:selected_twoway_inference}, shows that the full-sample
selected procedure does not exhibit over-rejection in the DGPs
considered.

\section{Empirical Exercise: The Effect of Ban-the-Box Policy on Occupational Upgrading}
\label{section: Empiricalexercise}

This section demonstrates the utility of the CCC test and the associated DID-INT specification through a novel empirical analysis. Specifically, I analyze how private-sector Ban-the-Box policies influence the income of employed individuals in the U.S. and investigate the mechanisms driving these effects. Furthermore, I compare the results from the DID-INT estimator with those obtained using more common methods, such as TWFE, CS-DID, and BJS, to assess how the choice of estimator affects the conclusions, particularly when CCC violations and staggered policy adoption are present. 

The aim of ``Ban-the-Box” policies is to improve the employment opportunities for people with criminal records. The “Box” refers to a common question on job applications that asks whether applicants have prior criminal convictions. ``Ban-the-Box” means a legal requirement to remove this checkbox from job applications to give people with a criminal conviction a fairer chance of getting a job. Put simply, Ban-the-Box policies mandate that employers delay inquiring about a job applicant’s criminal record until later stages of the hiring process, usually during a background check \citep{doleac2017moving}. BTB policies are typically divided into three main categories. The first category concerns public employers, such as government agencies. The second category pertains to private employers that hold government contracts. The third category includes all private employers \citep{doleac2020unintended}. 

Employers usually aim to hire workers who are highly productive, reliable, and pose low risk. Since productivity is not directly observable at the time of hiring, firms must rely on observable signals, such as education and experience, to evaluate applicants. However, for entry-level or low-skilled jobs, applicants tend to be younger, less experienced, and less educated. In these cases, employers may have fewer signals of applicant quality and may have to depend on alternative signals during screening. One common signal is criminal history, which employers often view as a negative indicator of worker productivity \citep{vuolo2017criminal}. 

Individuals with prior convictions face large penalties in employment and earnings as a result of those convictions \citep{rose2021does}. In addition, these individuals are likely to be less educated and less experienced than those without convictions. Former offenders are also more likely to exhibit violent, dishonest, or other antisocial behaviors \citep{doleac2020unintended}. Employers may also hesitate to hire individuals with criminal records due to various reasons. These include concerns about potential legal liability, unclear legal obligations, the perceived need for greater supervision and increased monitoring costs, and a general lack of trust in applicants with criminal backgrounds \citep{vuolo2017criminal}. The “Ban-the-Box” policy offers equal opportunities for applicants with criminal records by preventing early rejection and allowing candidates to demonstrate their qualifications, skills, and rehabilitation rather than being dismissed solely because of past convictions. 

Since Ban-the-Box policies are intended to improve employment prospects for individuals with criminal records, I begin by reviewing the existing evidence on their effects on key labor market outcomes. The literature generally finds mixed effects of Ban-the-Box policies on labor market outcomes. \cite{rose2021does} examines the overall effects of BTB on employment and earnings in Seattle using a DID approach, where King County, which includes Seattle, serves as the treated region while neighboring counties serve as the control regions. The analysis links administrative records for individuals supervised by the Washington State Department of Corrections to quarterly unemployment insurance earnings data from 1988 through 2016Q2, enabling tracking of employment and earnings over time. These records are also linked to arrest and court data, offering detailed insights into criminal justice involvement, including arrests, charges, convictions, offense dates, and sentencing periods. The analysis finds no statistically significant effect of the Ban-the-Box policy on the overall employment and earnings of individuals released from prison in King County.

A similar study by \cite{jackson2017effect} uses a DID design to examine the effect of Massachusetts’s Ban-the-Box policy on employment and earnings. The Ban-the-Box policy in Massachusetts was implemented in May 2012, removing the “box” for both public and private sector jobs. The treated group includes individuals with convictions whose records were searchable before the BTB policy was implemented but not afterward. Conversely, the control group comprises individuals with convictions whose records remained searchable both before and after the BTB policy was implemented. Their analysis combines criminal history records from the Massachusetts Department of Criminal Justice Information Services with unemployment insurance wage records from state labor agencies, covering the period from 2010Q1 to 2015Q3. They define the pre-period as 2010Q4 to 2012Q1 and the post-period as 2012Q2 to 2015Q3. Overall, they find that the Ban-the-Box policy had negative effects on both employment and earnings.

\cite{craigie2020ban} investigates how Ban-the-Box policies affect public-sector employment among individuals with convictions, using a DID approach. The study uses NLSY97 data from 2005 to 2015 and restricts the sample to individuals with convictions. Individuals living in counties with full or partial Ban-the-Box policies are considered the treated group, while those in counties without such policies are the control group. \cite{craigie2020ban} finds that BTB policies generally increase the likelihood of public-sector employment for individuals with convictions by about 30 percent. \cite{shoag2021ban} also perform a DID analysis to assess the effect of the BTB policy on employment. They draw on data from the National Employment Law Project and employment outcomes from the LEHD Origin Destination Employment Statistics from 2002 to 2013. Unlike \cite{craigie2020ban}, they use whether a person lives in a high-crime neighborhood as a proxy for increased exposure to criminal-record barriers, since individual criminal records are not observed from the data. High crime neighborhoods are defined using data from the National Neighborhood Crime Study. Their estimates suggest that “Ban-the-Box” policies increased employment among residents of high-crime neighborhoods by up to 4 percent.

Other research explores the unintended effects of Ban-the-Box policies. \cite{agan2018ban} conducted a field experiment on private sector Ban-the-Box laws in New Jersey and New York, in which they submitted about 15,000 fictitious online job applications in matched pairs of white and black applicants. The felony conviction status of each applicant was randomly assigned, while other characteristics of individuals in each pair remained unchanged. Using a triple-differences approach, they observed that without Ban-the-Box, applicants with felonies had lower callback rates than those without. After the Ban-the-Box policy was enacted, callback rates for Black male applicants declined, whereas those for White applicants rose. These results imply that Ban-the-Box might unintentionally increase statistical discrimination, as employers may depend more on race as a substitute for criminal history when direct felony information is unavailable. 

\cite{doleac2020unintended} also find that Ban-the-Box policies decrease employment among young, low-skilled Black and Hispanic men, based on data from the Current Population Survey from 2004 to 2014. The variation in treatment is coded at the state and local metropolitan-area levels. Their analysis specifically examines public-sector Ban-the-Box laws. 
\cite{kaestner2024ban} also investigate how Ban-the-Box policies influence racial statistical discrimination by building on \cite{doleac2020unintended} and extending the analysis through 2019, a period when these policies became more common. Unlike \cite{doleac2020unintended}, the study examines private-sector Ban-the-Box policies alongside public-sector ones. Using monthly CPS data from 2004 to 2019, they assess the effects of Ban-the-Box laws on employment rates among low-educated men from different racial backgrounds. Like \cite{doleac2020unintended}, they find that these laws decreased employment for young Black men between 2004 and 2014. However, this impact is no longer statistically significant when analyzing data through 2019. No notable statistically significant effects on employment are observed for young white and Hispanic men. Therefore, the authors conclude that Ban-the-Box laws do not lead to discrimination against young Black men. Both \cite{doleac2020unintended} and \cite{kaestner2024ban} use the TWFE method to assess the impact of Ban-the-Box policies, which are implemented in various metropolitan areas at different times. In cases with staggered adoption like this, TWFE can face challenges such as negative weighting and forbidden comparisons, which can lead to inconsistent estimates of the policy effect \citep{goodman2021difference}.

This study examines whether private-sector Ban-the-Box policies improve job quality for those already employed. Job quality is assessed using higher O*NET Job Zone rankings, which serve as a proxy for higher-quality and, therefore, higher-paying jobs. The idea is that Ban-the-Box policies could make initial screening of applicants with criminal records more challenging, potentially raising callback rates for these individuals. Although they might still be screened later in the process, they have a chance to showcase their skills, qualifications, and rehabilitation during interviews. This could boost their likelihood of securing higher-quality jobs. Consequently, access to better employment opportunities may also lead to higher earnings.

I contribute to the existing literature in three ways. First, while most studies on Ban-the-Box focus on employment outcomes and statistical discrimination, I examine whether private-sector Ban-the-Box policies improve access to higher-quality occupations by merging American Community Survey (ACS) data with O*NET Job Zone information, which allows me to assess whether BTB is associated with occupational upgrading. To the best of my knowledge, this is the first study that analyzes the effect of Ban-the-Box policies on occupational upgrading. Second, I move beyond the TWFE approach to more modern DID methods to analyze the effect of Ban-the-Box policies, including CS-DID \citep{callaway2021difference}, BJS \citep*{borusyak2024revisiting} and DID-INT \citep{karim2024good}. Most papers in the literature have used the conventional TWFE method to estimate the effect of Ban-the-Box policies, but this approach can lead to inconsistencies due to negative weights and forbidden comparisons \citep{goodman2021difference}. The CS-DID and DID-INT methods are robust to these issues. In addition, the DID-INT estimator is also robust to CCC violations \citep{karim2024good}. 

Third, I base my analysis on the Ban-the-Box policy dates reported by \cite{avery2021ban}. Table \ref{tab:btb_policy_adoption} presents the states with statewide private-sector Ban-the-Box laws, together with their announcement and implementation years. The main empirical analysis uses the implementation year as the treatment date, while the announcement year is used to define the pre-treatment matching window.

\begin{table}[htbp]
\centering
\caption{Private-Sector Ban-the-Box Policy Adoption by State}
\label{tab:btb_policy_adoption}
\small
\begin{tabular}{lcc}
\hline
State & Year of Announcement & Year of Implementation \\
\hline
California    & 2017 & 2018 \\
Colorado      & 2019 & 2019 \\
Connecticut   & 2016 & 2017 \\
Illinois      & 2014 & 2015 \\
Maine         & 2021 & 2021 \\
Maryland      & 2019 & 2020 \\
Massachusetts & 2010 & 2010 \\
Minnesota     & 2013 & 2014 \\
New Jersey    & 2014 & 2015 \\
New Mexico    & 2019 & 2019 \\
Oregon        & 2015 & 2016 \\
Rhode Island  & 2013 & 2014 \\
Vermont       & 2016 & 2017 \\
Washington    & 2018 & 2018 \\
Hawaii        & 1998 & 1998 \\
\hline
\end{tabular}
\end{table}

\FloatBarrier

\subsection{Data}

I use individual-level data from the American Community Survey (ACS), a repeated cross-sectional dataset on workers across the U.S. The main analysis uses the ACS survey years 2010 to 2023, after adjusting for the timing of reported income. This adjustment is necessary because the ACS reports income information for the previous calendar year. Therefore, the survey year is shifted back by one year so that the outcome aligns more closely with the period during which income is reported. Each person’s race and sex remain the same, but their age is decreased by one year, reflecting that they were likely a year younger in the previous survey year. Pre-treatment trends are analyzed separately using ACS data from 1999 to 2006 because the first relevant private-sector Ban-the-Box policy in the analysis sample was announced in 2013.

To ensure that the estimated effects are more closely tied to private-sector Ban-the-Box policies, the sample is restricted to employed individuals and private-sector workers. All currently employed government workers, self-employed individuals, unpaid family workers, and those working in nonprofit organizations are excluded. Since Ban-the-Box policies are most likely to benefit young workers entering the labor force or relatively early in their careers, the sample is further restricted to workers aged 18 to 34. Annual wage income is transformed to logged wages. Age groups are split into two categories: 18-23 for new entrants and 24-34 for early-career workers. Race and ethnicity are categorized into four mutually exclusive groups: non-Hispanic White, non-Hispanic Black, Hispanic, and other race/ethnicity. Sex is represented by a male indicator.

Rather than using all remaining states as control groups, I use a covariate-based matching procedure to select a set of suitable control states. This approach serves three key purposes. First, it ensures the control states are more comparable to the treated states in observable traits, thereby improving covariate balance. Second, it likely improves the plausibility of conditional or integrated parallel trends assumptions. Third, it enhances the practicality of using estimators such as CSDID, which are often computationally intensive.  I also restrict the sample to clean control states that do not adopt a statewide private-sector Ban-the-Box policy during the period considered. These include states that lack a private-sector Ban-the-Box policy altogether and states where Ban-the-Box policies apply only to particular local jurisdictions, such as a city. For instance, several local jurisdictions in New York State, such as New York City, Buffalo, and Rochester, adopted municipal-level private Ban-the-Box policies, although the policy is not enforced statewide \cite{avery2021ban}. A list of states that do not have a state-level private-sector Ban-the-Box implementation is included in column 3 of Table \ref{tab:control_state_classification}. Since the private-sector Ban-the-Box policies do not apply uniformly to all workers in these states, I exclude them from the analysis.  

I use the nearest-neighbor matching procedure based on Euclidean distance \citep[Section 13.2, pg. 459]{hastie2009elements} to select an equal number of treated and control states. The matching variables include average log wage, the share of new entrants aged 18 to 23, the male share, and the shares of non-Hispanic Black, Hispanic, and other race/ethnicity groups. The variables are first aggregated to the state level using ACS person weights and measured over each treated state’s relevant pre-announcement period. Before calculating the Euclidean distance using the formula shown in Equation \eqref{eq:matching_distance}, the matching variables are normalized using their cross-state means and standard deviations within the relevant matching window. Since states announce private-sector Ban-the-Box policies in different years, the matching window is state-specific: for each treated state, observations from 2010 through the year immediately preceding the policy announcement are used. Consequently, the matching procedure relies exclusively on pre-treatment information.
For treated state $s$ and candidate control state $c$, the Euclidean distance is calculated as:
\begin{equation}
\label{eq:matching_distance}
d_{s,c}
=
\sqrt{
\sum_{k=1}^{K}
\left(
Z_{k,s}-Z_{k,c}
\right)^2
},
\end{equation}
where $d_{s,c}$ denotes the Euclidean distance between treated state $s$ and candidate control state $c$, $K$ denotes the number of matching variables, and $Z_{k,s}$ and $Z_{k,c}$ denote the normalized values of matching variable $k$ for the treated and candidate control states, respectively. Matching is conducted without replacement, so that one treated state is matched to one control state. The resulting matched treated-control state pairs are reported in Table \ref{tab:matched_states}.

\begin{table}[htbp]
\centering
\caption{Matched Treated and Control States}
\label{tab:matched_states}
\begin{tabular}{llc}
\toprule
Treated State & Matched Control State & Control Rank \\
\midrule
California   & Nevada        & 1 \\
Colorado     & Nebraska      & 2 \\
Connecticut  & Wisconsin     & 2 \\
Illinois     & Virginia      & 1 \\
Maine        & Kentucky      & 1 \\
Maryland     & Delaware      & 2 \\
Minnesota    & South Dakota  & 3 \\
New Jersey   & Oklahoma      & 4 \\
New Mexico   & Arizona       & 1 \\
Oregon       & Kansas        & 1 \\
Rhode Island & Michigan      & 1 \\
Vermont      & Montana       & 1 \\
Washington   & Alaska        & 3 \\
\bottomrule
\end{tabular}
\end{table}

To assess covariate balance, I compare the normalized differences in means (Norm Diff.) before and after matching \citep[Section 4.1]{baker2026difference}. For matching variable $k$, the Norm Diff. is calculated as:
\begin{equation}
\label{eq:smd}
\text{Norm Diff.}_k
=
\frac{
\bar{X}_{k,T}-\bar{X}_{k,C}
}{
\sqrt{
\left(
s_{k,T}^{2}+s_{k,C}^{2}
\right)/2
}
},
\end{equation}
where $\bar{X}_{k,T}$ and $\bar{X}_{k,C}$ denote the treated and control means, respectively, and $s_{k,T}^{2}$ and $s_{k,C}^{2}$ denote their corresponding variances. The absolute mean Norm Diff. for all matching variables fell among the candidate control states compared to the overall sample, indicating improved covariate balance.

\begin{table}[htbp]
\centering
\caption{Covariate Balance Before and After Matching}
\label{tab:matching_balance}
\begin{tabular}{lcc}
\toprule
Variable
& Absolute Norm Diff. Before
& Absolute Norm Diff. After \\
\midrule
Mean log wage
& 0.670
& 0.389 \\

New entrant share (18--23)
& 0.511
& 0.476 \\

Male share
& 0.106
& 0.089 \\

Non-Hispanic Black share
& 0.527
& 0.140 \\

Hispanic share
& 0.826
& 0.573 \\

Other race/ethnicity share
& 0.648
& 0.061 \\
\midrule
Mean absolute SMD
& 0.548
& 0.288 \\
\bottomrule
\end{tabular}
\end{table}

After selecting the candidate control states, I construct the job quality measure using O*NET Job Zone data. This dataset sorts occupations into five levels based on the education, training, and experience required. I convert this into a $high_{jz}$ indicator, where Job Zones 1 and 2 are considered lower quality and Job Zones 3, 4, and 5 are considered higher quality. When an OCC2010 category maps to occupations on both sides of the High Job Zone threshold, the binary outcome is left missing rather than assigning the category arbitrarily to either group. Next, I merge the ACS and Job Zone datasets using harmonized occupation codes to ensure comparability across occupational classifications. This harmonization is necessary because the ACS reports occupations using the harmonized OCC2010 classification, whereas the O*NET Job Zone data use more recent SOC codes. To align the datasets, I use a manually compiled OCC2010 mapping, along with the official BLS 2010-to-2018 SOC crosswalk, to map ACS occupation codes to the 2018 SOC classification used to link occupations to O*NET Job Zones. Appendix \ref{sec:occ_harmonization} provides details of the harmonization.

\subsection{Threats to Identification}

The raw ACS data presents several identification threats. These threats may cause the estimated ATT to capture changes in outcomes that are not attributed to Ban-the-Box policies. First, ACS income does not correspond directly to the survey year. Instead, it reflects income received over the past 12 months, which may not fall within the survey's calendar year. Ignoring this timing issue can lead to wrong treatment assignment based on when the income was earned. For example, income reported by a respondent in a given ACS year may partly reflect the previous year's labor market conditions. To address this, I shift the ACS survey year back by one year, making the outcome align more closely with the income-earning period. 

Second, the early years of the sample align with changes in the federal minimum wage. Under the Fair Minimum Wage Act of 2007, the federal minimum wage increased in three stages. It rose to \$5.85 per hour effective July 24, 2007, to \$6.55 per hour effective July 24, 2008, and to \$7.25 per hour effective July 24, 2009 \citep{dol_minimum_wage}. The federal minimum wage has stayed at \$7.25 per hour for work done on or after July 24, 2009. These increases may affect wages independently, posing a threat to identification if wage growth driven by the federal minimum wage is wrongly attributed to Ban-the-Box adoption. 

To address this concern, I exclude observations from 2007 to 2009 after adjusting the ACS year to align with the income reference period. This restriction aligns with the long-difference structure of the empirical designs of CS-DID and DID-INT, in which the period immediately before treatment serves as the relevant pre-treatment period. By excluding the years directly affected by federal minimum wage increases, the analysis ensures that the pre-treatment period for each treated state begins after the federal minimum wage has reached its current level. Assuming the federal minimum wage changes affected states similarly reduces the risk that the estimated ATT reflects the impact of the federal wage increase rather than the Ban-the-Box policies. I also exclude Massachusetts, which was treated in 2010, from the analysis to guarantee that every treated state has a valid pre-treatment comparison period following the federal minimum wage changes. Hawaii is also excluded because its private-sector Ban-the-Box policy was implemented in 1998, well before the beginning of the main analysis period. I also exclude these years in the parallel trend diagnostics.

A separate identification issue arises from the staggered rollout of private-sector Ban-the-Box policies. Because different states adopted these policies in different years, standard TWFE estimators may not consistently estimate the ATT. Therefore, I use the CS-DID and BJS estimators, which avoid many of the issues that conventional TWFE faces in staggered adoption settings \citep*{callaway2021difference, borusyak2024revisiting}. Additionally, states vary considerably in their policy environments, legislative institutions, demographic makeup, and job market structure. These differences can cause the relationship between covariates and untreated potential outcomes to vary across states and over time, leading to a two-way CCC violation \citep{karim2024good}. Many existing DID estimators are unable to account for CCC violations of this kind. Therefore, I apply the CCC test to identify CCC violations and use the corresponding DID-INT specification for CCC violations. I also restrict the sample to clean control states that do not adopt a statewide private-sector Ban-the-Box policy during the period considered.   

\subsection{Methodology}

To estimate the effect of private-sector Ban-the-Box policy on earnings and occupational upgrading, I first estimate a conventional TWFE regression, shown below:
\begin{equation}
Y_{i,s,t} = \alpha_s + \lambda_t + \beta D_{s,t} + X_{i,s,t}' \gamma + \varepsilon_{ist},
\end{equation}

Here, $Y_{i,s,t}$ is the outcome for individual $i$ in state $s$ and year $t$. The outcome is the log wage  when estimating the effect of Ban-the-Box policies on earnings, and is the high Job Zone indicator, $high_{jz}$, when analyzing the effect on occupational upgrading. $D_{s,t}$ is the interaction term between the treatment indicator and the post indicator, which indicates an observation's exposure to the Ban-the-Box policy. $X_{i,s,t}$ are individual-level covariates, which include race, age, and sex. Race and ethnicity are categorized into four mutually exclusive groups: non-Hispanic White, non-Hispanic Black, Hispanic, and other race/ethnicity. Indicators for non-Hispanic Black, Hispanic, and other race/ethnicity are included, with non-Hispanic White serving as the omitted category. Sex is used to construct a male indicator. Age is grouped into two categories to distinguish new entrants (ages 18–23) from other early-career workers (ages 24–34). $\alpha_s$ and $\lambda_t$ are state and year fixed effects, respectively.

Since the TWFE estimator may be inconsistent in staggered treatment adoption settings, I also estimate the model using the CS-DID estimator with covariates. This approach estimates region and time-specific treatment effects separately, using the period immediately before treatment as the relevant pre-treatment period for each treated group. These state-time treatment effects are then aggregated to obtain the overall estimate of the ATT. Since covariates are included in the CS-DID specification, I use the default \texttt{drdid} method to estimate the ATT. This method implements a doubly robust inverse probability weighting approach by combining inverse probability weighting with regression adjustment. In practice, it reweights observations in the control states so that those with covariate values similar to the treated observations receive greater weight when constructing the counterfactual untreated outcome. Finally, I use the DID-INT estimator to account for CCC violations. In order to select the appropriate specification of DID-INT, I conduct the CCC test described in Section \ref{section: ccctestsection}. 

\subsection{Parallel Trends}

\begin{figure}
    \centering
    \includegraphics[width=1\linewidth]{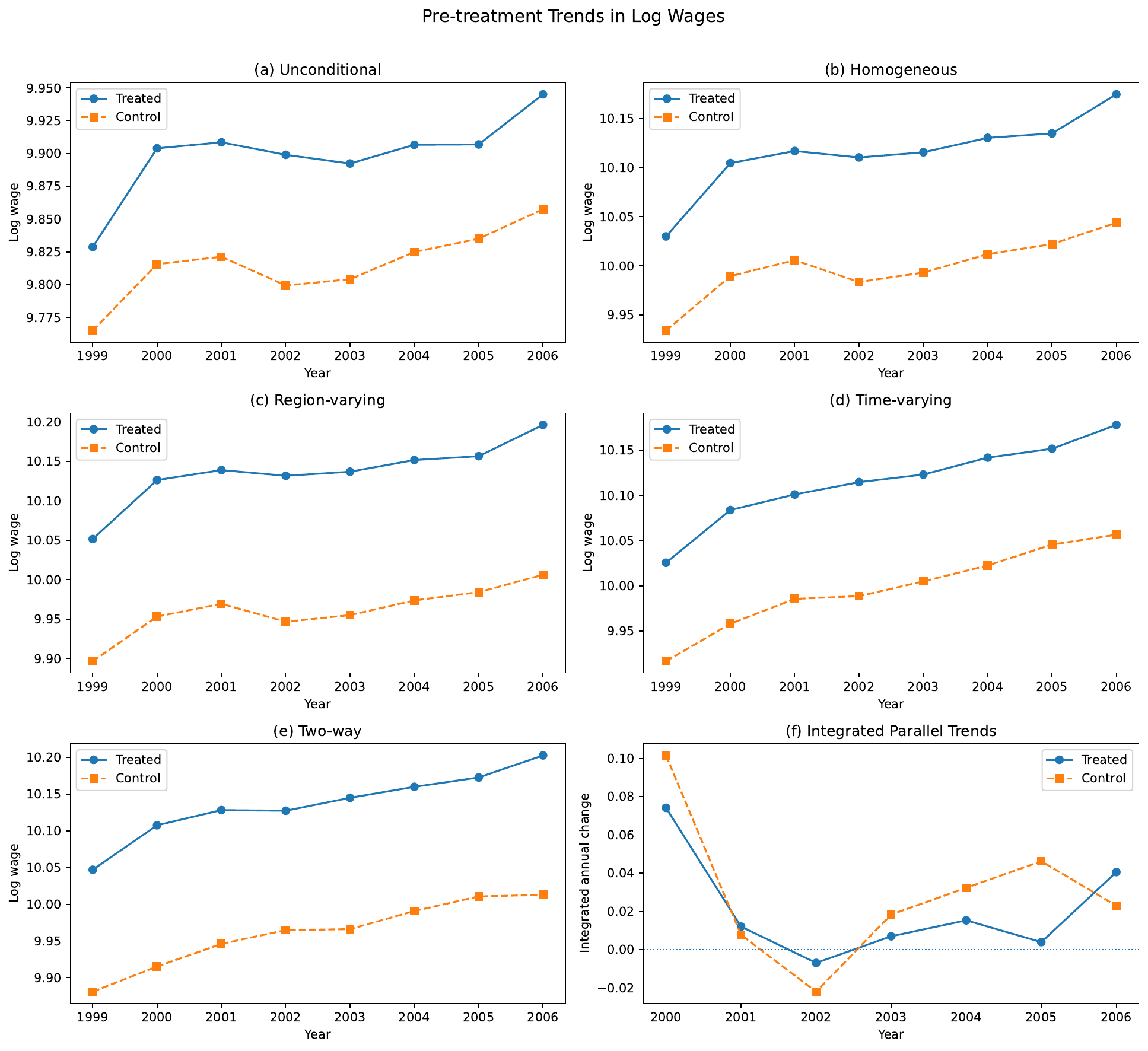}
    \caption{Pre-trend figures for lwage}
    \label{fig:lwagetrends}
\end{figure}

\FloatBarrier

\begin{figure}
    \centering
    \includegraphics[width=1\linewidth]{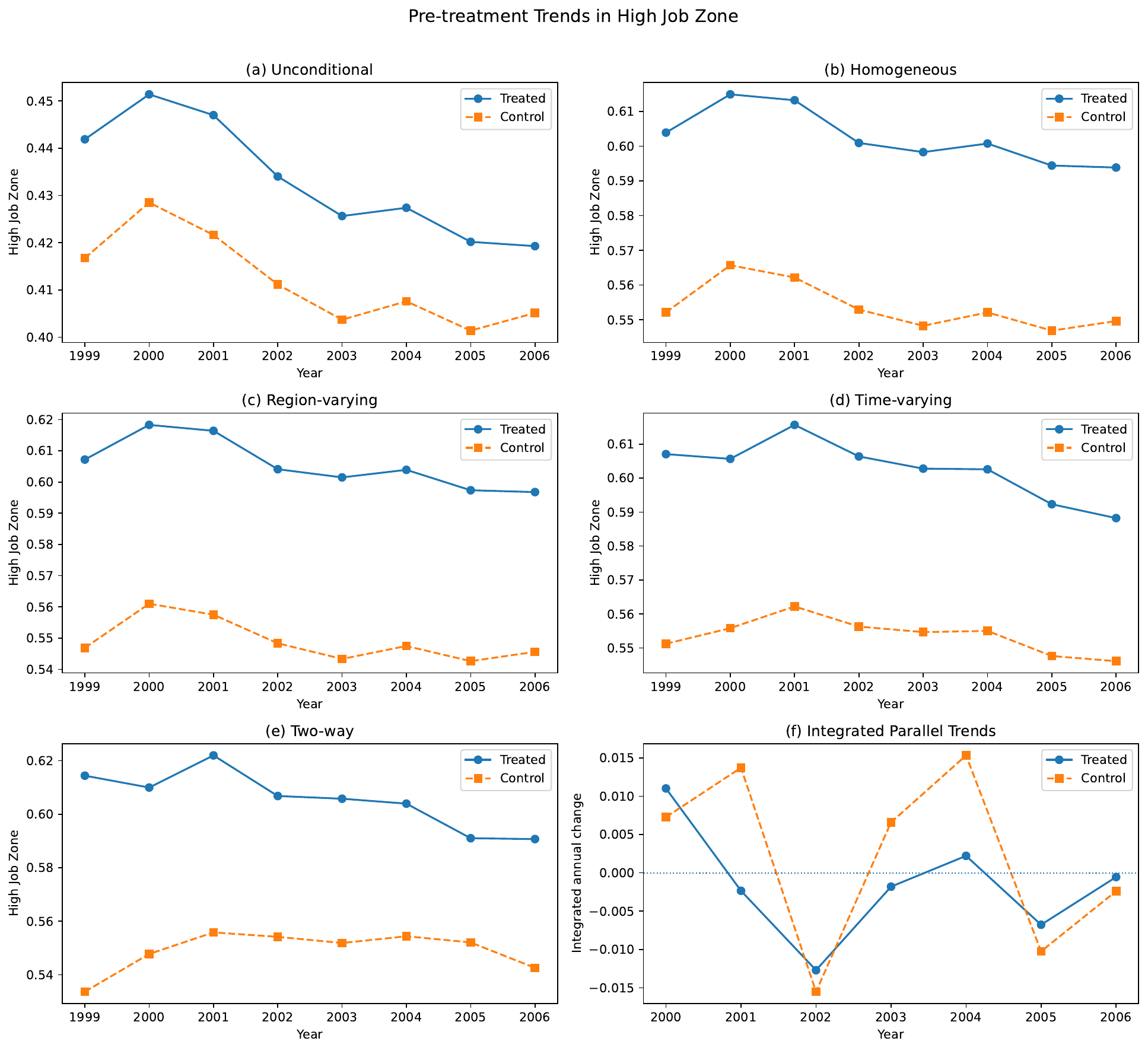}
    \caption{Pre-trend figures for High job zone}
    \label{fig:highjztrends}
\end{figure}

Figures \ref{fig:lwagetrends} and \ref{fig:highjztrends} illustrate the parallel trends under different model specifications for log wages and high Job Zone, respectively. For this pre-trend analysis, I download ACS data for survey years 2000 to 2007. Because the ACS income variables refer to income earned in the preceding year, I shift the survey year back by one year so that the outcomes are aligned with the year in which the income was earned. Therefore, the resulting analysis covers 1999 to 2006, omitting years when the general minimum wage laws were enacted. These figures provide a visual assessment of parallel trends, which makes the evaluation somewhat subjective. In both figures, Panel (a) presents the unconditional trends, while Panel (b) presents the conditional trends obtained by residualizing the outcome under a specification that imposes a common coefficient on the covariates. Panel (c) allows the covariate coefficients to vary across states by interacting the covariates with state indicators, whereas Panel (d) allows them to vary over time by interacting the covariates with year indicators. Panel (e) allows the covariate coefficients to vary jointly across states and over time by interacting the covariates with both state and year indicators. Across these alternative specifications, the treated and control trends appear reasonably parallel in the pre-treatment period. For ease of presentation, the treated states and matched control states are each aggregated into a common trend. Across the alternative specifications, the treated and control trends appear reasonably parallel. 

Panel (f) provides a graphical assessment of integrated parallel trends using the year-to-year differences in expected outcomes over the covariate distribution of the treated states. To construct this plot, I first regress the outcome on the covariates separately for each state and year:
\begin{equation}
Y_{i,s,t} = \theta_{s,t} + \gamma_{s,t}X_{i,s,t} + u_{i,s,t} \quad \forall s,t.
\end{equation}
For each treated state $s$ and its matched control state $s'$, I then use the estimated coefficients to predict outcomes for both states using the mean covariate values of the treated state, $\bar X_{s,t}$. This allows me to compare how predicted outcomes change from one year to the next in the treated and control states while evaluating both states over the same covariate distribution. The corresponding year-to-year change for the treated state is given by:
\begin{equation}
\Delta \widehat{Y}_{s,t}(\overline{X}_{s,t})
=
\widehat{\theta}_{s,t}
+ \widehat{\gamma}_{s,t} \overline{X}_{s,t}
- \widehat{\theta}_{s,t-1}
- \widehat{\gamma}_{s,t-1} \overline{X}_{s,t}.
\end{equation}
Similarly, the corresponding year-to-year change for the control state is given by:
\begin{equation}
\Delta \widehat{Y}_{s',t}(\overline{X}_{s,t})
= \widehat{\theta}_{s',t} + \widehat{\gamma}_{s',t} \overline{X}_{s,t} - \widehat{\theta}_{s',t-1}
- \widehat{\gamma}_{s',t-1} \overline{X}_{s,t}.
\end{equation}

These differences are aggregated across the treated and control states and plotted on a common figure. The trends in Panel (f) appear less parallel than those in the preceding panels. However, the scale of the vertical axis is substantially smaller, so relatively small deviations may appear more pronounced.

\subsection{Results}

Before the empirical analysis, I apply the CCC test to identify which CCC assumption the data support. Based on the results of the CCC test, I select the corresponding DID-INT specification to estimate the effect of the Ban-the-Box policies. The results are shown in Table \ref{tab:ccc_selection}. Columns 2 and 3 report the $\chi^2$ statistic and p-value from the first stage, which tests whether the covariate coefficients are equal across regions and periods. Because only a single joint hypothesis is tested in the first stage, the reported p-values are not adjusted using the Holm procedure. For both outcomes, this null hypothesis is strongly rejected (p-value $<0.001$). Consequently, I proceed to the second stage of the CCC test, which separately evaluates whether covariate effects are equal across regions and over time.
Columns 4 to 7 present the results from the second stage of the CCC test. Columns 4 and 5 report the test statistic and Holm-adjusted p-value for the region-varying restriction, respectively, while Columns 6 and 7 report the corresponding test statistic and Holm-adjusted p-value for the time-varying restriction. For both log wages and High Job Zone, both restrictions are rejected, with Holm-adjusted p-values below 0.001. Since both the region and time restrictions are rejected for these outcomes, the CCC test selects the two-way DID-INT specification.

\begin{table}[htbp]
\centering
\caption{Common Causality of Covariates Tests and DID-INT Specification Selection}
\label{tab:ccc_selection}
\resizebox{\textwidth}{!}{
\begin{tabular}{lccccccc}
\toprule
& \multicolumn{2}{c}{Stage 1}
& \multicolumn{4}{c}{Stage 2}
& \\
\cmidrule(lr){2-3}
\cmidrule(lr){4-7}
Outcome
& $\chi^2(df)$
& $p$-value
& State $\chi^2(df)$
& State $p$-value
& Time $\chi^2(df)$
& Time $p$-value
& Selected Specification \\
\midrule

Log wage
& 12555.790 (1370)
& $<0.001$
& 7182.924 (125)
& $<0.001$
& 554.415 (65)
& $<0.001$
& Two-way \\

High Job Zone
& 10154.866 (1370)
& $<0.001$
& 4569.419 (125)
& $<0.001$
& 190.510 (65)
& $<0.001$
& Two-way \\

\bottomrule
\end{tabular}
}
\end{table}

Table \ref{tab:main_results} presents the estimated ATT of private-sector Ban-the-Box policies on log wages in Row 1 and the probability of being employed in a high Job Zone occupation in Row 2. Column 1 shows the TWFE regression results for the full sample. The findings suggest that private-sector Ban-the-Box policies are associated with approximately a 3.4\% increase in wages, which is statistically significant at the 1\% level. However, no statistically significant effect on the probability of being employed in a high Job Zone occupation is observed. 

Column 2 shows the CS-DID estimates, which account for staggered treatment adoption and are therefore more appropriate than TWFE in this setting. The conventional CS-DID estimates use state-clustered standard errors. The results suggest that private-sector Ban-the-Box policies increase wages by approximately 2.0\%, which is statistically significant at the 10\% level. This is similar in magnitude to the TWFE estimate. However, as with TWFE, the CS-DID estimate for the probability of being employed in a high Job Zone occupation is not statistically significant. Column 3 reports the same CS-DID point estimates using state-level cluster jackknife inference, as proposed by \citet*{karim2026improved}. Under the cluster jackknife, the estimated wage effect is no longer statistically significant, while the High Job Zone estimate remains statistically insignificant. Column 4 reports the estimates from the BJS estimator. The estimated effect on log wages is approximately 3.5\% and is statistically significant at the 10\% level. The estimated effect on the probability of being employed in a high Job Zone occupation is positive, at approximately 0.7\%, but is not statistically significant.

Column 5 reports the results from the CCC-selected DID-INT specification, with standard errors calculated using the clustered jackknife method. The estimated effect on log wages is positive and statistically significant at the 5\% level, suggesting that private-sector Ban-the-Box policies increase wages by approximately 2.2\%. The estimated effect on the probability of being employed in a high Job Zone occupation is also positive, at approximately 0.7\%, and is statistically significant at the 10\% level. Table \ref{tab:didint_alternative} reports the DID-INT estimates across the alternative covariate specifications and shows that the estimated effects on both log wages and High Job Zone remain positive and statistically significant across all three specifications.

\begin{table}[htbp]
\centering
\caption{Estimated Effects of Private-Sector Ban-the-Box Policies}
\label{tab:main_results}

{\footnotesize
\begin{tabular}{lccccc}
\toprule
& TWFE & CS-DID & CS-DID Jackknife & BJS & CCC-Selected DID-INT \\
\midrule

\multicolumn{6}{l}{\textit{Panel A: Log Wage}} \\[0.3em]

ATT
& 0.0339**
& 0.0203*
& 0.0203
& 0.0346*
& 0.0223** \\
& (0.0139)
& (0.0112)
& (0.0144)
& (0.0175)
& (0.0112) \\

Observations
& 2,164,444
& 2,164,444
& 2,164,444
& 2,164,444
& 2,164,444 \\

\addlinespace

\multicolumn{6}{l}{\textit{Panel B: High Job Zone}} \\[0.3em]

ATT
& 0.0061
& 0.0061
& 0.0061
& 0.0073
& 0.0066* \\
& (0.0072)
& (0.0041)
& (0.0058)
& (0.0055)
& (0.0035) \\

Observations
& 1,923,354
& 1,923,354
& 1,923,354
& 1,923,354
& 1,923,354 \\

\bottomrule
\end{tabular}
}

\vspace{0.5em}
\begin{minipage}{0.90\textwidth}
\footnotesize
\textit{Notes:} Standard errors are reported in parentheses. Columns 1, 3, 4, and 5 report the cluster jackknife standard errors, while Column 2 uses conventional state-clustered standard errors.
*** $p<0.01$, ** $p<0.05$, * $p<0.10$.
\end{minipage}
\end{table}

The log-wage and High Job Zone outcomes reported in Table \ref{tab:main_results} are observed only among employed private-sector
workers. If Ban-the-Box policies also affect the probability of private-sector employment, the composition of the employed sample may
change following treatment. Therefore, I examine whether private-sector ``Ban-the-Box" policies cause a change in the employment margin. To do so, I expand the sample to
include unemployed individuals and estimate whether private-sector Ban-the-Box policies affect the probability of being employed in the
private-sector rather than unemployed. Next, I construct an unconditional High Job Zone outcome, coding unemployed individuals as
zero, to examine whether the policies affect the overall probability
of being employed in a high-Job-Zone private-sector occupation.
Finally, using only employed private-sector workers, I estimate the
effect on the continuous Job Zone measure to assess whether the
occupational-quality result is robust to using the full Job Zone scale
rather than a binary indicator. All estimates are obtained using the
CCC-selected two-way DID-INT specification.

The results are shown in Table \ref{tab:additional_outcomes}. Column 2 reports the estimated effect of Ban-the-Box policies on private-sector employment relative to unemployment. The policies are associated with a 0.8 \% decline in private-sector employment, and the estimate is statistically significant at the 5\% level. Column 3 reports the effect on the unconditional High Job Zone outcome, where unemployed individuals are coded as zero. The estimated effect is small and statistically insignificant, suggesting that the policies do not increase the overall probability of being employed in a high-Job-Zone private-sector occupation. Column 4 reports the effect on the continuous Job Zone measure among employed private-sector workers. The estimate is positive and statistically significant at the 5\% level, indicating an increase of approximately 0.018 Job Zone units. The results suggest that the effect of Ban-the-Box policies on occupational upgrading remains concentrated among individuals who are already employed in the private sector. However, this occupational upgrading among employed private-sector workers is accompanied by a small and statistically significant decline in private-sector employment relative to unemployment. The positive and statistically significant effect on the continuous Job Zone measure reinforces the earlier High Job Zone result, suggesting that occupational upgrading is not driven solely by the binary cutoff. This provides an additional robustness check on the high Job Zone results in Table \ref{tab:main_results}.

\begin{table}[htbp]
\centering
\caption{Employment and Occupational Outcomes}
\label{tab:additional_outcomes}
\begin{tabular}{lccc}
\toprule
& Private-Sector & Unconditional & Continuous \\
& Employment & High Job Zone & Job Zone \\
\midrule

ATT
& -0.0076**
& 0.0021
& 0.0180** \\
& (0.0036)
& (0.0040)
& (0.0080) \\

Observations
& 2,416,792
& 2,175,702
& 2,135,016 \\

\bottomrule
\end{tabular}

\vspace{0.5em}
\begin{minipage}{0.90\textwidth}
\footnotesize
\textit{Notes:} Cluster jackknife standard errors are reported in parentheses.
*** $p<0.01$, ** $p<0.05$, * $p<0.10$.
\end{minipage}
\end{table}

Motivated by the results in \cite{agan2018ban}, I next report the
estimated effects of private-sector Ban-the-Box policies separately
across racial and ethnic groups. To do so, I estimate the effects of Ban-the-Box policies on both wages and the binary High Job Zone indicator separately for non-Hispanic White, non-Hispanic Black, and Hispanic individuals. I continue to use sex and age as controls and the same employed private-sector sample as in the main analysis. For the employment and unconditional High Job Zone outcomes, I instead use the expanded sample, which includes both unemployed individuals and employed private-sector workers.

Table \ref{tab:race_heterogeneity} presents the estimated effects separately by race and ethnicity. Column 1 reports the estimates for non-Hispanic White individuals, Column 2 reports the estimates for non-Hispanic Black individuals, and Column 3 reports the estimates for Hispanic individuals. For non-Hispanic White individuals, there is weak evidence of an increase in wages and a decline in private-sector employment, with both estimates statistically significant at the 10\% level. For non-Hispanic Black and Hispanic individuals, none of the estimated effects are statistically significant. Most importantly, the High Job Zone and unconditional High Job Zone estimates are statistically insignificant across all three groups, providing no clear evidence that the occupational-upgrading effect is concentrated within a particular racial or ethnic group.

\begin{table}[htbp]
\centering
\caption{Heterogeneity by Race and Ethnicity}
\label{tab:race_heterogeneity}
\begin{tabular}{lccc}
\toprule
& Non-Hispanic White & Non-Hispanic Black & Hispanic \\
\midrule

\multicolumn{4}{l}{\textit{Panel A: Employed Private-Sector Sample}} \\[0.3em]

Log Wage
& 0.0217*
& -0.0101
& 0.0161 \\
& (0.0122)
& (0.0300)
& (0.0210) \\[0.4em]

High Job Zone
& 0.0058
& 0.0079
& 0.0030 \\
& (0.0053)
& (0.0116)
& (0.0165) \\

\addlinespace
\multicolumn{4}{l}{\textit{Panel B: Expanded Labor-Force Sample}} \\[0.3em]

Private-Sector Employment
& -0.0064*
& 0.0022
& -0.0084 \\
& (0.0036)
& (0.0084)
& (0.0101) \\[0.4em]

Unconditional High Job Zone
& 0.0019
& 0.0056
& -0.0005 \\
& (0.0048)
& (0.0121)
& (0.0175) \\

\bottomrule
\end{tabular}

\vspace{0.5em}
\begin{minipage}{0.90\textwidth}
\footnotesize
\textit{Notes:} Cluster jackknife standard errors are reported in parentheses. *** $p<0.01$, ** $p<0.05$, * $p<0.10$.
\end{minipage}
\end{table}
\FloatBarrier

\section{Conclusion}

This paper emphasizes the importance of systematically evaluating how covariates are incorporated into DID models. While researchers often use visual checks of conditional parallel trends, these methods are subjective and may be limited when different covariate specifications produce similar pre-treatment patterns. The CCC test complements these visual checks by offering a structured approach for determining whether covariate effects remain stable across regions, periods, or both. Although the CCC test does not directly test parallel trends, it can guide researchers in choosing the appropriate covariate specification in DID and help avoid overly restrictive assumptions about covariates.

The theoretical results show that imposing the correct covariate structure can improve efficiency under homoskedasticity, whereas imposing overly restrictive structures can introduce misspecification bias. The Monte Carlo simulations further show that the CCC test has good size and power and can help select the appropriate DID-INT specification. The empirical application to private-sector Ban-the-Box policies illustrates the practical importance of this procedure. I first assess whether these policies resulted in higher wages for employed workers, then explore whether these wage increases correspond to moves into higher-ranked occupations. Both the TWFE and CS-DID methods indicate a positive and statistically significant effect on wages. The CCC test reveals a two-way CCC violation, leading to the use of the two-way DID-INT specification. Using this specification, the estimated wage effect is positive and statistically significant at the 5\% level. For high Job Zones, the TWFE and CS-DID estimates are not statistically significant, whereas the CCC-selected DID-INT estimate is positive and significant at the 10\% level, implying that Ban-the-Box policies may increase access to better jobs. Additional analyses that expand the sample to include both unemployed individuals and private-sector workers show a small and statistically significant decline in private-sector employment and no significant change in the unconditional likelihood of working in a high Job Zone, although the positive effect on occupational quality persists when using the continuous Job Zone measure. Subgroup analyses provide limited evidence that these effects are concentrated within specific racial or ethnic groups.

\clearpage

\bibliography{mybib}

\appendix

\section{Proof of Theorem \ref{theorem: nesting}}
\label{app3: nesting}
\begin{proof}[Proof of Theorem \ref{theorem: nesting}]
Since the region-period intercept component is common across the alternative DID-INT specifications, it is sufficient to establish nesting by considering their respective covariate components. The covariate component of the homogeneous DID-INT specification is given by:
\begin{equation}
\label{eq: homcovcomp}
    \sum_{k=1}^K \gamma^{*}_{k} X^k_{i,s,t}.
\end{equation}
Since the homogeneous model is correctly specified, $\gamma^{\circ}_{k} = \gamma^{*}_{k}$, where $\gamma^{*}_{k}$ is the probability limit of the estimated coefficient on the $k^{th}$ covariate. Similarly, the covariate component of the region-varying DID-INT specification is given by:
\begin{equation}
\label{eq: regioncovcomp}
    \sum_{k=1}^K \sum_{s=1}^S \gamma_{k,s} I(s)X^k_{i,s,t}.
\end{equation}
Under Assumption \ref{assumption: twowayCCC}, and the maintained regularity conditions, the homogeneous coefficient can be recovered as a special case of the region-varying specification, such that:
\begin{equation}
    \gamma_{k,s} = \gamma^{*}_{k} \quad \forall k, s.
\end{equation}
Under this restriction, Equation \eqref{eq: regioncovcomp} reduces to:
\begin{equation}
\label{eq: regioncovcomp2}
    \sum_{k=1}^K \gamma^{*}_{k} \biggr[\sum_{s=1}^S I(s)\biggr]X^k_{i,s,t}.
\end{equation}
Since each observation belongs to exactly one region:
\begin{equation}
    \biggr[\sum_{s=1}^S I(s)\biggr] = 1.
\end{equation}
Therefore, Equation \eqref{eq: regioncovcomp2} can be simplified to:
\begin{equation}
\label{eq: regioncovcomp3}
    \sum_{k=1}^K \gamma^{*}_{k} X^k_{i,s,t}.
\end{equation}
Equation \eqref{eq: regioncovcomp3} is exactly the covariate component of the homogeneous specification, as shown in Equation \eqref{eq: homcovcomp}. Hence,
\begin{equation}
\label{eq: statement1}
\Gamma_H \subset \Gamma_R.
\end{equation}

The covariate component of the two-way DID-INT specification is given by:
\begin{equation}
\label{eq: twowaycovcomp}
    \sum_{k=1}^K \sum_{s=1}^S \sum_{t=1}^T \gamma_{k,s,t} I(s)I(t)X^k_{i,s,t}.
\end{equation}
Since the region-varying DID-INT specification contains the homogeneous coefficient structure as a special case under Assumption \ref{assumption: twowayCCC}, $\gamma_{k,s} = \gamma^*_{k} = \gamma^{\circ}_{k}$. The region-varying specification can also be recovered as a special case of the two-way specification by imposing:
\begin{equation}
    \gamma_{k,s,t} = \gamma_{k,s} \quad \forall k,s,t.
\end{equation}
Under this restriction, Equation \eqref{eq: twowaycovcomp} reduces to:
\begin{equation}
\label{eq: twowaycovcomp2}
    \sum_{k=1}^K \sum_{s=1}^S  \gamma_{k,s} I(s) \biggr[\sum_{t=1}^T I(t)\biggr] X^k_{i,s,t}.
\end{equation}

Since each observation belongs to exactly one period:
\begin{equation}
    \biggr[\sum_{t=1}^T I(t)\biggr] = 1.
\end{equation}
Therefore, Equation \eqref{eq: twowaycovcomp2} can be simplified to:
\begin{equation}
\label{eq: twowaycovcomp3}
    \sum_{k=1}^K \sum_{s=1}^S  \gamma_{k,s} I(s) X^k_{i,s,t}.
\end{equation}
Equation \eqref{eq: twowaycovcomp3} is exactly equal to the covariate component of the region-varying specification, as shown in Equation \eqref{eq: regioncovcomp}. Hence,
\begin{equation}
\label{eq: statement2}
\Gamma_R \subset \Gamma_{TW}.
\end{equation}
Combining equations \eqref{eq: statement1} and \eqref{eq: statement2} yields:
\begin{equation}
\Gamma_H \subset \Gamma_R \subset \Gamma_{TW}.
\end{equation}
The second nesting relationship can be derived using the same steps as shown above, since the time-varying DID-INT specification can also be obtained as a special case of the two-way DID-INT specification by restricting the covariate coefficients to vary only across periods and not across regions. Hence,
\begin{equation}
\Gamma_H \subset \Gamma_T \subset \Gamma_{TW}.
\end{equation}
\end{proof}

\section{Proof of the Mapping Between the Homogeneous and Region-Varying DID-INT Specifications}
\label{appendix: A}

For each observation $i$, the region-varying regressor vector contains the region-specific covariate terms
\[
\left(
1\{S_i=1\}X_{1,i},\ldots,1\{S_i=S\}X_{K,i}
\right)'.
\]
Since each observation belongs to exactly one region,
\[
\sum_{s=1}^{S}1\{S_i=s\}X_{k,i}
=
X_{k,i},
\qquad k=1,\ldots,K.
\]
Therefore, the replication matrix $\boldsymbol{M}$ leaves the $S \times T$ region-by-period indicators unchanged and sums the region-specific covariate columns into the $K$ common covariate columns. Hence,
\[
(\boldsymbol{X}_i^R)'\boldsymbol{M}
=
(\boldsymbol{X}_i^H)'.
\]
Stacking over all $N$ observations gives
\[
\boxed{
\boldsymbol{X}^H
=
\boldsymbol{X}^R\boldsymbol{M}.
}
\]

\section{Results with Heteroskedasticity}

\begin{table}[H]
\centering
\caption{True DID-INT Specification versus Two-Way DID-INT under Heteroskedasticity}
\label{tab:oracle_twoway_hetero}
\resizebox{\textwidth}{!}{
\begin{tabular}{llccccccc}
\toprule
& & \multicolumn{2}{c}{Bias}
& \multicolumn{2}{c}{Monte Carlo SD}
& \multicolumn{2}{c}{RMSE}
& \\
\cmidrule(lr){3-4}
\cmidrule(lr){5-6}
\cmidrule(lr){7-8}
DGP
& $n_{\text{cell}}$
& True
& Two-way
& True
& Two-way
& True
& Two-way
& Relative Efficiency \\
\midrule

\multirow{5}{*}{Homogeneous}
& 25  &  0.043 &  0.034 & 0.762 & 0.849 & 0.763 & 0.849 & 1.241 \\
& 50  & -0.011 & -0.008 & 0.539 & 0.594 & 0.539 & 0.594 & 1.214 \\
& 100 &  0.007 &  0.005 & 0.378 & 0.409 & 0.378 & 0.408 & 1.165 \\
& 200 &  0.007 &  0.009 & 0.284 & 0.317 & 0.284 & 0.317 & 1.245 \\
& 500 & -0.002 & -0.001 & 0.176 & 0.189 & 0.176 & 0.188 & 1.151 \\

\midrule

\multirow{5}{*}{Region-varying}
& 25  &  0.041 &  0.034 & 0.765 & 0.849 & 0.766 & 0.849 & 1.230 \\
& 50  & -0.013 & -0.008 & 0.541 & 0.594 & 0.541 & 0.594 & 1.207 \\
& 100 &  0.005 &  0.005 & 0.380 & 0.409 & 0.380 & 0.408 & 1.156 \\
& 200 &  0.005 &  0.009 & 0.285 & 0.317 & 0.285 & 0.317 & 1.240 \\
& 500 & -0.003 & -0.001 & 0.176 & 0.189 & 0.176 & 0.188 & 1.143 \\

\midrule

\multirow{5}{*}{Time-varying}
& 25  &  0.042 &  0.034 & 0.765 & 0.849 & 0.766 & 0.849 & 1.230 \\
& 50  & -0.010 & -0.008 & 0.543 & 0.594 & 0.543 & 0.594 & 1.198 \\
& 100 &  0.008 &  0.005 & 0.378 & 0.409 & 0.378 & 0.408 & 1.168 \\
& 200 &  0.006 &  0.009 & 0.284 & 0.317 & 0.284 & 0.317 & 1.246 \\
& 500 & -0.002 & -0.001 & 0.176 & 0.189 & 0.176 & 0.188 & 1.151 \\

\bottomrule
\end{tabular}
}
\end{table}

\section{Additional CCC Test Results}
\label{app:additional_ccc_power}

\begin{figure}[H]
    \centering

    \begin{subfigure}[t]{0.47\linewidth}
        \centering
        \includegraphics[width=\linewidth]{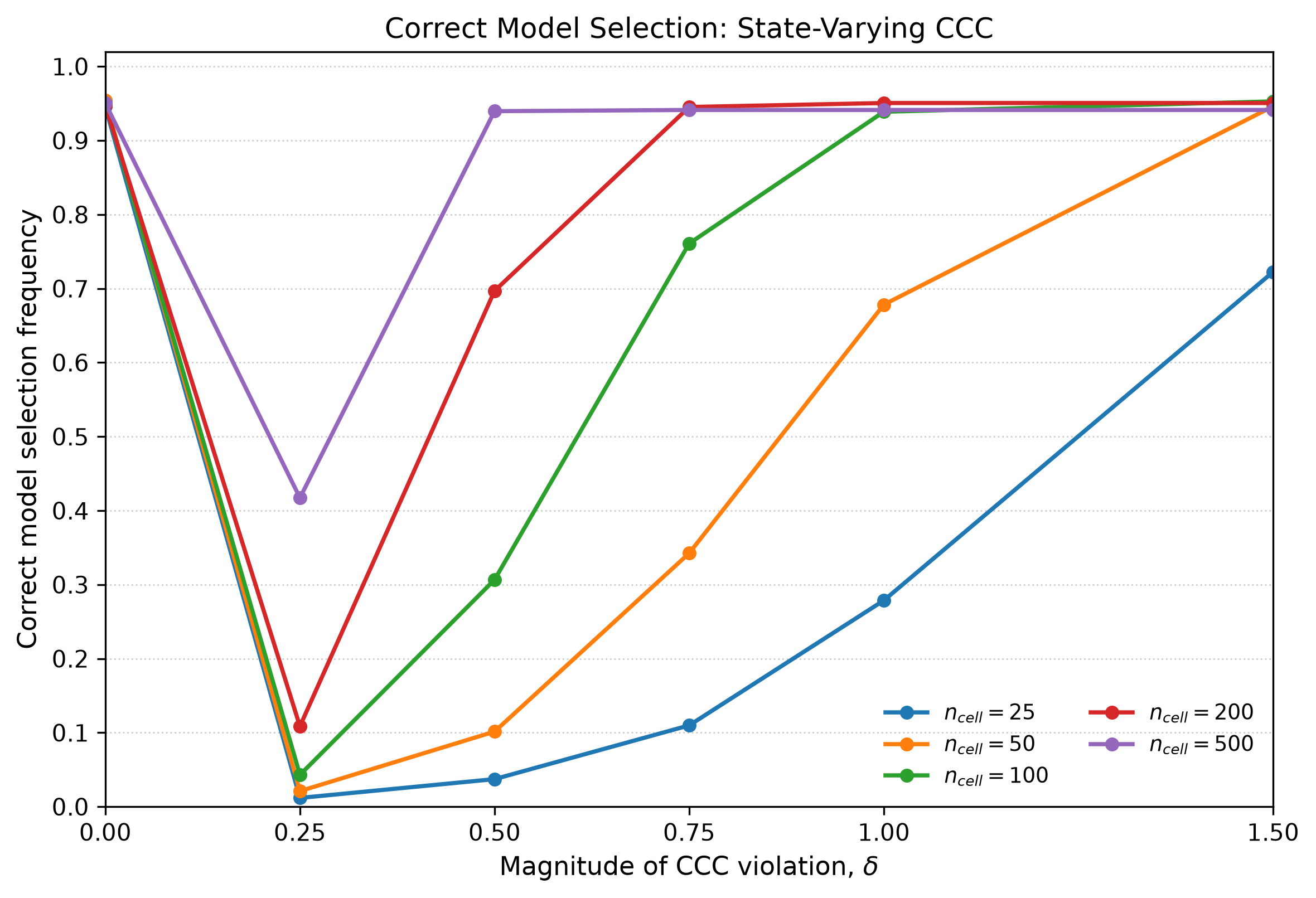}
        \caption{Region-varying DGP}
        \label{fig:correctselection_state}
    \end{subfigure}
    \hfill
    \begin{subfigure}[t]{0.47\linewidth}
        \centering
        \includegraphics[width=\linewidth]{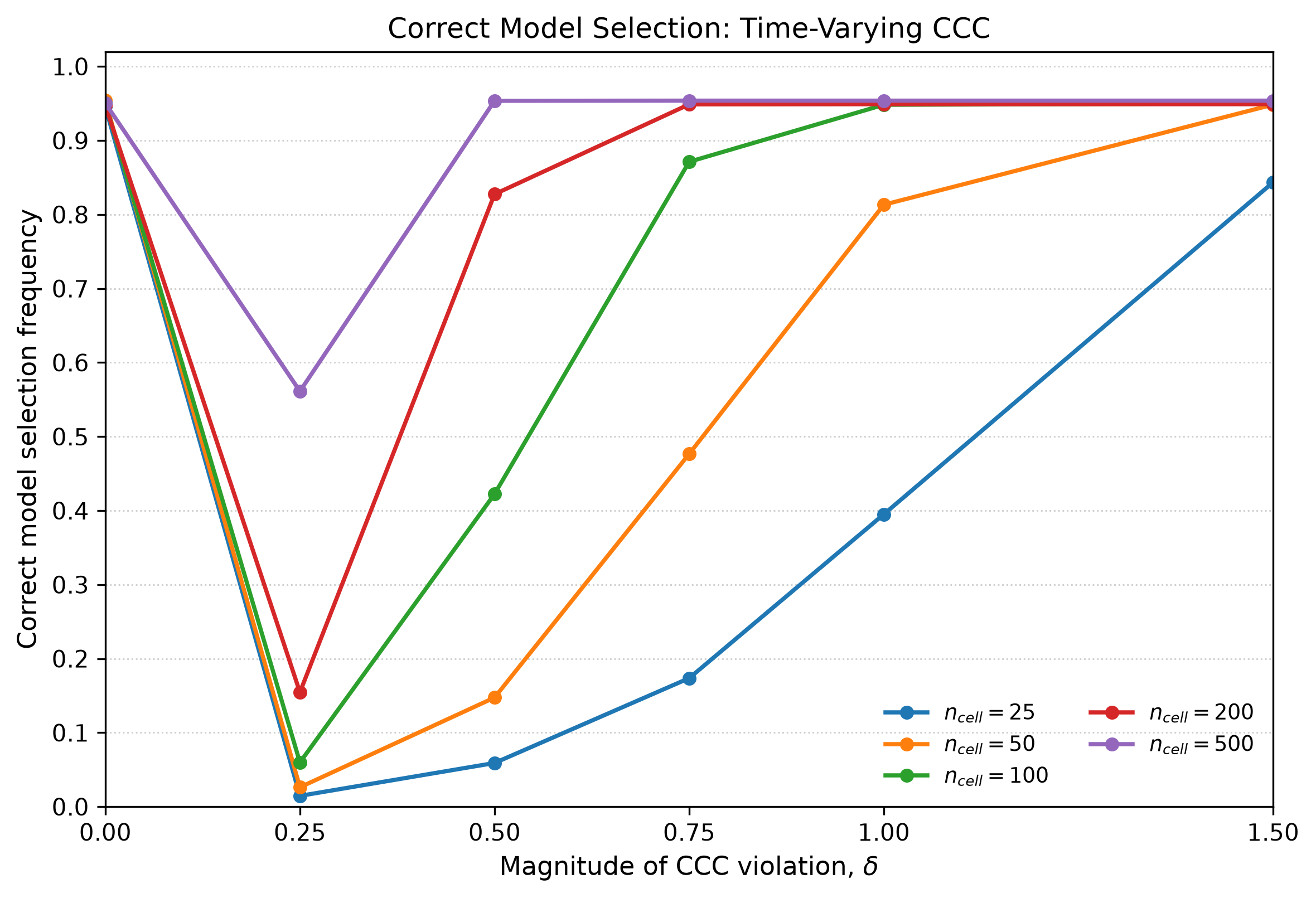}
        \caption{Time-varying DGP}
        \label{fig:correctselection_time}
    \end{subfigure}

    \caption{Frequency of Correct CCC Specification Selection: Region- and Time-Varying DGPs}
    \label{fig:correctselection_appendix}
\end{figure}

Figure \ref{fig:correctselection_appendix} reports the correct model selection frequency for the region-varying and time-varying DGPs. This pattern arises because the CCC test only evaluates the region ($H^0_{R}$) or time nulls ($H^0_{T}$) after rejecting the two-way null ($H^0_{TW}$) in the first stage. When the coefficients vary by region, increased power leads to rejecting $H^0_{R}$ with probability approaching one. The time null $H^0_{T}$ remains true and is therefore falsely rejected only 5 \% of the time. Consequently, the region-varying model is correctly selected 95 \% of the time. Similarly, when the coefficients vary by period, the region null $H^0_{R}$ is true and is falsely rejected 5 \% of the time, while the probability of rejecting the false time null $H^0_{T}$ approaches one. Consequently, the time-varying model is correctly selected approximately 95 \% of the time.

\begin{figure}[H]
    \centering

    \begin{subfigure}[t]{0.47\linewidth}
        \centering
        \includegraphics[width=\linewidth]{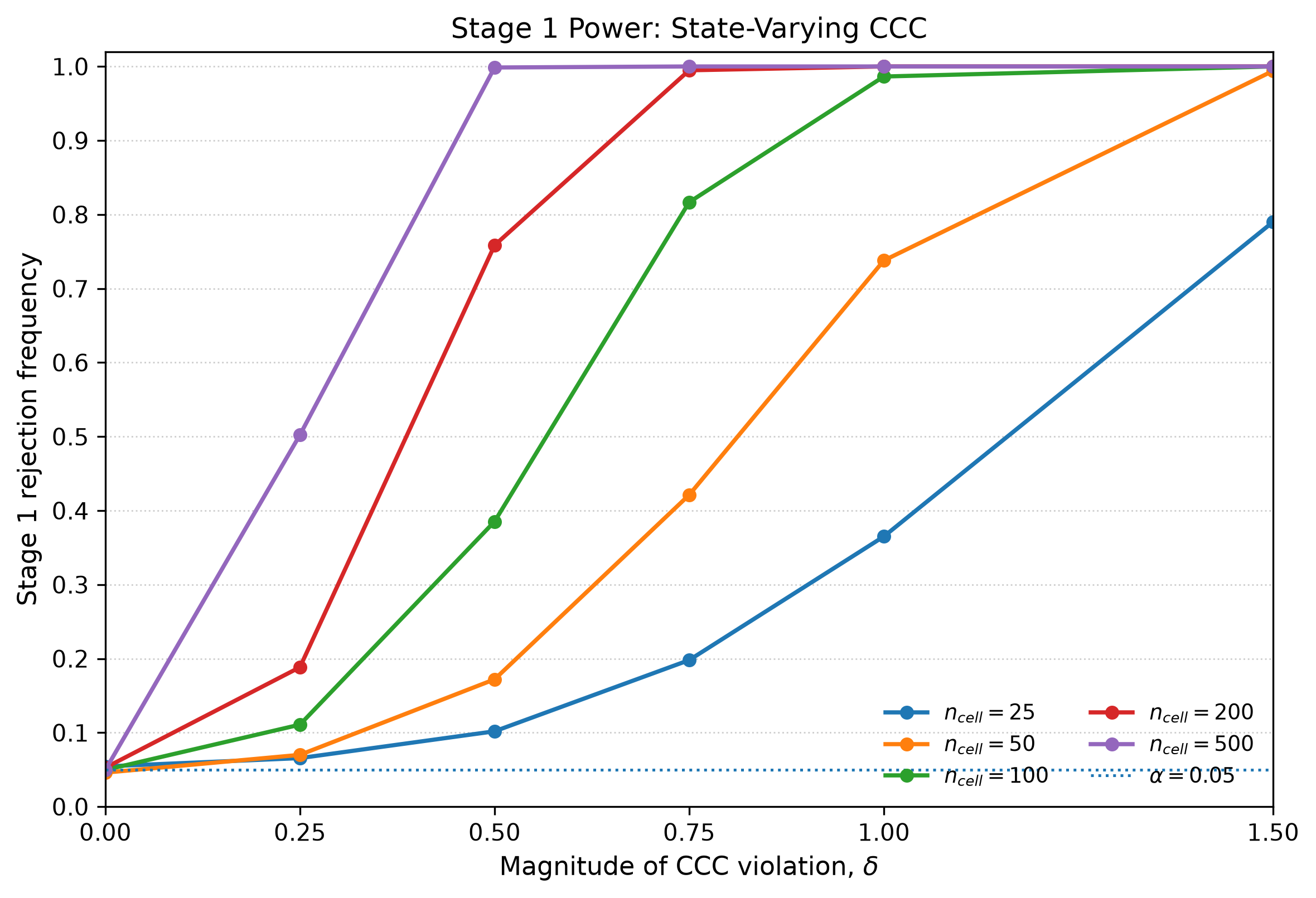}
        \caption{Stage 1: Region-varying DGP}
        \label{fig:stage1_state}
    \end{subfigure}
    \hfill
    \begin{subfigure}[t]{0.47\linewidth}
        \centering
        \includegraphics[width=\linewidth]{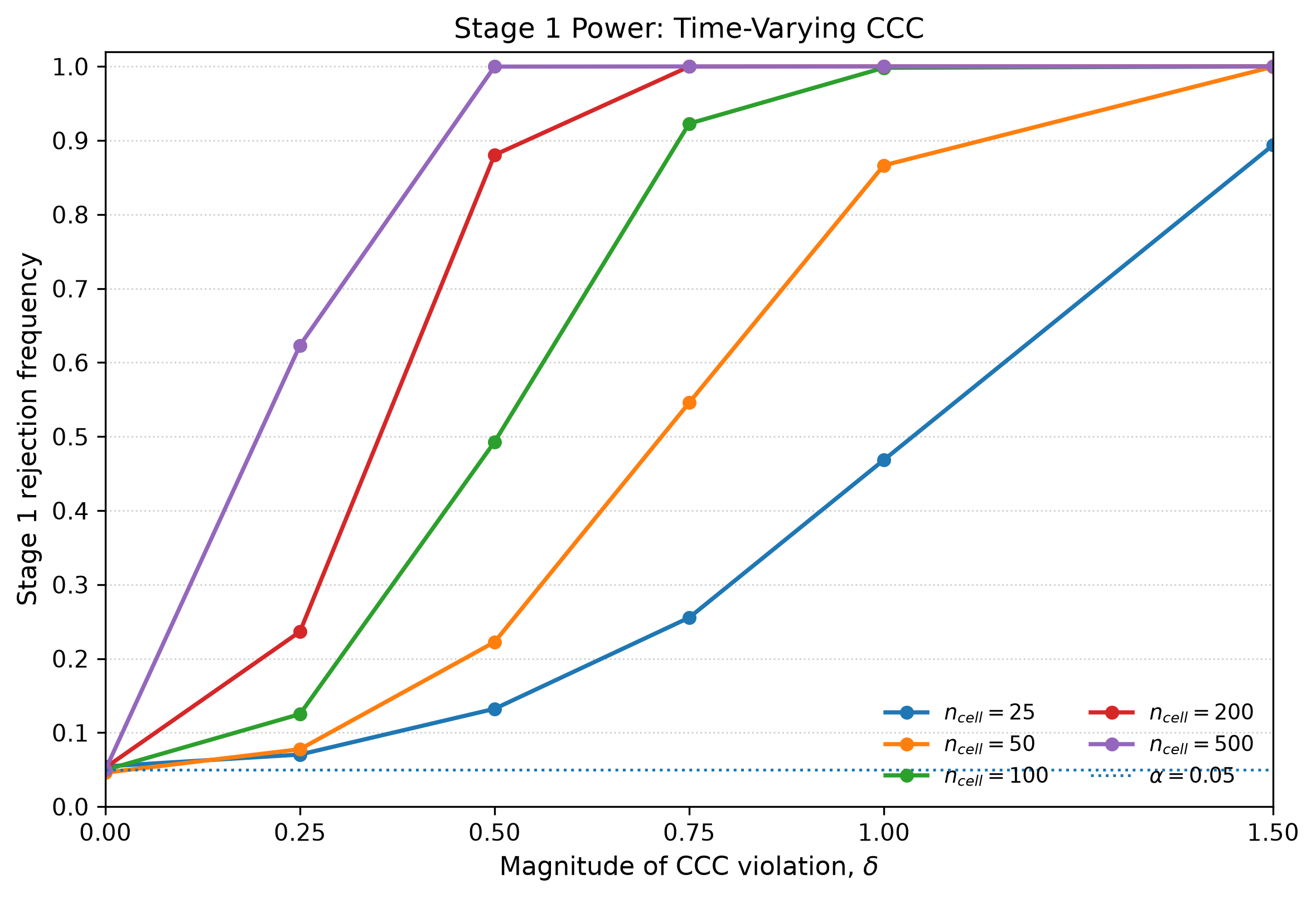}
        \caption{Stage 1: Time-varying DGP}
        \label{fig:stage1_time}
    \end{subfigure}

    \vspace{0.5cm}

    \begin{subfigure}[t]{0.47\linewidth}
        \centering
        \includegraphics[width=\linewidth]{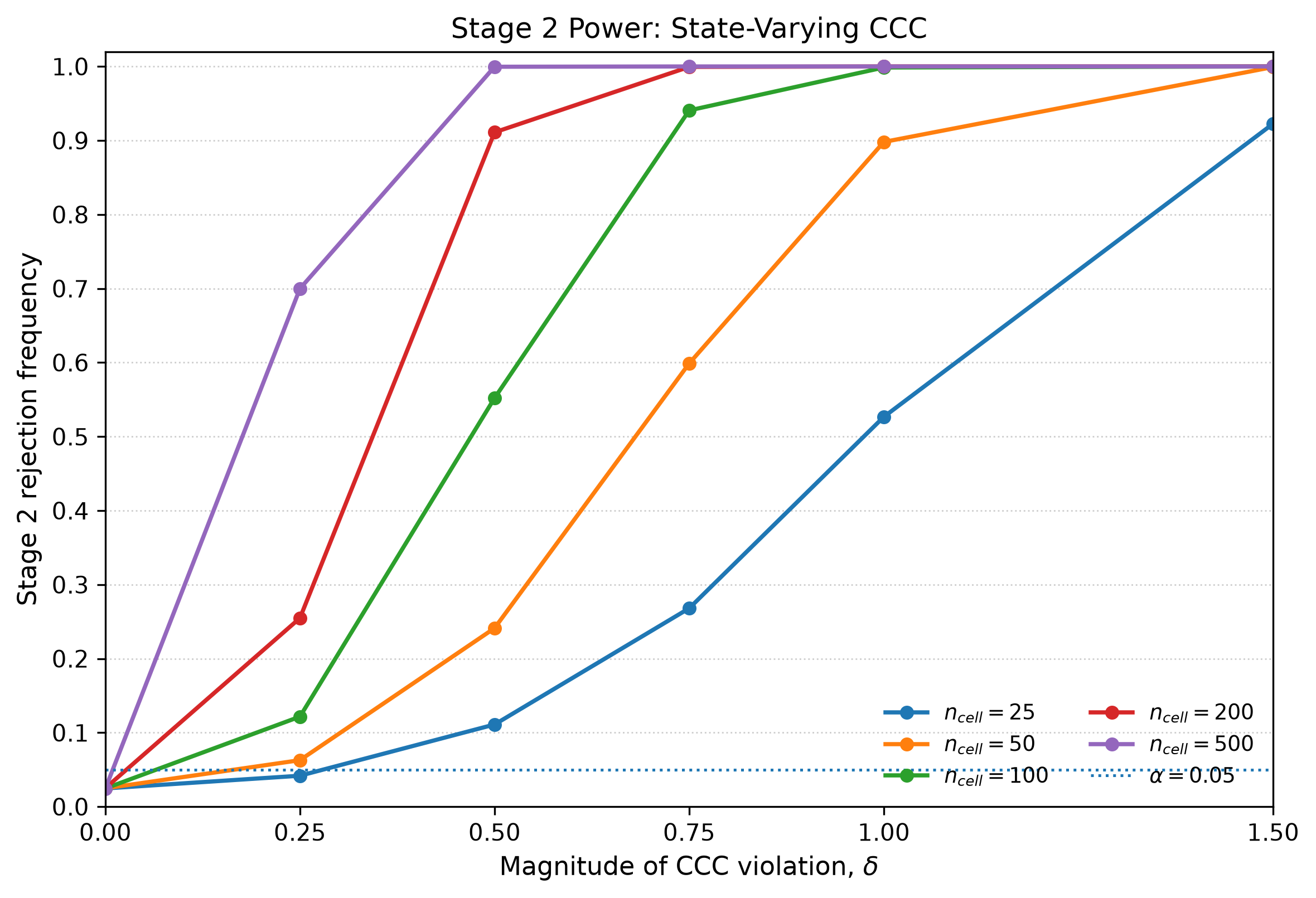}
        \caption{Stage 2: Region-varying DGP}
        \label{fig:stage2_state}
    \end{subfigure}
    \hfill
    \begin{subfigure}[t]{0.47\linewidth}
        \centering
        \includegraphics[width=\linewidth]{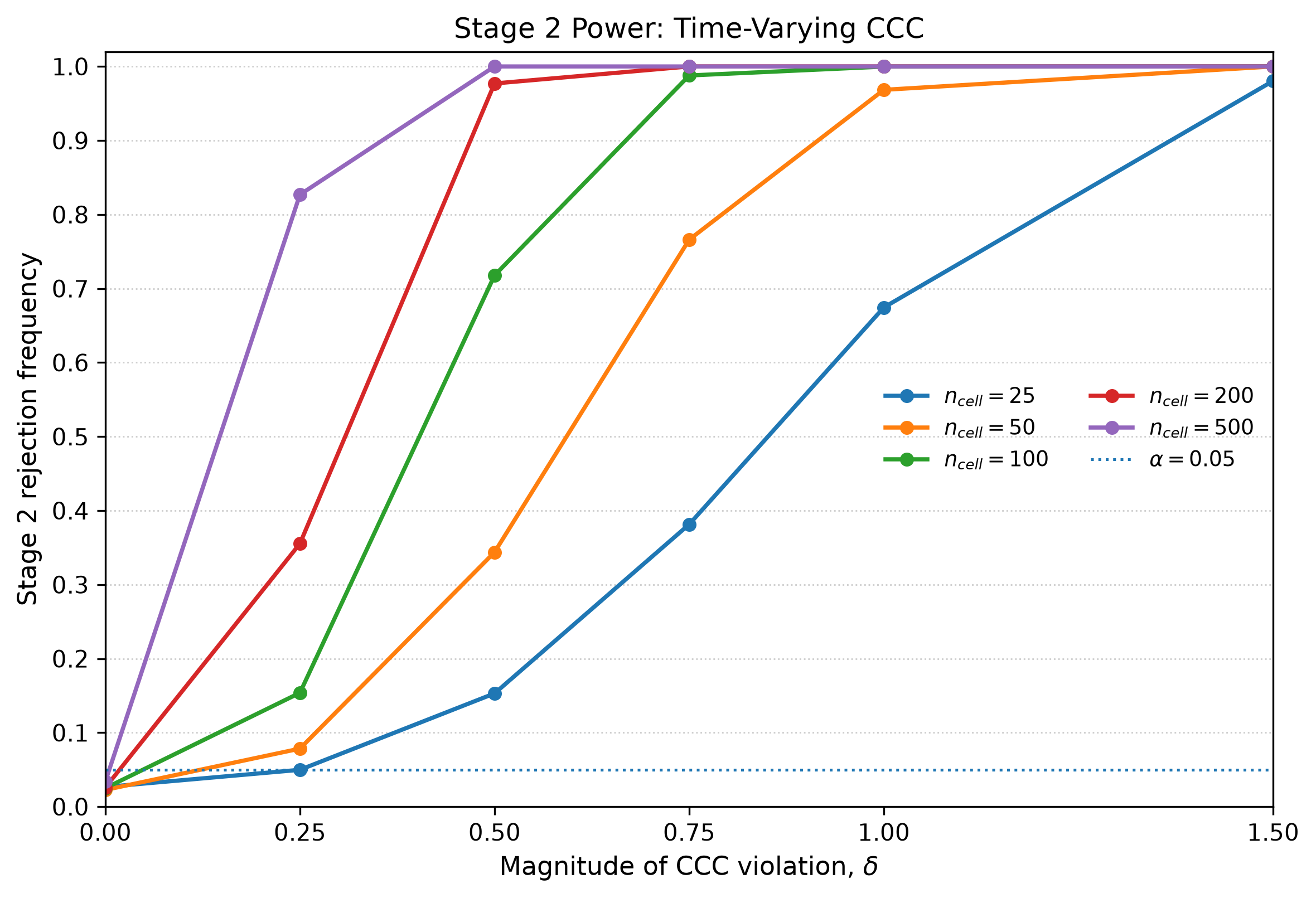}
        \caption{Stage 2: Time-varying DGP}
        \label{fig:stage2_time}
    \end{subfigure}

    \caption{Power of the Sequential CCC Tests: Region- and Time-Varying DGPs}
    \label{fig:ccc_power_appendix}
\end{figure}

Figure \ref{fig:ccc_power_appendix} illustrates the power curves for Stage 1 and Stage 2 CCC tests across the region-varying and time-varying DGPs. The power curves for Stage 1 and Stage 2 display the expected pattern. In Stage 1, the test evaluates whether the two-way null hypothesis $H^0_{TW}$ holds across both DGPs. Since this null is false under each of these DGPs, the probability of rejection approaches one as sample size and degree of CCC violation increase. In Stage 2, the tests determine whether the violation arises from region variation or time variation. For the region-varying DGP, $H^0_{R}$ is false, and for the time-varying DGP, $H^0_{T}$ is false. Consequently, the corresponding rejection probabilities approach one as sample size and degree of CCC violation increase. In both stages, larger CCC violations increase the ability of the tests to detect violations of the null. As a result, the power curves slope upward and approach one as $\delta$ increases. In addition, larger sample sizes allow weaker violations to be detected more reliably, causing the power curves to shift upward and to the left as the number of observations per region-by-period cell increases.

\section{OCC2010 to O*NET Job Zone Harmonization}
\label{sec:occ_harmonization}

To construct the occupational quality measure, I first map the ACS occupation variable, \texttt{occ2010}, to the 2018 Standard Occupational Classification (SOC) system, which is used to link occupations to O*NET Job Zone data. I use the OCC2010-to-SOC mapping and the official Bureau of Labor Statistics 2010-to-2018 SOC crosswalk, while leaving the original ACS \texttt{occ2010} variable unchanged. Since some OCC2010 categories correspond to multiple 2018 SOC occupations, I expand the crosswalk to include a separate entry for each corresponding occupation rather than selecting only one. I then merge the resulting 2018 SOC codes with O*NET Job Zone data. O*NET groups occupations into one of five Job Zones according to the education, experience, and training they generally require. For the main outcome, I classify Zones 1 and 2 as low-Job-Zone occupations and Zones 3, 4, and 5 as high-Job-Zone occupations.

When an OCC2010 category maps to multiple 2018 SOC occupations, I use the Job Zones of all matched occupations to determine its binary classification. If every matched occupation falls in Job Zones 3 through 5, I classify the category as high Job Zone. If every matched occupation falls in Job Zone 1 or 2, I classify it as low Job Zone. If the matched occupations span both high and low Job Zones, I leave the binary classification missing rather than arbitrarily assigning the category to either group. 

Table \ref{tab:occ_harmonization_qc} summarizes the resulting crosswalk. Of the 481 OCC2010 codes observed in the analysis data, 479 have at least one SOC mapping. Among these, 334 map to a single 2018 SOC occupation and 145 map to multiple 2018 SOC occupations. Overall, 431 of the 481 active OCC2010 codes receive a usable binary high or low Job Zone classification. The remaining 50 codes consist of 32 categories that contain both high and low Job Zone occupations, 16 with no corresponding O*NET Job Zone, and two with no occupation mapping.

\begin{table}[htbp]
\centering
\caption{Quality-Control Results for OCC2010 Harmonization}
\label{tab:occ_harmonization_qc}
\begin{tabular}{lr}
\toprule
Result & Number of OCC2010 Codes \\
\midrule
Active codes in the analysis data & 481 \\
Active codes with at least one SOC mapping & 479 \\
One OCC2010 to one 2018 SOC & 334 \\
One OCC2010 to multiple 2018 SOC codes & 145 \\
Exact numeric Job Zone consensus & 375 \\
Binary consensus despite mixed numeric Job Zones & 56 \\
Mixed high and low Job Zones & 32 \\
No O*NET Job Zone match & 16 \\
No occupation mapping & 2 \\
Usable binary consensus values & 431 \\
\bottomrule
\end{tabular}
\end{table}

The resulting OCC2010-level classification is then merged back to the individual-level ACS data using \texttt{occ2010}. The final binary outcome equals one when an individual's OCC2010 category is classified as high Job Zone, equals zero when it is classified as low Job Zone, and remains missing when no binary consensus can be established. Among the employed private-sector observations used in the analysis, 1,923,354 of 2,164,444 observations have a usable High Job Zone classification. I used generative AI tools (ChatGPT) to help verify portions of the occupation-code crosswalk and check the consistency of mappings I had originally constructed. I subsequently reviewed the results to ensure that the final mappings were correct.

\section{List of Clean control states}
Obtained from \cite{avery2021ban}

\begin{table}[htbp]
\centering
\caption{Classification of States by Private-Sector Ban-the-Box Exposure}
\label{tab:control_state_classification}
\footnotesize
\setlength{\tabcolsep}{4pt}
\renewcommand{\arraystretch}{0.90}
\begin{tabular}{lll}
\toprule
Clean States
& Statewide Private-Sector BTB
& Local Private-Sector BTB Only \\
\midrule
Alabama        & California    & Florida \\
Alaska         & Colorado      & Iowa \\
Arizona        & Connecticut   & Missouri \\
Arkansas       & Hawaii        & New York \\
Delaware       & Illinois      & Pennsylvania \\
Georgia        & Maine         & Texas \\
Idaho          & Maryland      & \\
Indiana        & Massachusetts & \\
Kansas         & Minnesota     & \\
Kentucky       & New Jersey    & \\
Louisiana      & New Mexico    & \\
Michigan       & Oregon        & \\
Mississippi    & Rhode Island  & \\
Montana        & Vermont       & \\
Nebraska       & Washington    & \\
Nevada         &               & \\
New Hampshire  &               & \\
North Carolina &               & \\
North Dakota   &               & \\
Ohio           &               & \\
Oklahoma       &               & \\
South Carolina &               & \\
South Dakota   &               & \\
Tennessee      &               & \\
Utah           &               & \\
Virginia       &               & \\
West Virginia  &               & \\
Wisconsin      &               & \\
Wyoming        &               & \\
\bottomrule
\end{tabular}
\end{table}

\clearpage

\section{Robustness checks}

\begin{table}[htbp]
\centering
\caption{DID-INT Estimates Across Alternative Covariate Specifications}
\label{tab:didint_alternative}
\begin{tabular}{lccc}
\toprule
& Homogeneous & Region-Varying & Time-Varying \\
\midrule

\multicolumn{4}{l}{\textit{Panel A: Log Wage}} \\[0.3em]

ATT
& 0.0203**
& 0.0184*
& 0.0212** \\
& (0.0100)
& (0.0105)
& (0.0102) \\

Observations
& 2,164,444
& 2,164,444
& 2,164,444 \\

\addlinespace

\multicolumn{4}{l}{\textit{Panel B: High Job Zone}} \\[0.3em]

ATT
& 0.0074**
& 0.0064*
& 0.0075** \\
& (0.0036)
& (0.0034)
& (0.0031) \\

Observations
& 1,923,354
& 1,923,354
& 1,923,354 \\

\bottomrule
\end{tabular}

\vspace{0.5em}
\begin{minipage}{0.90\textwidth}
\footnotesize
\textit{Notes:} Standard errors are reported in parentheses and are calculated using the cluster jackknife.
*** $p<0.01$, ** $p<0.05$, * $p<0.10$.
\end{minipage}
\end{table}

\end{document}